\documentclass[floatfix,aps,prd,preprintnumbers,footinbib,bibnotes,superscriptaddress,letterpaper,twocolumn]{revtex4-1}
\usepackage{graphicx}  
\usepackage{dcolumn}   
\usepackage{bm}        
\usepackage{amssymb}   
\usepackage{xcolor}   
\def\Dslash{{\cal D}\!\!\!\!/}

\usepackage{hyperref}

\begin{document}
\title{Dynamical Generation of Elementary Fermion Mass: First Lattice Evidence}\vspace*{0.4cm}
\newcommand{\Frank}{\affiliation{Goethe U. Frankfurt, Inst.\ f.\ Theor.\ Phys, Max-von-Laue-Str. 1, 60438 Frankfurt am Main, Germany}}
\newcommand{\ToV}{\affiliation{Dipartimento di Fisica, Universit\`a di Roma Tor Vergata, Via della Ricerca Scientifica, 00133 Roma, Italy}}
\newcommand{\Fermi}{\affiliation{Centro Fermi - Museo Storico della Fisica e Centro Studi e Ricerche Enrico Fermi, 00184 Rome, Italy}}
\newcommand{\INFN}{\affiliation{INFN, Sezione di Roma Tor Vergata, Via della Ricerca Scientifica, 00133 Roma, Italy}}
 \newcommand{\HISKP}{\affiliation{HISKP (Theory), Rheinische Friedrich-Wilhelms-Universit\"at Bonn, Nussallee 14-16, 53115 Bonn, Germany}}

\author{S.~Capitani}
\Frank
\author{P.~Dimopoulos}
\Fermi
\ToV
\author{R.~Frezzotti}
\ToV
\INFN
\author{M.~Garofalo}
\INFN
\author{B.~Kostrzewa}
\HISKP
\author{F.~Pittler}
\HISKP
\author{G.C.~Rossi}
\Fermi
\ToV
\INFN
\author{C.~Urbach}
\HISKP

                             

\vspace{0.2cm}
\begin{abstract}

Using lattice simulations we demonstrate from first principles the existence of a non-perturbative mechanism for elementary particle mass generation in models with gauge fields, fermions and scalars, if an exact invariance forbids power divergent fermion masses and fermionic chiral symmetries broken at UV scale are maximally restored. We show that in the Nambu-Goldstone phase a fermion mass term, unrelated to the Yukawa operator, is dynamically generated. In models with electro-weak interactions weak boson masses are also generated opening new scenarios for beyond the Standard Model physics.
\end{abstract}

\maketitle


\vspace{0.4cm}

\noindent {\em 1: Introduction}\hspace{0.1cm}--\hspace{0.1cm}In spite of its impressive phenomenological success, the 
Standard Model (SM) of particle physics   
is believed to represent only an effective low energy theory,  
as it neither accounts for dark matter and quantum aspects of gravity nor
provides enough CP-violation for baryogenesis. Fermions and electro-weak (EW)  
bosons masses are described in terms 
of a well established symmetry breaking pattern~\cite{We67}, 
but the SM is by construction unable to shed light 
on the problems of EW scale naturalness~\cite{THOOFT} 
and fermion mass hierarchy~\cite{FrogNiel}.

There have been numerous attempts to build phenomenologically viable models 
where the EW scale is stable under quantum corrections, either because the basic 
theory enjoys an approximate symmetry larger than in the SM or because EW and 
Higgs mass scales are related in a fixed way to a fundamental dynamical scale.
Examples of the first kind are the many models based on 
SuperSymmetry (SUSY)~\cite{Nilles:1983ge,WeinbergBook2000} which,
besides having the problem of explaining SUSY
breaking, are presently disfavoured owing to the experimental 
exclusion of SUSY particles with mass up to a few TeV~\cite{SUSYexp}. 
Approaches of the second kind assume the existence of some new interaction that gets strong around or above the EW scale, and of new fermionic particles. 
The original TechniColor idea~\cite{Weinberg:1975gm,Susskind:1978ms} 
could account for the EW boson masses, but attempts to understand heavy 
fermion masses in Extended TechniColor 
(ETC) models~\cite{DimoSuss1979,EichLane1980} face severe problems to comply 
with experimental constraints on flavour changing neutral currents, even in 
subsequently developed Walking ETC models~\cite{Holdom:1984sk,Yamawaki:1985zg,AppeWije1986,PhysRevD.36.568}. 
Other ways to comply with experimental constraints and possibly 
address the flavour hierarchy problem are represented by the partially 
composite Higgs models~\cite{Kaplan:1991dc,PaniWulz} and by models  
with extra dimensions~\cite{ArkaniHamed:1998rs,Randall:1999ee,Gherghetta:2000qt}.

At variance with previous attempts, a novel intrinsically non-perturbative (NP) 
mechanism for {\em elementary fermion mass generation} was 
{\em conjectured} in~\cite{Frezzotti:2014wja}.
This mechanism is expected to be at work in non-Abelian gauge models 
where (as usual) 1) chiral transformations acting
on fermions and scalars are exact symmetries, but (deviating from common assumptions)
2) purely fermionic chiral symmetries undergo an explicit breaking at the UV scale.
When bare parameters are ``naturally'' tuned so as to minimize fermion chiral breaking,
in the effective Lagrangian (EL - generating functional of proper vertices) no Yukawa term occurs, but operators of NP origin that violate fermion chiral symmetries, 
among which a fermion mass term, should appear, if the 
scalar potential is such that the theory lives in its Nambu--Goldstone (NG) phase.
Upon introducing EW interactions, 
the same mechanism also yields massive $W^{\pm}$,
$Z^0$ bosons and a composite Higgs boson in the $W^+W^-/Z^0Z^0$ 
channel~\cite{FR18}. 

In this paper we employ lattice simulations (lacking
analytical methods) to provide evidence 
from first principles for 
the occurrence of the NP mass generation mechanism
of Ref.~\cite{Frezzotti:2014wja}, within the 
simplest $d=4$ model 
where it could take place.

\vspace{0.4cm}
\noindent {\em 2: Mass generation in a toy model}\hspace{0.1cm}--\hspace{0.1cm}
The Lagrangian of the ``toy'' (yet non-trivial) model 
of interest here~\cite{Frezzotti:2014wja} reads
\begin{equation}
\hspace{-.06cm}
{\cal L}_{\rm{toy}} \! =\! {\cal L}_{k}(Q,A,\Phi)\!+\!{\cal V}(\Phi)\!+\!{\cal L}_{W}(Q,A,\Phi)\!+\! {\cal L}_{Y}(Q,\Phi) \label{TOYLAG}
\end{equation}
with ${\cal L}_{k}$ and ${\cal V}$ representing standard kinetic terms and scalar potential. 
${\cal L}_{\rm{toy}}$ includes an SU(3) gauge field, $A_\mu$,  
with bare (renormalized) coupling $g_0$ ($g_s$), a Dirac doublet, $Q=(u,d)^T$, transforming as a triplet under SU(3) and a complex scalar doublet, $\varphi =(\varphi_0 + i\varphi_3, -\varphi_2 + i\varphi_1)^T$, singlet under SU(3). For the latter we use the $2\times 2$ matrix notation $\Phi = [\,\varphi\,|\! -i\tau^2\varphi^*]$. The model has an UV cutoff $\Lambda_{UV} \sim b^{-1}$ and includes 
a Yukawa term, ${\cal L}_{Y}(Q,\Phi) = \eta 
\big{(} \bar Q_L\Phi Q_R+\bar Q_R \Phi^\dagger Q_L\big{)}$, as well as a non-standard term
\begin{equation} 
\hspace{-.2cm}{\cal L}_{W}(Q,\!A,\!\Phi) \! = \! \frac{b^2}{2}\rho\big{(} \bar Q_L{\overleftarrow{\cal D}}_{\!\mu}\Phi {\cal D}_{\!\mu} Q_R + \bar Q_R \overleftarrow{\cal D}_{\!\mu} \Phi^\dagger {\cal D}_{\!\mu} Q_L\big{)} \, . \label{LWIL}
\end{equation}
The latter is a $\Lambda_{UV}^{-2} \times d=6$ operator that
leaves the model power-counting renormalizable~\cite{Frezzotti:2014wja}, like it 
happens for the Wilson term in lattice QCD~\cite{WilsonLQCD,KarsSmit}.
Neither ${\cal L}_{W}$ nor ${\cal L}_{Y}$ are invariant under purely 
fermionic chiral transformations.

Among other symmetries, the Lagrangian~(\ref{TOYLAG}) is
invariant under the global transformations ($\Omega_{L/R} \in {\mbox{SU}}(2)$)
\begin{eqnarray}
\hspace{-.5cm}&&\chi_L\times \chi_R =  [\tilde\chi_L\times (\Phi\to\Omega_L\Phi)]\times [\tilde\chi_R\times (\Phi\to\Phi\Omega_R^\dagger)] \, ,\label{CHIL}\\
\hspace{-.5cm}&&\tilde\chi_{L/R} : Q_{L/R}\rightarrow\Omega_{L/R} Q_{L/R} \, ,\quad \bar Q_{L/R}\rightarrow \bar Q_{L/R}\Omega_{L/R}^\dagger \, . \label{GTWT}
\end{eqnarray}
No power divergent fermion masses can be generated as a term like $\Lambda_{UV}  (\bar Q_L Q_R+ \bar Q_R Q_L)$ is not $\chi_L\times\chi_R$ invariant.

\vspace{0.2cm}
\noindent {\em 2A: Wigner phase and fermion chirality 
restoration}\hspace{0.1cm}--\hspace{0.1cm}${\cal L}_{\rm{toy}} $ is not invariant under the
purely fermionic chiral transformations $\tilde\chi_L\times\tilde\chi_R$.
However, as shown in~\cite{Frezzotti:2014wja}, in the phase
with positive renormalized squared scalar mass
($\hat\mu^2_\phi>0$), where the $\chi_L\times\chi_R$ symmetry is realized 
{\it \`a la} Wigner, a critical value of the Yukawa coupling, $\eta_{cr}$, 
exists at which (up to O($b^2$) corrections) 
the effective Yukawa term vanishes. 
The renormalized Schwinger--Dyson equations (SDE), say for the 
$\tilde\chi_{L}$ transformations read (no sum over $i=1,2,3$, $|x| \gg b$)
\begin{eqnarray}
\hspace{-.3cm}&&\partial_\mu \langle Z_{\tilde J} \tilde J^{L\, i}_\mu(x) \,\hat O^i(0)\rangle 
\!=\! (\bar\eta \!-\!\eta) \langle \big{(} 
 \tilde{D}_L^i(x) \hat O^i(0)\rangle \!+\! {\mbox{O}}(b^2) \, ,\label{CTLTI} \\
\hspace{-.3cm}&& \tilde{D}_L^i = \bar Q_L\frac{\tau^i}{2}\Phi Q_R\!-\!\bar Q_R\Phi^\dagger\frac{\tau^i}{2}Q_L 
\label{DLJ} \, ,
\end{eqnarray}
with $\hat O^i$ any $\tilde\chi_L$ covariant operator.
The current $\tilde J^{L\, i}_\mu$ is given in~\cite{Frezzotti:2014wja}.
Owing to parity, similar SDE hold for $\tilde\chi_R$. 
At the value $\eta=\eta_{cr}(g_0^2,\rho,\lambda_0)$ that solves the equation 
$\eta - \bar\eta(\eta; g_0^2, \rho, \lambda_0)=0$ the SDE take the form of 
Ward--Takahashi identities (WTI) and the fermionic chiral 
transformations $\tilde\chi_L\times\tilde\chi_R$ become 
symmetries~\cite{Bochicchio:1985xa} 
of the model~(\ref{TOYLAG}) up to O($b^2$) terms. 
In~(\ref{CTLTI}) the dimensionless coefficient 
$\bar\eta = \bar\eta(\eta; g_0^2, \rho, \lambda_0)$ stems from the mixing of the 
$\tilde\chi_L$-variations of ${\cal L}_W$ and ${\cal L}_Y$,
while $Z_{\tilde J}$ is a multiplicative renormalization factor (free of logarithmic UV divergencies at $\eta=\eta_{cr}$). 
Symmetries constrain the expression 
of the EL so that in the Wigner phase its $d=4$ piece is analogous 
in form to ${\cal L}_{\rm toy}$, namely
\begin{eqnarray}
\hspace{-.2cm}&&\Gamma_4^{Wig}\equiv \Gamma_{\hat \mu_\phi^2 > 0}  = \Gamma_k(A,Q,\Phi)+
\label{L4Wig}\\
\hspace{-.2cm}&&
 \quad+ [\eta -\bar\eta(\eta; g_0^2,\rho,\lambda_0)] \big{(} \bar Q_L\Phi Q_R+\bar Q_R \Phi^\dagger Q_L\big{)}+
\hat {\cal V}(\Phi)\, ,\nonumber\\
\hspace{-.2cm}&&\Gamma_k\!=\!\frac{1}{4}(F F)\!+\!\bar Q_L\Dslash Q_L\!+\!\bar Q_R\Dslash \,Q_R \!+\!\frac{1}{2}{\mbox{Tr}}\big{[}\partial_\mu\Phi^\dagger\partial_\mu\Phi\big{]}\, .\label{KIN}
\end{eqnarray}

The Wigner phase is thus well suited to determine the critical value of $\eta$, 
where the effective 
Yukawa term disappears from Eq.~(\ref{L4Wig}). 
We stress that, neglecting O($b^2$) artifacts, $\bar{\eta}$ and $\eta_{cr}$ 
are independent of the subtracted scalar mass $\hat\mu_\phi^2$ (see 
Supplemental Material \cite{supplemental_material}, Sect.~III) 
and thus equal in the Wigner and NG phase.
From Eq.~(\ref{CTLTI}) $\eta_{cr}$ can be determined, 
e.g.\ as the value of $\eta$ where (no sum over $i=1,2,3$, $x \neq 0$)
\begin{eqnarray}
\hspace{-.4cm}&& 
\frac{\partial_\mu\langle \tilde A_{\mu}^i(x) \tilde D^i_P(0)\rangle}{\langle  \tilde D^i_P(x) \tilde  D^i_P(0) \rangle} =0\, , \quad \tilde A_{\mu}^i\!=\!\tilde J^{L\, i}_\mu\!-\!\tilde J^{R\, i}_\mu \, ,
\label{RAWI}\\
\hspace{-.4cm}&&\tilde D^{i}_P\! =\bar Q_L\left\{\Phi,\frac{\tau^{i}}{2}\right\}Q_R  - \bar Q_R \left\{\frac{\tau^{i}}{2},\Phi^\dagger\right\}Q_L \, .
\label{DSI}
\end{eqnarray}
An equivalent, but statistically less noisy, condition is discussed
below ({\em ``Lattice study and results''}). We stress that the
existence of an $\eta_{cr}$, where the $\tilde\chi_{L}\times\tilde\chi_{R}$ 
transformations become symmetries of the theory (up to O($b^2$)),  
is a general property of the Wigner phase independently 
of the specific form of ${\cal L}_W$. 
Owing to renormalizability and universality, changing ${\cal L}_W$ 
would only modify the values of $\eta_{cr}$, $Z_{\tilde{J}}$ and O($b^2$) artifacts.

\vspace{0.2cm}
\noindent {\em 2B: Nambu--Goldstone phase and ``NP anomaly''}\hspace{0.1cm}--\hspace{0.1cm}Most interesting is the case $\hat\mu^2_\phi<0$ where 
${\cal V}(\Phi)$ has a double-well shape with 
{\mbox{$\langle : \! \Phi^\dagger\Phi \!: \rangle= v^2 1\!\!1\, , v\neq 0$}}, so that the 
$\chi_L\times\chi_R$ symmetry is realized {\it \`a la} NG. 
In large volume under any, even infinitesimal, symmetry breaking perturbation 
the $\chi_L\times\chi_R$ symmetry will be spontaneously broken to SU(2)$_V$. 
Moreover, 
at $\eta=\eta_{cr}$ residual O($b^2v$) 
$\tilde\chi_L\times\tilde\chi_R$ violating terms will polarize the degenerate vacuum 
resulting from the spontaneous $\tilde\chi_L\times\tilde\chi_R$ symmetry breaking 
brought about by strong interactions.

Realization of the $\chi_L\times\chi_R$ invariance {\it \`a la} NG has 
a key impact on low energy physics. Three elementary massless Goldstone   
bosons appear in the spectrum which need to be included in the EL.
In Ref.~\cite{Frezzotti:2014wja}
it was argued that at $\eta=\eta_{cr}$ the EL describing the model~(\ref{TOYLAG}) 
should include $\tilde\chi_L\times\tilde\chi_R$ violating terms of NP origin, 
among which a fermion mass term. 

This conjecture can be checked by studying the 
SDE associated with the $\tilde\chi_L\times\tilde\chi_R$ transformations (see 
e.g.\ Eq.~(\ref{CTLTI})) in the NG phase. Exploiting parity invariance, we more conveniently 
evaluate the effective PCAC mass from the axial SDE 
(no sum over $i=1,2,3$, $x_0 \gg b$) 
\begin{equation}
\hspace{-.4cm}
\frac{Z_{\tilde A}}{Z_{P}} \, m_{AWI}
\!\equiv\! \frac{Z_{\tilde A} \sum_{\bf x}\partial_0\langle \tilde A_{0}^i(x) P^i(0)\rangle}{2Z_{P} \sum_{\bf x} \langle P^i(x)P^i(0)\rangle} \Big{|}_{\eta_{cr}}\!, \;\; P^i\!\!=\!\bar Q\gamma_5 \frac{\tau^i}{2}Q\, , \!\!
\label{MAWIR}
\end{equation} 
where $Z_{\tilde A} = Z_{\tilde J}|_{\eta_{cr}}$ and $Z_{P}$ are renormalization factors.
If NP $\tilde\chi_L\times\tilde\chi_R$ violating terms were absent in 
the critical EL, one should find $\frac{Z_{\tilde A}}{Z_{P}}m_{AWI} \!\to\!~0$
as $b \!\sim \!\Lambda_{UV}^{-1}\! \to\!~0$. Lattice simulations (see {\em``Lattice study and 
results''}) show that  $\frac{Z_{\tilde A}}{Z_{P}}m_{AWI}$ is not vanishing 
in the continuum limit.

In the NG phase the EL describing the model~(\ref{TOYLAG}) 
is expressed in terms of {\em effective} fermion, gauge and scalar fields. 
The latter, in view of $v \neq 0$, are conveniently rewritten~\cite{EffThParam} 
introducing Goldstone ($\zeta_{1,2,3}$) and massive ($\zeta_0$) scalar fields in the form 
\begin{equation}
\Phi = R U\; , \quad R= (v+ \zeta_0)\; , \quad  U = \exp [i v^{-1} \tau^k\zeta_k ] \, ,
\label{ULAM}
\end{equation}
where $U$ is a {\em dimensionless} field transforming as $U \rightarrow \Omega_L U \Omega_R^\dagger$ 
under $\chi_L \times \chi_R$. The non-analytic field $U$ only makes sense if $v\neq 0$.
At $\eta=\eta_{cr}$ the $d\leq 4$ EL sector reads~\cite{Frezzotti:2014wja} 
(see Eq.~(\ref{L4Wig}))
\begin{eqnarray}
\hspace{-.0 cm}&&\Gamma_4^{NG} \! =  c_2 \Lambda_S^2 {\mbox{Tr}}(\partial_\mu U^\dagger\partial_\mu U)  + c_1 \Lambda_S [\bar Q_L U Q_R + {\rm h.c.}]  + \nonumber \\
\hspace{-.0 cm}&& + \tilde c \, \Lambda_S R {\mbox{Tr}}(\partial_\mu U^\dagger\partial_\mu U)
 + \Gamma_{\hat \mu_\phi^2 < 0} + {\rm O}(\Lambda_S^2/v^2) \; , \label{L4NG}
\end{eqnarray}
where $\Lambda_S$ is the renormalization group invariant (RGI) 
scale and the term $\propto c_1$ 
describes the NP breaking of $\tilde\chi_L \times \tilde\chi_R$.
Upon expanding $U$ around the identity one gets 
\begin{equation}
c_1 \Lambda_S [\bar Q_L U Q_R + \bar Q_R U^\dagger Q_L] =c_1 \Lambda_S \bar Q Q [1 
+{\mbox{O}}(\zeta/v)]\, ,
\label{NPM}
\end{equation} 
thus a fermion mass term plus a host of more complicated, non-polynomial 
$\bar Q\!-\!\zeta_{1,2,3}\,'s\!-\!Q$ vertices.

The NP term~(\ref{NPM}) was conjectured in~\cite{Frezzotti:2014wja} to arise 
dynamically in the EL~\ref{L4NG} of the theory~(\ref{TOYLAG}) 
from the interplay of strong interactions and 
the breaking of $\tilde\chi_L \times \tilde\chi_R$ at the UV scale. 
The occurrence of this $\tilde\chi_L \times \tilde\chi_R$ violating 
term in the EL 
is essential to account for the non-zero value we find for $Z_{\tilde A}\sum_{\bf x}\partial_0\langle \tilde A_{0}^i(x) P^i(0)\rangle|_{\eta=\eta_{cr}}$ (Eq.~(\ref{MAWIR})) despite the fact that the operators $\tilde A_{0}^i$ and 
$P^i$ transform differently under $\tilde\chi_L \times \tilde\chi_R$.
The coefficient $c_1$ in Eq.~(\ref{NPM}) has been argued
in~\cite{Frezzotti:2014wja} to be an O$(g_{s}^4)$ 
odd function of $\rho$. 
As for its dependence on the scalar squared mass, 
$c_1$ is expected to stay finite in the phenomenologically
interesting limit $-\hat\mu_\phi^2 \gg \Lambda_S^2$~\cite{Frezzotti:2014wja,FR18} 
and be non-zero only for $\hat\mu_\phi^2 <0$. Indeed in our interpretation
the NP term~(\ref{NPM}) arises from the spontaneous
$\tilde\chi_L \times \tilde\chi_R$ breaking which
is effective only in the NG phase where the degenerate vacuum gets
polarized by residual O($b^2 v$) $\tilde\chi_L \times \tilde\chi_R$ breaking
effects.

A proper understanding of the NP terms in the expression~(\ref{L4NG})
requires considering what is the natural extension of the 
$\tilde\chi_L \times \tilde\chi_R$ symmetry in the presence of EW 
interactions~\cite{FR18}. In that context one finds that maximal restoration of 
the chiral fermion symmetry entails the
vanishing of the coefficient $\tilde{c}$ in~(\ref{L4NG}), leaving only the NP terms
with coefficients $c_1$ and $c_2$, which are responsible for the dynamical
mass of fermions and weak bosons, respectively~\cite{Frezzotti:2014wja,FR18}. 

At $\eta=\eta_{cr}$ all the NP terms 
occurring in the r.h.s.\ of the $\tilde\chi_L \times \tilde\chi_R$ SDE 
must be RGI, as the l.h.s.\ of the SDE is. Conservation up to O($b^2$) of the $Z_{\tilde J} \tilde J_{\mu}^{L(R)i}$
currents makes the l.h.s.\ of the $\tilde\chi_L \times \tilde\chi_R$ SDE scale 
invariant independently of $\hat\mu_\phi^2$. 

The full NG phase EL, $\Gamma^{NG} \supset \Gamma_4^{NG}$
contains of course an infinite number of local terms of arbitrarily 
high dimension, among which terms of NP origin that break the  
$\tilde\chi_L \times \tilde\chi_R$ symmetry. The occurrence of these
NP and RGI terms in the EL will be referred to as a {\em ``NP anomaly''} 
in the restoration of $\tilde\chi_L \times \tilde\chi_R$ symmetry. 

\vspace{0.4cm}
\noindent {\em 3: Lattice study and results}\hspace{0.1cm}--\hspace{0.1cm}We 
describe the main steps and results of our numerical study of 
the model~(\ref{TOYLAG}) in a lattice regularization consistent with the 
exact $\chi_L \times \chi_R$ symmetry. 
Lattice $\chi_L \times \chi_R$ invariance entails 
relations between renormalization constants 
(e.g.\ $Z_{\tilde A} = Z_{\tilde V}$) and  
discretization errors of O($b^{2n}$) only ($n$ integer).
The arguments of Ref.~\cite{Frezzotti:2014wja} imply that the ``NP anomaly''
leading to elementary fermion mass generation occurs even if
fermion loops are neglected. Thus in this first investigation we decided to
work in the quenched approximation~\cite{quenching}. 
Quenching brings key simplifications as scalar and gauge fields can be generated and renormalized independently from each other. 
In Supplemental Material \cite{supplemental_material} we give details on our lattice setup
and data analysis. 

\begin{figure}[hpt]
\begin{center}
\includegraphics[scale=0.5]{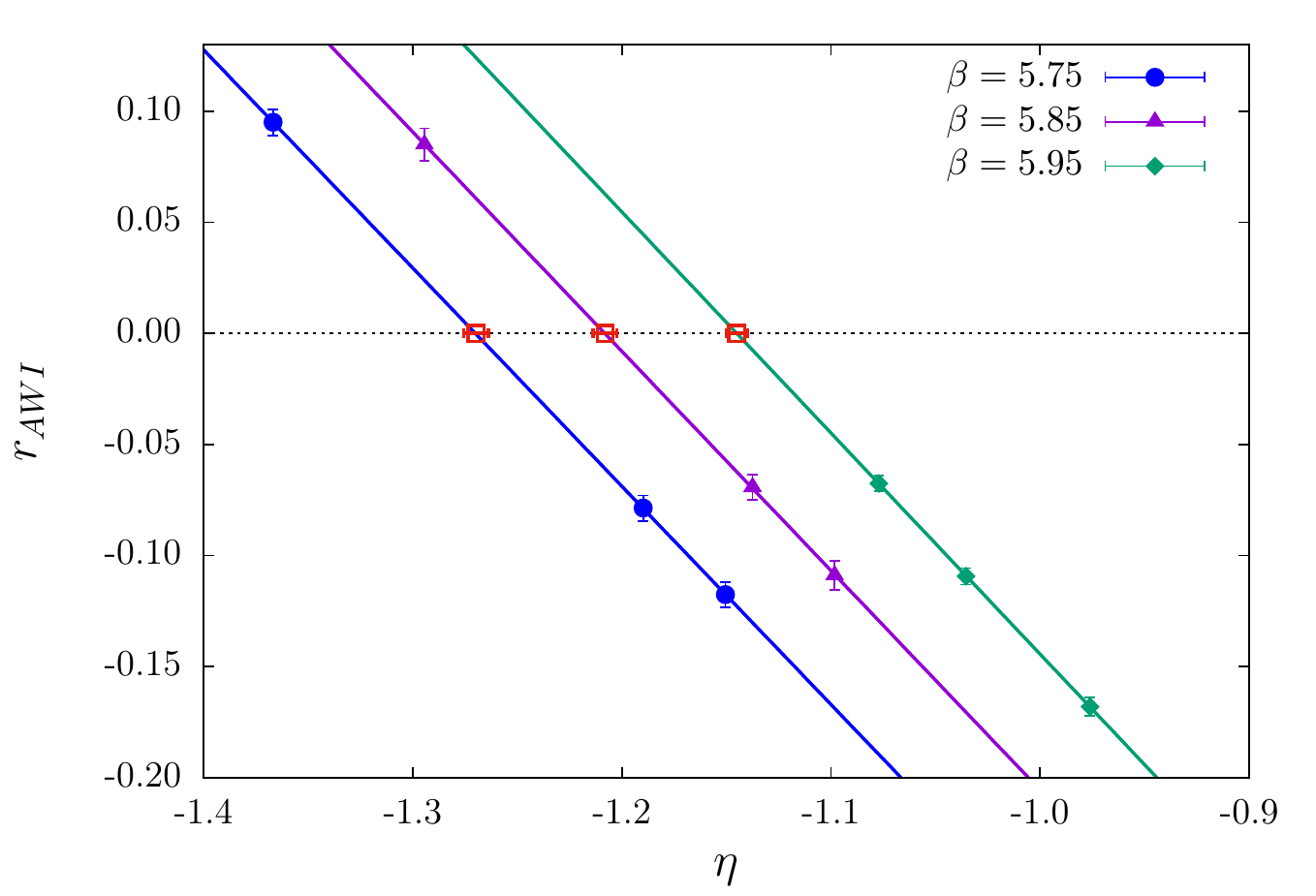}
\end{center}
\vspace{-0.3cm}
\caption{\small{$r_{AWI}$ as a function of $\eta$ at $\beta=5.75$, 5.85 and 5.95. 
Red squares denote the values of $\eta_{cr}$, at which $r_{AWI}=0$.}}
\label{fig:figRAWI}
\end{figure}

For a given choice of bare gauge coupling ($\beta = 6/g_0^2$), scalar potential 
parameters ($\lambda_0$, $m_\phi^2- m_{cr}^2= \hat\mu_\phi^2 / Z_{m_\phi^2} $) 
and strength ($\rho$) of the ${\cal L}_W$ term, we tune the bare  
coupling, $\eta$, so as to restore the fermionic chiral symmetry 
$\tilde\chi_L\times\tilde\chi_R$. This task is conveniently carried out 
in the Wigner phase by looking for the value of $\eta$ at which
\begin{eqnarray}
\hspace{-.7cm}&&r_{AWI}(\eta_{cr},\rho,\lambda_0,g_0^2)\! = \!\frac{
{\cal N}(x_0,y_0; \eta,\rho,\lambda_0,g_0^2)}{{\cal D}(x_0,y_0; \eta,\rho,\lambda_0,g_0^2)}\Big{|}_{\eta=\eta_{cr}} \hspace*{-0.4cm} = 0 \, ,\label{NDR}\\
\hspace{-.7cm}&&{\cal N}(x_0,y_0;\!...)\! = \!b^6 \sum_{\bf x,\bf y} \langle P^1(0) 
\partial_{0}^{FW} \tilde{A}_0^{1,BW}(x) \, \varphi_0(y) \rangle \, ,\label{NR}\\
\hspace{-.7cm}&&{\cal D}(x_0,y_0;\!...)\! =\!  b^6\! \sum_{\bf x, \bf y} \langle P^1(0)
\tilde{D}_P^1 (x) \varphi_0(y) \rangle , 
\;\varphi_0\!=\!\frac{ {\mbox{Tr}}[\Phi] }{2} \, ,\label{NDRB}
\label{DEF}
\end{eqnarray}
where $\partial_{0}^{FW}$ is a forward lattice derivative and
$\tilde A_0^{1,BW}$ the backward one-point-split lattice version of
$\bar Q \gamma_0 \gamma_5 \frac{ \tau^1}{2} Q$. $\tilde{D}_P^1$ and $P^1$ 
are given in Eqs.~(\ref{DSI}) and~(\ref{MAWIR}), respectively. 
The time distances 
$x_0$ and $y_0-x_0$ are separately optimized so as 
to isolate the lowest-lying pseudoscalar (PS) meson and one-$\Phi$ particle states. 
In Fig.~\ref{fig:figRAWI} we show $r_{AWI}$ vs.\ $\eta$ 
at three values of the bare gauge coupling, $\beta=5.75$, 5.85 and 5.95. 
In the quenched approximation they correspond to lattice spacings of about 0.15, 0.12 and 0.10~fm, respectively, if we {\em conventionally} assume for the Sommer scale $r_0 = 0.5$~fm (as in QCD). The values of 
$\eta_{cr}(6/\beta,\rho,\lambda_0)|_{\rho=1.96}$ are denoted by red squares. 

In the NG phase we work at $\eta=\eta_{cr}$, taking into account its uncertainty.
First we compute the effective PCAC mass~(\ref{MAWIR}). For convenience
$Z_P^{-1}$ is evaluated in a hadronic scheme defined   
in the Wigner phase by taking $Z_P^{-1}r_0^{-2} = G_{PS}^W=\langle 0| P^1 |{\rm PS\;meson} \rangle^W$ at the subtraction point, ${M_{PS}}$, given by the PS meson mass (see below). As for $Z_{\tilde{A}}$, we exploit the equality 
$Z_{\tilde{A}}=Z_{\tilde{V}}$, entailed by the $\chi_L\times\chi_R$ 
invariance, and evaluate $Z_{\tilde{V}}$ from an exact WTI.

\begin{figure}[hpt]
\begin{center}
\includegraphics[scale=0.5]{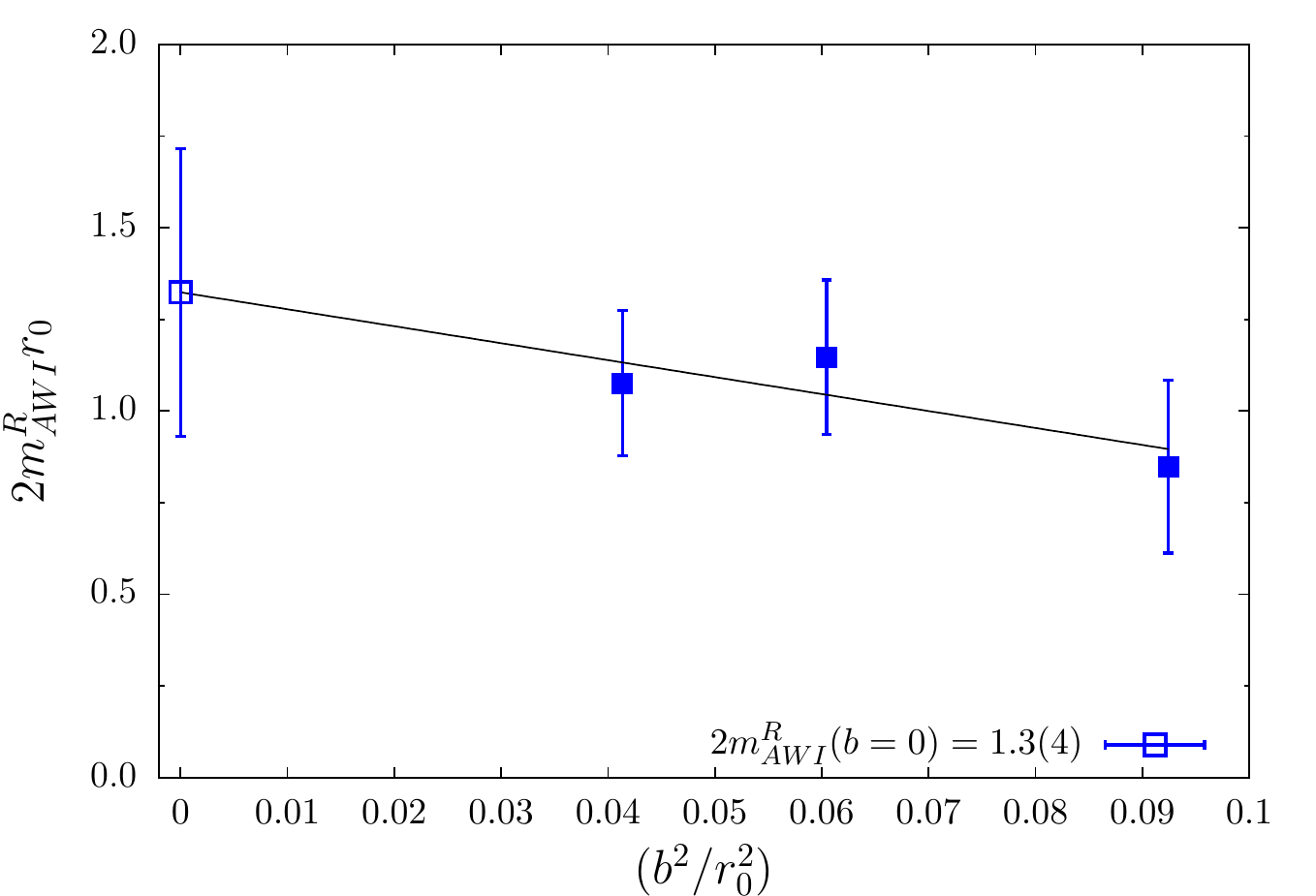}
\end{center}
\vspace{-0.3cm}
\caption{\small{$2m_{AWI}^R r_0 \equiv 2r_0 m_{AWI} Z_{\tilde V} Z_P^{-1}$ 
vs.\ $(b/r_0)^2$ in the NG phase and its linear 
extrapolation to the continuum limit.}}
\label{fig:figMAWI}
\end{figure}

In Fig.~\ref{fig:figMAWI} we plot the renormalized quantity 
$2r_0 m_{AWI} Z_{\tilde V} Z_P^{-1}$ vs.\ $(b/r_0)^2$. 
A linear extrapolation shows that its continuum limit lies about three 
standard deviations away from zero.

\begin{figure}[hpt]
\begin{center}
\includegraphics[scale=0.5]{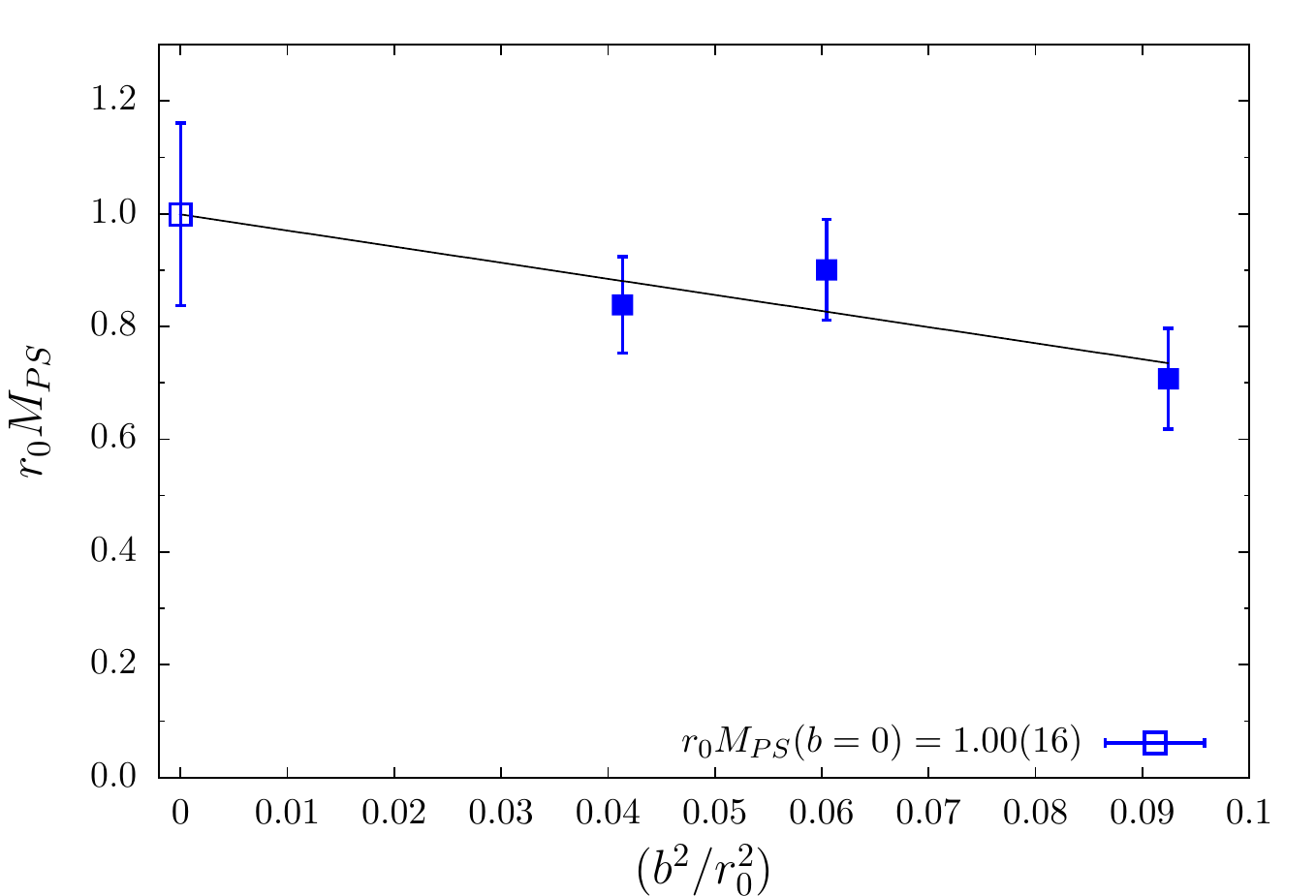}
\end{center}
\vspace{-0.3cm}
\caption{\small{$r_0 M_{PS}$ in the NG phase vs.\ $(b/r_0)^2$ 
and its linear extrapolation to the continuum limit.}}
\label{fig:fig_M_PS}
\end{figure}

Secondly, as another check that the effective PCAC mass
$\frac{Z_{\tilde A}}{Z_P} m_{AWI}$ is non-zero, indicating
the presence in the EL of a fermion mass term, we have computed
from the $\sum_{\bf x} \langle P^1(0) P^1(x) \rangle$ correlator 
the mass of the lowest lying PS meson. Our data for $r_0 M_{PS}$ are plotted in 
Fig.~\ref{fig:fig_M_PS} as a function of $(b/r_0)^2$ together with the best fit 
linear extrapolation to vanishing lattice spacing.
The figure shows that the continuum limit of $r_0 M_{PS}$ lies above zero 
by more than five standard deviations. On the other hand, if $r_0 M_{PS}$
had a vanishing continuum limit, one should see $r_0^2 M_{PS}^2$ approaching
zero as $b^2 \to 0$ with a $b^4$-rate. On a $(b/r_0)^4$ scale our $r_0^2 M_{PS}^2$
data lie very close to the continuum limit and are not compatible with a``no mechanism'' hypothesis (see Supplemental Material \cite{supplemental_material}).     

Finally we checked that at $\eta=\eta_{cr}(\rho)$ the magnitude 
of $\frac{Z_{\tilde A}}{Z_P}m_{AWI}$
(and $M_{PS}^2$) is actually controlled by the strength of fermionic chiral 
breaking as measured by the 
parameter $\rho$ in front of ${\cal L}_W$~(\ref{LWIL}). 
An increase of $\rho$ by a factor 1.5 yields
an increase of $\frac{Z_{\tilde A}}{Z_P} m_{AWI}$  
by a factor of about $2.4$ (see Supplemental
Material \cite{supplemental_material}) in good agreement with expectations 
from~\cite{Frezzotti:2014wja} and section {\em ``Mass generation in a toy model''}. 
For a correct interpretation of this finding one should notice that in more
realistic models supporting the mass generation mechanism under discussion
and accounting for EW interactions, parameters (like $\rho$) that control the 
strength of fermionic chiral breaking, and hence the NP mass terms, are not
completely free, as they are constrained by the conditions of maximal
restoration of the $\tilde\chi_L \times \tilde\chi_R$ symmetry 
(see~\cite{FR18} and Sect.~VI of Supplemental Material \cite{supplemental_material}).

\vspace{0.4cm}
\noindent {\em 4: Conclusions and outlook}\hspace{0.1cm}--\hspace{0.1cm}
In this paper by means of pioneering lattice simulations we have demonstrated from 
first principles the occurrence of an elementary fermion mass term as a ``NP anomaly'' 
in the EL of a renormalizable SU(3) gauge model (Eq.~(\ref{TOYLAG})). In this
model a doublet of strongly interacting fermions is coupled to a colourless 
complex scalar doublet via chiral breaking Yukawa and higher dimensional 
(${\cal L}_W$ in Eq.~(\ref{LWIL})) operators. Indeed, once bare parameters 
are chosen so as to ensure (maximal) restoration of fermionic chiral symmetries,
a fermion mass of the order of the RGI scale ($\Lambda_S$) is generated 
in the phase where the exact symmetry acting on fermions and scalars is 
spontaneously broken.  

This result represents a ground-breaking progress in quantum field theory as 
it provides evidence for the occurrence of a NP obstruction (``anomaly'') 
to the recovery of broken fermionic chiral symmetries giving rise 
to a dynamically generated fermion mass term.

From a phenomenological viewpoint 
EW interactions can be included without introducing tree-level flavour changing 
neutral currents by promoting the exact 
$\chi_L \times U(1)_Y$ invariance to a gauge symmetry.
Since the $\tilde\chi_L\times \tilde \chi_R$ transformations now 
act both on fermions and weak bosons, one can show that 
i) the requirement of maximal $\tilde\chi_L$ restoration 
leads to stringent constraints on 
the chiral breaking parameters and ii) a unique NP mechanism 
generates fermion and weak boson mass terms~\cite{FR18}. 

This mass generation phenomenon is alternative to the Higgs mechanism and 
provides an interesting starting point for  
beyond the SM models. As all masses 
are parametrically of the order of the RGI scale,
the latter must be much larger than
$\Lambda_{QCD}\! \sim \! 300$~MeV, if the 
masses of the top quark and the EW bosons have to be explained.  
This observation suggests the existence 
of a new non-Abelian gauge interaction that gets strong at a scale
$\Lambda_T\!\gg\!\Lambda_{QCD}$, and of new elementary fermions 
with O($\Lambda_T$) NP masses. 
Crude estimates~\cite{Frezzotti:2014wja,FR18} 
hint at $\Lambda_T\!=\! {\rm O}({\mbox{a few TeV}})$. Since the condition
of $\tilde\chi_L$ restoration implies the decoupling of $\zeta_0$ (the isosinglet
component of $\Phi$), one 
ends up with models of the composite Higgs~\cite{Kaplan:1991dc,PaniWulz} type, 
where the 125~GeV resonance is 
a bound state~\cite{Frezzotti:2014wja,FR18} in the $WW\!+\!ZZ$ channel 
formed owing to the new strong force. 

In the theoretical framework sketched above we see also a chance of
understanding the observed fermion mass hierarchy. Denoting by
$c_{1,f}\Lambda_T$ the dynamical mass of the SM fermion $f$ and by $g$
the gauge coupling of the strongest gauge interaction which 
$f$ is subjected to, it turns out~\cite{Frezzotti:2014wja} that 
$c_{1,f}$ is O($g^4$) for the heaviest fermion generation and possibly 
of higher order for the other generations. 
In this way one can understand e.g.\ the top to $\tau$ mass ratio~\cite{FR18}. 
Further remarks about extending the model~(\ref{TOYLAG}) to
a phenomenologically sensible theory 
are deferred to the Supplemental Material \cite{supplemental_material}.

In summary, by combining the condition of maximal restoration of fermionic chiral 
symmetry explicitly broken at the UV scale (a weak form of 't~Hooft 
naturalness) with the assumption 
of the existence of a new non-Abelian gauge interaction with 
$\Lambda_T\!=\!{\rm O}({\mbox{a few TeV}})$, we find a novel
mechanism that gives mass to elementary fermions and EW gauge bosons. 
In models based on this mechanism the EW scale can be related
to the scale of new physics at which 
new resonances should be detected in accelerator experiments.

\vspace{0.4cm}
{\em Acknowledgements}\hspace{0.1cm}---\hspace{0.1cm}We are 
grateful to J.~Kuti, M.~L\"uscher, G.~Martinelli, N.~Tantalo 
and M.~Testa for very valuable discussions.
We thank G.M. de Divitiis, B. Knippschild and 
M. Schr\"ock for their 
participation in the early stages of this project.
CPU time on Galileo and Marconi clusters was provided by the
INFN group LQCD123 under the 2017 and 2018 CINECA-INFN agreements.
This work was supported in part by the DFG in the Sino--German
research center CRC110.


\vspace{0.2cm}

\bibliography{bibliography}

\end{document}


\title{Supplemental Material for the Letter \\ 
A novel mechanism for dynamical generation \\ of elementary fermion  
mass: lattice evidence}\vspace*{0.4cm}
\newcommand{\Frank}{\affiliation{Goethe U. Frankfurt, Inst.\ f.\ Theor.\ Phys, Max-von-Laue-Str. 1, 60438 Frankfurt am Main, Germany}}
\newcommand{\ToV}{\affiliation{Dipartimento di Fisica, Universit\`a di Roma Tor Vergata, Via della Ricerca Scientifica, 00133 Roma, Italy}}
\newcommand{\Fermi}{\affiliation{Centro Fermi - Museo Storico della Fisica e Centro Studi e Ricerche Enrico Fermi, 00184 Rome, Italy}}
\newcommand{\INFN}{\affiliation{INFN, Sezione di Roma Tor Vergata, Via della Ricerca Scientifica, 00133 Roma, Italy}}
 \newcommand{\HISKP}{\affiliation{HISKP (Theory), Rheinische Friedrich-Wilhelms-Universit\"at Bonn, Nussallee 14-16, 53115 Bonn, Germany}}

\author{S.~Capitani}
\Frank
\author{P.~Dimopoulos}
\Fermi
\ToV
\author{R.~Frezzotti}
\ToV
\INFN
\author{M.~Garofalo}
\INFN
\author{B.~Kostrzewa}
\HISKP
\author{F.~Pittler}
\HISKP
\author{G.C.~Rossi}
\Fermi
\ToV
\INFN
\author{C.~Urbach}
\HISKP

                             

\vspace{0.2cm}

\maketitle
\tableofcontents

\section{Lattice formulation}
\label{sec:LATREG}

Lattice simulations of models with gauge, fermion and scalar fields are rather
uncommon in the literature. See, however, Ref.~\cite{De:1987bt} for a quenched 
study where, unlike here, the scalar field belongs to the fundamental representation of 
the SU(2) gauge group but no $\tilde\chi_L \times \tilde\chi_R$ symmetry violating 
terms are included.

For our purposes it is crucial to adopt a lattice formulation preserving
the exact $\chi_L \times \chi_R$ symmetry, while including terms breaking the
purely fermionic $\tilde\chi_L \times \tilde\chi_R$ invariance. We have thus
chosen the lattice regularized action of the model~(L1) to be of the form 
\begin{eqnarray}
\hspace{-.5cm}&&{{S}_{{L}}\!=\!b^4\!\sum_x\!\Big{\{} {\cal L}_k^g[U_{.}]\!+\!{\cal L}_{k}^{s}(\Phi)\!+\!{\cal V}(\Phi)}{+ \bar Q  { D_{L}[U_{.},\Phi]} Q\Big{\}}}\, , \;
\label{ACTLAT}\\ 
\hspace{-.5cm}&&{\cal L}_k^g[U_{.}]\,\,{\mbox{: SU($3$) plaquette action}}\, , \nonumber\\
\hspace{-.5cm}&&{\cal L}_{k}^{s}(\Phi)+{\cal V}(\Phi)=\nonumber\\
\hspace{-.5cm}&&=\frac{1}{2}{\tr}[\Phi^\dagger(-\partial_\mu^*\partial_\mu)\Phi]+\frac{m_\phi^2}{2}{\tr}[\Phi^\dagger\Phi]+\frac{\lambda_0}{4}({\tr}[\Phi^\dagger\Phi])^2 \, ,\nonumber
\end{eqnarray}
where the na\"ive lattice discretization of fermions was used. We set $\Phi = \varphi_01\!\!1+i \varphi_j\tau^j$ and denote by the symbol $U_{.}$ the set of gauge links. Here and in the following by ``(Ln)'',
with n integer, we shall refer to the equation (n) of the Letter, also
called ``main text''. 
Setting $F(x) \equiv [\varphi_0 1\!\!1+i\gamma_5\tau^j\varphi_j](x)$ 
the lattice Dirac operator reads 
\begin{eqnarray} \label{DLATT}
\hspace{-.4cm}&& (D_{L}[U_{.},\Phi] Q)(x)=
\gamma_\mu \widetilde\nabla_\mu  Q(x) + \eta F(x)Q(x)+\nonumber\\
\hspace{-.4cm}&&-b^2\rho \frac{1}{2}F(x)
\widetilde\nabla_\mu \widetilde\nabla_\mu  Q(x) +\nonumber\\ 
\hspace{-.4cm}&& - b^2\rho \frac{1}{4} \Big{[} (\partial_\mu F)(x) U_\mu(x)
\widetilde\nabla_\mu  Q(x+\hat\mu) +\nonumber\\
\hspace{-.4cm}&&+ (\partial_\mu^*F)(x) U_\mu^\dagger(x-\hat\mu) \widetilde\nabla_\mu  Q(x-\hat\mu) \Big{]} \, .
\end{eqnarray}
Gauge covariant lattice derivatives are defined via
\begin{eqnarray}
&&\widetilde\nabla_\mu f(x)\!\equiv\! \frac{1}{2}(\nabla_\mu+\nabla_\mu^*)f(x) \, , \label{TILDE}\\
&&b\nabla_\mu f(x)\equiv U_\mu(x)f(x+\hat\mu)-f(x)\, ,\label{DEV} \nonumber \\
&& b\nabla_\mu^* f(x)\!\equiv \!f(x)-U_\mu^\dagger(x-\hat\mu)f(x-\hat\mu) \, .
\end{eqnarray}
The partial derivatives $\partial_\mu$, $\partial_\mu^*$ are the forward, backward
derivatives $\nabla_\mu$, $\nabla_\mu^*$, respectively, with $U_\mu\!=\!1\!\!1$. 
We use periodic boundary conditions for the boson (gauge and scalar) fields, while
for fermions we impose antiperiodic boundary conditions in time and periodic ones in
three-dimensional space.

The lattice action~(\ref{ACTLAT}) describes 2 (isospin) $\times$ 16 (doublers) 
fermionic flavours, even in the $b\rightarrow 0$ limit. In fact, the fermion
Lagrangian term proportional to $\rho$, which is a lattice regularization of
the ${\cal L}_W$ term of Eq.~(L2), being the product of $b^2$ times
a (dimension six) fermion bilinear operator with two derivatives, does not remove 
the doublers in the continuum limit. 
However, for the goals of the present {\em quenched} study, the presence of doublers 
in the {\em valence} sector, which does not alter the $\beta$-function and hence 
preserves the asymptotic
freedom of the model, is harmless. With respect to the unquenched case, 
the quenched approximation is expected to entail only some systematic error 
(typically of order 10--20\% based on Lattice QCD experience) in the value of
the coefficient $c_1$ of the NP fermion mass term (see Eq.~(L13)) 
that is possibly generated in the NG phase EL. 
For analogous unquenched studies a lattice Dirac operator $D_L[U_{.},\Phi]$ with no 
doublers would of course be needed, which could e.g.\ be built by using the 
domain-wall~\cite{DW} or overlap~\cite{Neuberger} fermion kernel for two flavours 
with the addition of suitable Yukawa and ${\cal L}_W$-like terms.

\vspace*{-0.2cm}
\subsection{Fermionic regularization}
\label{Fermions discretization}

Following closely Refs.~\cite{KlubergStern:1983dg},
\cite{Patel:1992vu} and~\cite{Luo:1996vt} for staggered fermions,
one can analyze the fermion flavour content of the action (\ref{ACTLAT}).
We first rewrite its fermionic sector in terms of the field
$\chi(x)={\cal A}_x^{-1}Q(x)$ with  ${\cal A}_x=\gamma_1^{x_1}\gamma_2^{x_2}\gamma_3^{x_3}\gamma_4^{x_4}$, then we pass to the so-called flavour basis
\begin{equation}
q^B_{\alpha,a}\!(y)\!=\!\sum_\xi \! \overline{U}(2y,2y+\xi)[{\cal A}_\xi]_{\alpha,a}
\! \frac{(1\!-\! b\xi_\mu\widetilde\nabla_{\mu})}{8}\chi^B\!(2y+\xi), 
\label{FLABAS}
\end{equation}
where $\alpha$, $a$ and $B$, all taking values in $\{1,2,3,4\}$, denote Dirac, taste 
and replica indices. While the $\chi^B(x)$ fields are defined on the fine lattice 
$x_\mu=2y_\mu+\xi_\mu$, $\xi_\mu=0,1$ with $\overline{U}(2y,2y+\xi)$ 
denoting the average of link products along the shortest paths from $2y$ to
$2y + \xi$, the flavour basis fields $q^B_{\alpha,a}(y)$ live on the coarse lattice 
with coordinate $y_\mu$ (here we have set $b=1$). In terms of these fields  
close to the continuum limit the lattice fermionic action reads
 \begin{equation}\label{S_latt[q]}
 S_{L}^{F} \! = \! \sum_{y,B} \!\bar q^B(y)\!\left\{\! \sum_\mu (\gamma_\mu\! \otimes \! 1\!\!1)D_\mu \!+\!\eta{\cal F}^B(y)\! \right\}\! q^B(y)\!+\!{\mbox{O}}(b^2) \, ,
\end{equation}
where for each $B$ value the matrix  
$ {\cal F}^B(y)=\varphi_0(2y)(1\!\!1 \otimes 1\!\!1)+s_Bi\tau^i\varphi_i(2y)(\gamma_5 \otimes t_5)$, $s_{1,2} = 1 = -s_{3,4}$,  
has a ``Dirac $\otimes$ taste'' structure with $t_\mu=\gamma_\mu^*$ taste matrices. 
Unlike in the original
$Q(x)$ basis, in the flavour basis $q(y)$ defined via Eq.~(\ref{FLABAS})
space-time and flavour degrees of freedom are
completely disentangled from each other. As the action~(\ref{S_latt[q]}) 
is diagonal in taste and replica indices up to $O(b^2)$, it
obviously describes 32 fermion species, namely 4 replicas of 
the 4 tastes of the isospin doublet $Q^T=(u,d)$. 

The quark bilinears expressed in the $Q$ basis, once summed over one
$2^4$ hypercube of the fine lattice, have well defined quantum numbers 
in the classical continuum limit, as one checks by rewriting them 
in the flavour $q(y)$ basis.
For example the isotriplet pseudoscalar density
\begin{eqnarray} \label{latt_PSdens}
\hspace{-.4cm} &&P^{i}(x)=\overline Q(x)\gamma_5\frac{\tau^i}{2}Q(x) 
\,,\quad x_\mu=2y_\mu+\xi_\mu 
\end{eqnarray}
summed over the hypercube ($\xi_\mu=0,1$) yields in flavour basis
the corresponding isotriplet, taste diagonal density 
\begin{equation}\label{J_Pdens}
\hspace{-.1cm} \frac{1}{16} \! \sum_\xi \!  P^{i}(2y+\xi) \!=\! \sum_{B=1}^4 s_B \overline 
q^{B}(y)(\gamma_5\otimes t_5 )\frac{\tau^i}{2} q^B(y)\!+\!{\mbox{O}}(b^2) \, .
\end{equation}
Similarly for the lattice backward isotriplet axial current
\begin{eqnarray} \label{latt_AXcurr}
\hspace{-.4cm} &&\tilde{A}_\mu^{i,BW}(x)=
\frac{1}{2} \overline Q(x-\hat\mu)\gamma_\mu\gamma_5\frac{\tau^i}{2}U_\mu(x-\hat\mu)Q(x) + \\
\hspace{-.4cm}&& 
+ \frac{1}{2} \overline Q(x)\gamma_\mu\gamma_5\frac{\tau^i}{2}U_\mu^\dagger(x-\hat\mu) Q(x-\hat\mu)\,,\quad x=2y+\xi \nonumber
\end{eqnarray}
summing over the hypercube one finds in the flavour basis
the corresponding isotriplet, taste diagonal current, 
\begin{equation}
\hspace{-.1cm} \frac{1}{16}\! \sum_\xi \!  \tilde{A}_\mu^{i,BW} \!(2y+\xi) 
\!\!  =\! \!\sum_{B=1}^4 \! s_B \overline 
q^{B}\!(y)(\gamma_\mu\gamma_5\otimes t_5 )\frac{\tau^i}{2} q^B\!(y)+{\mbox{O}}(b^2) \, .
\label{J_AXcurr}
\end{equation}

\vspace*{-0.2cm}
\subsection{Renormalizability and other technicalities}

Beyond tree level, loop effects do not generate $d\leq 4$ 
terms besides the operators that are already  
present in the lattice action~(\ref{ACTLAT}). This follows from the symmetries (gauge,
$H(4)$, $\chi_L \times \chi_R$, C, P, T) of the
model~(\ref{ACTLAT}) which were discussed in Ref.~\cite{Frezzotti:2014wja}
and are preserved by the lattice regularization. 
The latter is chosen so as to respect the ``spectrum doubling 
symmetry''~\cite{Montvay} of na\"ive fermions, implying that the action~(\ref{ACTLAT}) 
is invariant under the transformations
\begin{eqnarray}
\hspace{-.4cm} && Q(x) \to  Q'(x)=e^{-ix\cdot\pi_H}M_H Q(x)  \nonumber\\
\hspace{-.4cm} &&\overline Q(x)\to \overline Q'(x)=\overline Q(x)M_H^\dagger e^{ix\cdot\pi_H}\, ,
\end{eqnarray}
where $H$ is an ordered set of four-vector indices $H\equiv\{\mu_1,...,\mu_h\},\,(\mu_1<\mu_2<...<\mu_h)$. For $1 \leq h \leq 4$ there are 16 four-vectors $\pi_H$,  with $\pi_{H,\mu}=\pi$ if $\mu\in H$ or $\pi_{H,\mu}=0$ otherwise and 16 matrices
$M_H\equiv (i\gamma_5\gamma_{\mu_1})...(i\gamma_5\gamma_{\mu_h})$. The use
of symmetric (gauge covariant) lattice derivatives $\widetilde \nabla_\mu$ acting on 
$Q$ in the ${\cal L}_W$ term is dictated by~the need of preserving the
spectrum doubling symmetry. The latter, being an exact symmetry of $S_{L}$, 
is an invariance of the lattice effective action $\Gamma_{L}[U_{.},\Phi, Q]$. Hence
all fermion species, including doublers, enter in a symmetric way in the 
effective action, in particular in the non-irrelevant local terms
$\sum_{y} \sum_B \overline q^B(y)(\gamma_\mu\otimes 1\!\!1) D_\mu q^B(y)$
and $(\eta - \bar{\eta}) \sum_{y} \sum_B \overline q^B(y){\cal F}^B(y) q^B(y)$
(in the flavour basis).

These nice renormalization properties imply in particular that $\eta_{cr}$, 
as determined from the vanishing of $r_{AWI}$ (Eq.~(L15)), 
is unique for all fermion species.

As the model (\ref{ACTLAT}) is studied in the quenched approximation,
problems with exceptional configurations of gauge and scalar fields 
supporting very small eigenvalues of $D_{L}$ plague, as expected, 
the Monte Carlo sampling of fermionic observables. In order to avoid
such problems and have a robust IR cutoff in the fermionic propagators
we include in the lattice action~(\ref{ACTLAT}) an additional
``twisted mass'' term~\cite{Frezzotti:2000nk,Frezzotti:2003ni}
 \begin{equation} \label{ACT-TM}
S_L^{tm} = b^4\mu \sum_x \bar Q(x) i\gamma_5 \tau^3 Q(x).
 \end{equation}
This term does not spoil power counting renormalizability, it only breaks in
a soft way the $\chi_L \times \chi_R$ symmetry and can thus be safely removed 
at the end of the calculations by taking (even at fixed lattice spacing) the limit $\mu\to 0$. 

\begin{figure}[htbp]
\vspace*{0.2cm}
\centerline{\hspace{0.2cm} 
\includegraphics[scale=0.440,angle=0]{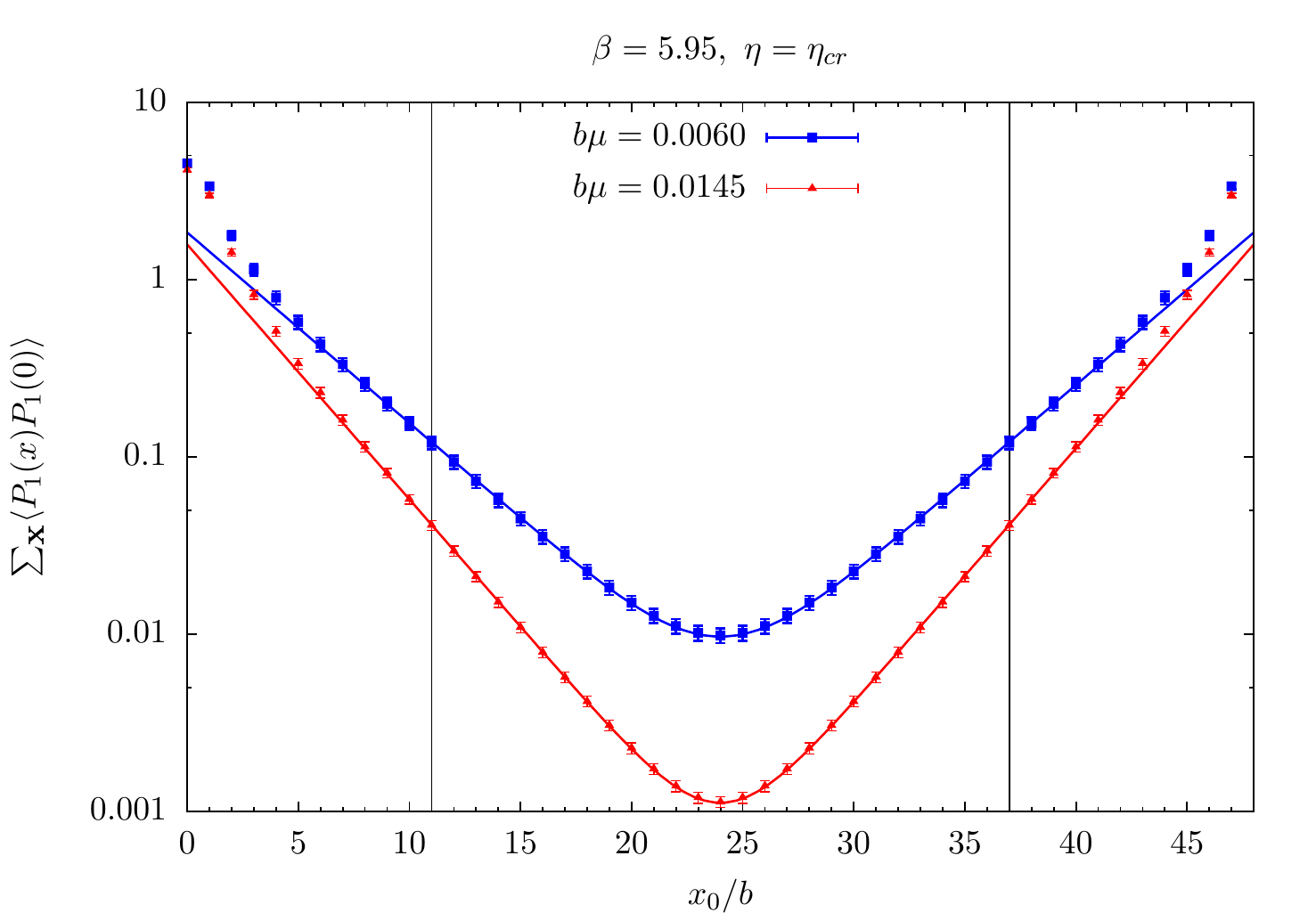} 
           }
\caption{PS-meson two--points (zero three-momentum) correlators as a function of the time
separation $x_0/b$ in the NG phase at $\beta=5.95$, $\rho=1.96$, $\eta=\eta_{cr}$, on a 
periodic $L^3 T$ lattice with $L/b=T/2b=24$ and for twisted quark mass $b\mu=(0.0060,~0.0145)$: 
data points with statistical error and the best fit curve corresponding to one-particle 
state dominance in the interval $11\leq x_0/b \leq 24$ are shown.}
\label{fig:NGcorr}
\end{figure}

\begin{figure}[htbp]
\vspace*{0.2cm}
\centerline{\hspace{0.2cm} 
\includegraphics[scale=0.440,angle=0]{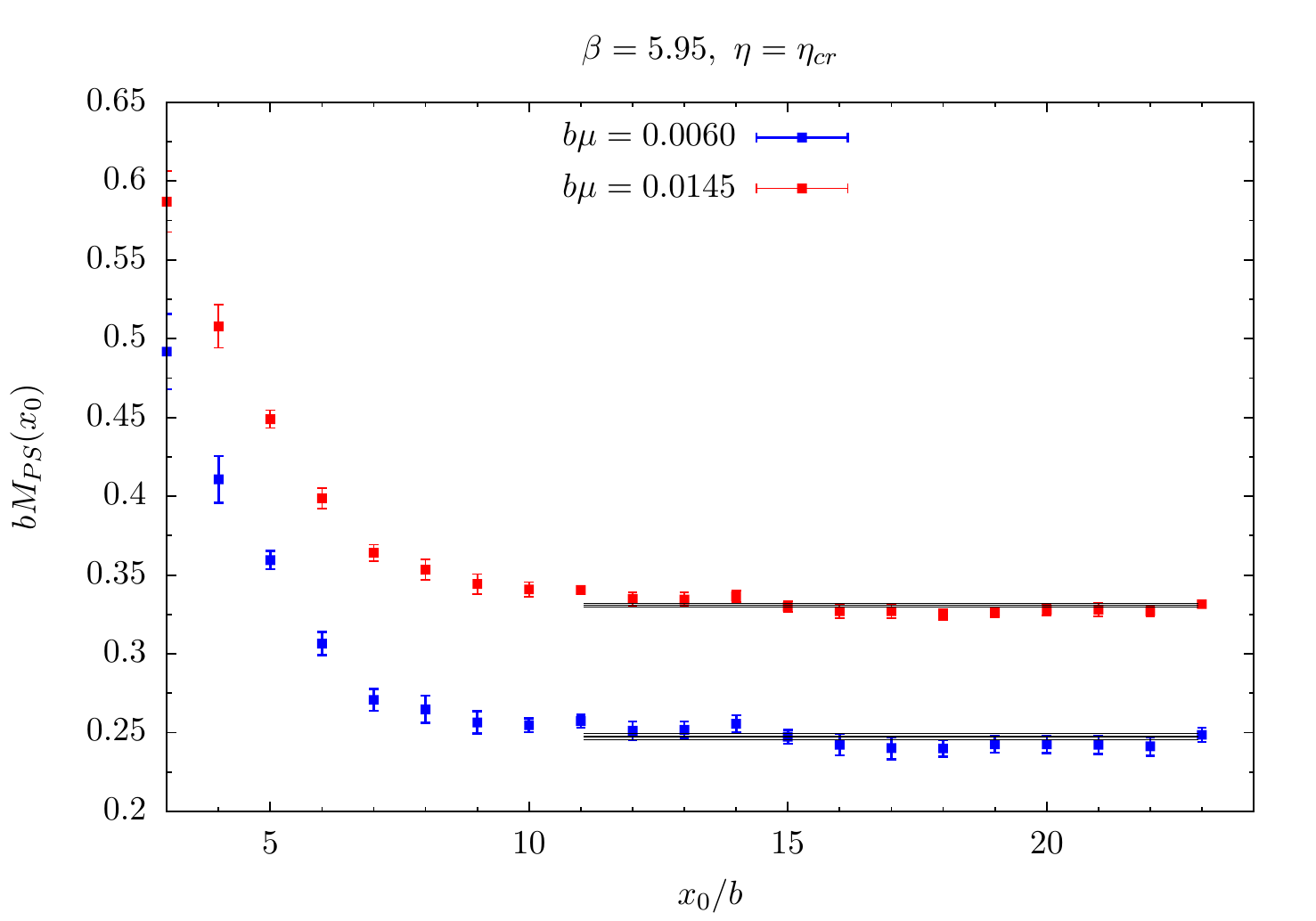}
           }
\caption{For the same correlators as in Fig.~\ref{fig:NGcorr}: effective mass values and 
the PS-meson mass estimate obtained by fitting them to a constant in $11\leq x_0/b \leq 24$
are shown.}
\label{fig:NGmeff}
\end{figure}

In quenched lattice QCD there is overwhelming numerical evidence (see e.g.\
Refs.~\cite{Butler:1994em,Bernard:1998db,Aoki:1999yr,Blum:2000kn}) that
at large enough Euclidean time separations the correlators (vacuum
expectation values of local fields) with fixed spatial momentum admit a 
spectral decomposition in terms of intermediate states with definite 
quantum numbers. The products of matrix elements entering this spectral
representation (in particular those involving excited states) may at worst take values
that are not consistent with time-reflection positivity, because the quenched model,
owing to the omission of virtual quark contributions, is not unitary. 
We shall here assume that the same holds true in our quenched model~(\ref{ACTLAT}).
The validity of this assumption about the lowest-lying states of the spectrum
is confirmed by strong numerical evidence for all the correlators we actually
evaluated -- as we routinely checked in our data analyses. As a typical example,
we show i) in Fig.~\ref{fig:NGcorr} the PS-meson two--points correlator in the NG~phase 
of the critical model for $\beta=5.95$ and two different twisted quark mass values (for more 
details on simulation parameters see Table~\ref{tab:sim_NG}), together with its best fit to the 
exponential time dependence that is expected from one-particle state dominance, and
ii) in Fig.~\ref{fig:NGmeff} the corresponding effective mass plot, from the plateau 
of which the PS-meson mass is eventually estimated.


\vspace*{-0.2cm}
\subsection{Ignoring fermionic disconnected diagrams}

In our quenched study of the lattice model~(\ref{ACTLAT}) the calculation
of the correlators appearing in Eq.~(L11) and Eqs.~(L16)-(L17)
was carried out by consistently neglecting all the fermionic disconnected 
Wick contractions. The latter arise in correlators with isospin non-singlet 
operators because of the occurrence of isospin changing fermion-antifermion-$\Phi$ 
vertices and are statistically quite noisy. Neglecting fermionic disconnected diagrams 
is, however, justified in our case as it affects neither the results for the mixing coefficient $\bar\eta$ (which determines $\eta_{cr}$, see Eq.~(L5)) nor the pseudoscalar isotriplet meson mass, $M_{PS}$. It entails only an immaterial (and numerically small) modification of the renormalized matrix element ratio $\frac{Z_{\tilde A}}{Z_{P}} m_{AWI}$, see Eq.~(L11).

\begin{figure}[htbp]
\vspace*{0.2cm}
\centerline{\hspace{0.2cm} \includegraphics[scale=0.890,angle=0]{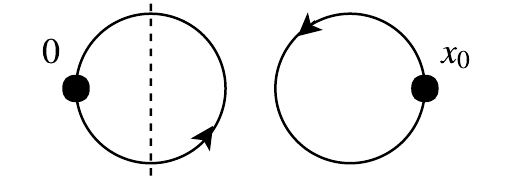} 
\hspace{-0.4cm} \includegraphics[scale=0.890,angle=0]{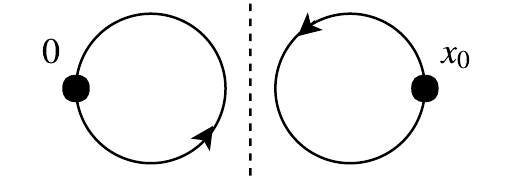} }
\caption{The typical two-boundaries diagram neglected in our work, with
fermionic bilinear operators inserted at times $(y_{\,\rm source})_0=0$
and $x_0 > 0$. The meaning of the cuts through one of the valence fermion loops 
(left panel) or just between them (right panel) is discussed below Eq.~(\ref{P1P1C}).}
\label{fig:2BD}
\end{figure}

The arguments behind the statements above are given in this section. 
Following~\cite{Veneziano:1976wm}, we call ``one-boundary'' diagrams 
the fermionic connected diagrams in which valence fermions localized at
$y_{\, \rm source}=0$ are contracted with the valence quarks sitting at 
$x \neq y_{\, \rm source}$, 
and ``two-boundaries'' diagrams those in which valence quarks and antiquarks 
are contracted with each other separately at $y_{\, \rm source}=0$ and 
$x\neq y_{\, \rm source}$.
The valence fermion structure of the two-boundaries Wick contractions that
were neglected in evaluating all the correlators considered in this paper
is shown in the diagrams of Fig.~\ref{fig:2BD} .

Concerning the evaluation of $\bar\eta - \eta$ the key observation is that the
bare SDE of $\tilde\chi_L$ and/or $\tilde\chi_R$ transformations (see
Eqs. (3.15)--(3.16) of Ref.~\cite{Frezzotti:2014wja}), as well as their
renormalized counterparts, see e.g.\ Eq.~(L5) for the $\tilde\chi_L$ SDE,  
can be split in two independent identities according to the number of diagram 
boundaries ensuing from the way valence quarks are contracted.
The reason is that the one-boundary and the two-boundaries diagrams have different functional dependencies on $x_0$, which must be separately matched 
between the l.h.s.\ and the r.h.s.\ of the SDE. In fact, for large enough time
separations, the one-boundary Wick contractions are dominated (as found in many 
numerical studies) by the lowest--lying isospin non-singlet intermediate state, while 
the two-boundaries diagrams are expected to be saturated by different states 
involving an odd number of $\Phi$-particles and possibly gluons. 
The quantity $\bar\eta -\eta$, being an operator mixing coefficient, is a factor 
common to both one-boundary and two-boundaries diagrams. In other words, $\bar\eta$, 
and thus $\eta_{cr}$, can be in principle computed equally well 
by restricting attention to either the one-boundary or the two-boundaries Wick 
contractions relevant for the SDE of $\tilde\chi_{L,R}$ transformations. We also
recall that such SDE make perfectly sense in the quenched approximation, because 
it can be defined as a renormalizable Lagrangian theory with ghost fields 
\`a la Morel--Bernard--Golterman (see Refs.~\cite{Morel:1987xk,Bernard:1992mk}).

The mass $M_{PS}$ of the isotriplet pseudoscalar meson, as obtained, say, from the 
correlator (here $x = ({\bf x},x_0)$)
\begin{equation}
C^{11}(x_0) = b^3\sum_{\bf x} \langle P^1(y_{\,\rm source}) P^1(x)\rangle|_{y_{\,\rm source}=0} \;\, ,
\label{P1P1C}
\end{equation}
can also be evaluated by just restricting attention to the one-boundary Wick contractions. 
To see that two-boundaries diagrams do not affect the value of $M_{PS}$, let us 
examine them in terms of their possible cuts, see Fig.~\ref{fig:2BD}. 
One can either cut through one of the two fermion 
loops stemming from the Wick contractions at $y_{\,\rm source}=0$ and at 
$x \neq y_{\,\rm source}$, or one can cut in the middle of the two-boundaries diagram 
(no fermionic lines are cut in this case). 
In the first case one exposes terms decreasing exponentially in Euclidean time 
with $M_{PS}$, but one also recognizes from the cut diagram topology that these 
terms merely contribute to the renormalization of either $P^1(x)$ (as in left
panel of Fig.~\ref{fig:2BD}) or $P^1(y_{\,\rm source})$ without affecting the
value of $M_{PS}$. In the other case (right panel of Fig.~\ref{fig:2BD}) one obtains 
terms that, owing to the quenched approximation, are expected to be dominated by a set 
of intermediate states that have nothing to do with a pseudoscalar meson, rather they 
involve an odd number of scalar $\Phi$-particles plus gluons.


Unlike $\bar\eta$ and $M_{PS}$, the value of bare correlator ratio $m_{AWI}$ 
(Eq.~(L11)) gets instead modified if two-boundaries diagrams are 
neglected. The reason is that two-boundaries diagrams contribute, as we 
just saw discussing the case of $M_{PS}$, to the renormalization of the 
isotriplet pseudoscalar quark density and the axial current four-divergence, 
as well as to mixing of these quark bilinears with other operators involving 
scalar fields but no fermions. In the {\em quenched} 
lattice ${\cal L}_{\rm toy}$ model of interest to us, however, the properly 
renormalized ratio, $m_{AWI}^R \equiv \frac{Z_{\tilde A}}{Z_{P}} m_{AWI}$, 
is obtained (at $\eta=\eta_{cr}$) if one consistently ignores all two-boundaries diagrams in the
evaluation of the renormalization constants $Z_{\tilde A}$ and $Z_{P}$, 
which is what we have done in our numerical study.

A proof of the above statement can be given as follows. Let us consider in 
the quenched approximation the model ${\cal L}_{\rm toy}$ (see Eq.~(\ref{ACTLAT}))
and its extension to two generations of fermion doublets, i.e. an analogous model
with fermion sector action $S_L^F = \overline Q D_{L} Q + \overline Q' D_{L} Q'$, 
with $Q=(u,d)$ and $Q'=(u',d')$ entering with identical $\tilde\chi$ violating
parameters, $\rho$ and $\eta$. Owing to fermion quenching, the latter model will 
admit an EL description analogous to the one of ${\cal L}_{\rm toy}$ (see Eq.~(L13)),
with the fermion sector simply replicated for the $Q$- and the $Q'$- generations.
In particular the $\tilde\chi$ violating NP terms in the EL will occur multiplied by
coefficients $c_1$, $c_2$, \dots that are generation-independent and have the same value 
as for ${\cal L}_{\rm toy}$. In the two generations lattice model introduced above
the non-vanishing of $c_1$ is equivalent to the non-vanishing of the renormalized
correlator ratio $(\frac{Z_{\tilde A}}{Z_{P}} m_{AWI})|^{\rm gen.\,off-diag}$, 
where
\begin{equation}
m_{AWI}^{\rm gen.\,off-diag} =
\frac{ \sum_{\bf x} \partial_0 \langle 
\bar Q \gamma_0\gamma_5 \frac{\tau^1}{2} Q' (x) \bar Q' \gamma_5 \frac{\tau^1}{2} Q (0) 
\rangle }
{ 2 \sum_{\bf x} \langle \bar Q \gamma_5 \frac{\tau^1}{2} Q' (x) \bar Q' \gamma_5 
\frac{\tau^1}{2} Q (0) \rangle }\Big|_{\eta_{cr}}  \label{TWOGENAXCUR}
\end{equation}
and $\bar Q \gamma_\mu\gamma_5 \frac{\tau^1}{2} Q'$ is the current of a 
$\tilde\chi$-axial {\em generation off-diagonal} transformation.
But this renormalized correlator ratio of the two generations model, to which
no two-boundaries diagrams contribute owing to the generation off-diagonal nature
of the axial current, is just equal to the ratio 
$\frac{Z_{\tilde A}}{Z_{P}} m_{AWI}$ that is computed in the one generation 
model ${\cal L}_{\rm toy}$ by neglecting two-boundaries diagrams both in $m_{AWI}$ and
in $\frac{Z_{\tilde A}}{Z_{P}}$. The latter ratio (at $\eta=\eta_{cr}$) is hence  
a renormalized and $\tilde\chi$ covariant quantity
that is non-vanishing if and only if the coefficient $c_1$ in the EL 
expression~(L13) is non-zero.

\vspace*{-0.2cm}
\subsection{Gauge and scalar sector renormalization} \label{Sec:Gauge-Scal}

Owing to the quenched approximation, gauge and scalar configurations can be independently generated. Similarly, action parameter renormalization can be carried out independently for gauge and scalar fields.
The gauge coupling is renormalized by keeping the Sommer length scale $r_0$~\cite{Guagnelli:1998ud,Necco:2001xg} fixed in physical units. In the following for mere orientation 
we take its value from calculations of the static quark potential in QCD and 
set $r_0=0.5$~fm\,$\sim 1/(394~{\rm MeV})$.

The choice of the phase in which the theory lives is made by taking 
$m_\phi^2$ either larger (Wigner) or smaller (NG phase) than 
the critical value $m_{cr}^2$ at which 
the scalar susceptibility diverges (in practice it has a sharp peak
on a large enough space-time volume).
The values of the critical mass are listed in Tab.~\ref{Tab:ren_cond_NG}.

In our study of $m_{AWI}$ and $M_{PS}$ in the NG phase
the bare scalar parameters in~(\ref{ACTLAT}) at each value of the lattice 
spacing are determined by the renormalization conditions  
\begin{eqnarray}
M^2_\sigma r_0^2\!= \!1.284  \quad\mbox{and}\quad \lambda_{R}\!\equiv\!\frac{M_\sigma^2}{2v_R^2}\!=\!0.441  \, .\label{SCRC}
\end{eqnarray}
Moreover the renormalized scalar vev, $v_R = Z_\varphi^{1/2} v$, 
where $Z_\varphi$ denotes the scalar field ($\varphi$) renormalization constant, 
is fixed in physical units by the condition $v_R^2 r_0^2\!=\!1.458$.
The $\varphi$ field renormalization constant is computed from  
\begin{equation}
Z_\varphi^{1/2}=\Big[M_\sigma^2 \Big(b^4\sum_y \langle\varphi_0(y)\varphi_0(0) \rangle \,
- \, V_4v ^2\Big)\Big]^{-1}\, ,
\end{equation}
where $V_4\!=\!L^3\!\times\! T$ is the lattice four-volume and the mass $M_\sigma$ of 
the one-$\varphi_0$ state is extracted from the time decay of the correlator 
$b^3 \sum_{\bf x} \langle\varphi_0(x_0,{\bf x})\varphi_0(0, {\bf 0})\rangle$. 
Once $M_\sigma$, $Z_\varphi$
and the bare scalar vev $v$ are known, $\lambda_R$ is easily evaluated.

In a numerical simulation on a finite lattice the scalar vev is always zero, even
if $m_\phi^2 < m_{cr}^2$. Hence following Refs.~\cite{Bulava:2012rb,Gockeler:1991ty}
the technique of  ``fixing'' the axial part of the global $\chi_L\times\chi_R$ 
symmetry was exploited in order to get $\langle \Phi \rangle = v 1\!\!1\ \neq 0$
in the NG phase. As this technique yields the
correct NG phase results up to finite size effects,
by simulating on lattices of different spatial size $L$ we checked that
for the observables discussed in the section {\em ``Lattice study and results''}
the finite-$L$ systematic errors on the quoted values at $L\simeq 2.4$~fm are well below
the uncertainties shown in Figs.~2 and~3 of the main text, or
those quoted in Sect.~\ref{sec:GLAN}. Indeed, as detailed
in Sect.~\ref{sec:FSE}, simulations at three $L$-values in the range 
$2.0~{\rm fm} < L < 3.4~{\rm fm}$  
show that the deviations of $M_{PS}$ and $m_{AWI}$ evaluated at $L\simeq 2.4$~fm 
from their infinite volume counterparts can be conservatively estimated to be
around 1-2\% and 5-7\%, respectively, and are hence marginally significant 
as compared to the corresponding purely statistical uncertainties of 
about 1.5\% and 3\%.
\begin{table}[h]
\begin{tabular}{|ccccccc|}
\hline
$b$/fm & $b^2m_{cr}^2$    & $r_0^2M_{\sigma}^2$  & $\lambda_{R}$ &$\beta$ & $b^2 m_\phi^2$ & $\lambda_0$   \\
\hline \hline
0.152 ~ & -0.5269(12)  & 1.278(6)  & 0.437(4)  & 5.75  & -0.5941  & 0.5807 \\
\hline
0.123 ~ &  -0.5357(11) & 1.286(6)  & 0.441(4)  & 5.85 & -0.5805 & 0.5917  \\
\hline
0.102 ~ &  -0.5460(10) & 1.290(7)  & 0.444(5) & 5.95  & -0.5756 & 0.6022  \\
\hline
\end{tabular}
\caption{Gauge and scalar bare parameters $\beta$, $m_\phi^2$, $\lambda_0$ 
and the corresponding measured value of the renormalized quantities
$r_0^2M_{\sigma}^2$ and $\lambda_{R}$ for the three lattice spacings we have 
considered. The value of the critical mass $b^2 m_{cr}^2$ which defines the phase 
transition point is also given.}
\label{Tab:ren_cond_NG}
\end{table}

The bare parameters used in the NG-phase simulations are listed
in Tab.~\ref{Tab:ren_cond_NG}.
Concerning the simulations in the Wigner phase aimed at determining $\eta_{cr}$, 
we have worked at the same lattice resolution (i.e.\ $\beta$) and the same bare 
quartic scalar coupling ($\lambda_0$) as employed in the NG phase. For each lattice
resolution we have chosen the value of the bare scalar mass parameter $b^2 m_\phi^2$
so as to obtain $(m_\phi^2- m_{cr}^2)r_0^2=1.22$ within errors.
The latter condition entails a small and uniform O($b^2 (m_\phi^2- m_{cr}^2)$) 
contribution to the lattice artifacts in 
the resulting $\eta_{cr}$ values at the three lattice spacings. These UV cutoff
artifacts are then propagated in the analysis of NG observables (see 
Sect.~\ref{sec:data_analysis}) and safely removed by extrapolating to the
continuum limit.

A comment about the triviality of the scalar quartic coupling is in order here.
Triviality implies that in order to really take the limit $\Lambda_{UV} \sim b^{-1}
\to \infty$ while keeping fixed the aforementioned low energy renormalization 
conditions, among which the one on $\lambda_R$ (see Eq.~(\ref{SCRC})), the bare
quartic coupling $\lambda_0$ should be made to diverge logarithmically 
with increasing $\Lambda_{UV}$. 
In practice however we shall estimate the limit $b^{-1} \to \infty$ of renormalized
observables by considering their values at a few values of $1/b^2$ which differ
from each other by a factor of about two and correspond to values of $\lambda_0$
and $g_0^2 $ that change only by a few percent in the range $\lambda_0 \lesssim 0.6$
and $g_0^2 \gtrsim 1.0$ . Under these circumstances it is a very plausible assumption 
that scaling violations behave effectively as polynomials in $b^2$ (about the absence 
of odd-powers of $b$ see Sect.~\ref{sec:RADA}), because terms O($b^2 \log^k(b^2)$) --
with any integer $k$ -- can not be resolved from genuine O($b^2$) terms within numerical 
errors and in the limited range of $1/b^2$-values that we explore. This assumption
that we make in our continuum limit analyses i) makes practically harmless the problem
of triviality and ii) is fully analogous to the assumption that are made (and well checked) 
when taking the continuum limit in all the numerical studies of lattice QCD, and more 
recently lattice QCD+QED.

\section{Data analysis details}
\label{sec:data_analysis}

\subsection{Analysis of section ``Lattice study and results''}
\label{sec:analybasic}

In the Wigner phase  we compute $r_{AWI}$ from the ratio~(L15) 
for the values  of the bare parameters $\eta$ and $\mu$ listed in 
Tab.~\ref{tab:sim_Wigner} and for a value of $\tau \equiv y_0-x_0 =0.6$~fm
chosen conveniently (see below) and kept fixed for the three lattice spacings. 
We fit the dependence of $r_{AWI}$ on $\eta$ and $\mu$ through the linear ansatz 
$r_{AWI}=K_0+K_1\eta+K_2\mu$, which turns out to describe very well our data. We 
thus identify $\eta_{cr}$ with $-K_0/K_1$. The resulting values of $\eta_{cr}$ for the 
three lattice spacings corresponding to $\beta=5.75, 5.85, 5.95$ 
are given in Tab.~\ref{tab:eta_cr}. There we also give the values of the
Sommer scale $r_0$ and of the
inverse of the pseudoscalar density renormalization constant $Z_P$, which
is defined in a hadronic scheme in the Wigner phase as $Z_P^{-1} = r_0^2 
\langle 0 | P^1 | {\rm PS~meson} \rangle^W$ (see main text, Sect.~{\em ``Lattice 
study and results''} ). The data for $\langle 0 | P^1 | {\rm PS~meson} \rangle^W$
at fixed lattice spacing are within statistical errors well linear in $\mu$ and in $\eta$
and were extrapolated to $\mu=0$ and $\eta=\eta_{cr}$ in order to obtain the quoted 
results for $Z_P^{-1}$. 

A remark is in order about the physics of the model in the Wigner phase, i.e. 
for $m_\phi^2 - m_{cr}^2>0$. Owing to the presence in the lattice action of the 
twisted mass term $\propto \mu$ (see Eq.~(\ref{ACT-TM})), 
which is needed to give a robust IR cutoff
to all observables, here, even if $\Phi$ takes zero vacuum expectation value, 
strong interactions induce anyway, just like in QCD, the spontaneous breaking 
of the $\tilde\chi$ symmetries. At $\eta=\eta_{cr}$ the latter get restored
up to negligible O($b^2$), the scalar field decouple from quark and gluons and
(in large volume) one observes pseudoscalar mesons with squared mass $M_{PS}^2
\sim \mu$ -- see the discussion in Sect.~\ref{sec:MPS_Wig}. In this respect the 
name of Wigner phase may be slightly abusive. The lattice spatial size $L$ is
large enough that the pseudoscalar meson mass at all our ``Wigner phase'' 
simulation points satisfies $M_{PS}(\eta,\mu) L \geq 5.3$. 
\begin{table}[hpt]
\begin{tabular}{|cccc|}
\hline
$b$/fm  &  $(L^3 \times T)/b^4$ & $\eta$      &  $b\mu$     \\
\hline
 0.152       &  $16^3 \times 32$ &  -1.1505          & 0.0180  0.0280   0.0480   \\
($\beta=5.75$)&     &    -1.1898        &      0.0180  0.0280   0.0480            \\
                          &     &    -1.3668        &      0.0180  0.0280   0.0480\\
\hline
 0.123     &  $16^3 \times 40$ &  -1.0983    &   0.0224   0.0316    0.0387     \\    
($\beta=5.85$) &  &  -1.1375    & 0.0120  0.0224 0.0316 0.0387  \\ 
                           &  &  -1.2944    & 0.0224       0.0387   \\ 
\hline
   0.102       &  $20^3 \times 48$ & -0.9761 &    0.0186, 0.0321    \\
($\beta=5.95$) &      & -1.0354 &     0.0186, 0.0321        \\
                          &      & -1.0771 &   0.0100  0.0186, 0.0321       \\
\hline
\end{tabular}
\caption{Simulation parameters in the Wigner phase. We used 480 gauge and 
scalar paired configurations at each simulation point. Each of these bosonic 
configurations is different from any other by either the scalar field or both
the scalar and the gauge ones.}

\label{tab:sim_Wigner}
\end{table}
\begin{table}[hpt]
\begin{center}
\begin{tabular}{|ccccc|}
\hline
 $\beta$ & $\lambda_0$ &$\eta_{cr}$ & $Z_P^{-1}$ & $r_0/b$ \\
\hline
~ 5.75 ~ & ~~ 0.5807 ~   & ~~ -1.271(10) ~ & ~~ 20.6(7)  ~  & ~~ 3.291(16) ~ \\
\hline
 5.85 & 0.5917 & -1.207(8) & 20.1(7)  & 4.060(22)   \\
\hline
 5.95 & 0.6022& -1.145(6) & 20.7(9)  & 4.912(29)    \\
\hline
\end{tabular}
%
%
\caption{The values of $\eta_{cr}$ and $Z_P^{-1}$ at our three lattice 
spacings for $\rho = 1.96$, the listed values of $\rho$ and $\lambda_0$ 
and $b^2m_\phi^2$ such that $r_0^2(m_\phi^2-m_{cr}^2) \simeq 1.22$. 
We also quote the values of $r_0$ in lattice units, taken from 
Eq.~(2.6) of \cite{Necco:2001xg}, that we use in the present work.}  
\label{tab:eta_cr}
\end{center}
\end{table}

\begin{figure}[hpt]
\begin{center}
\includegraphics[scale=0.5]{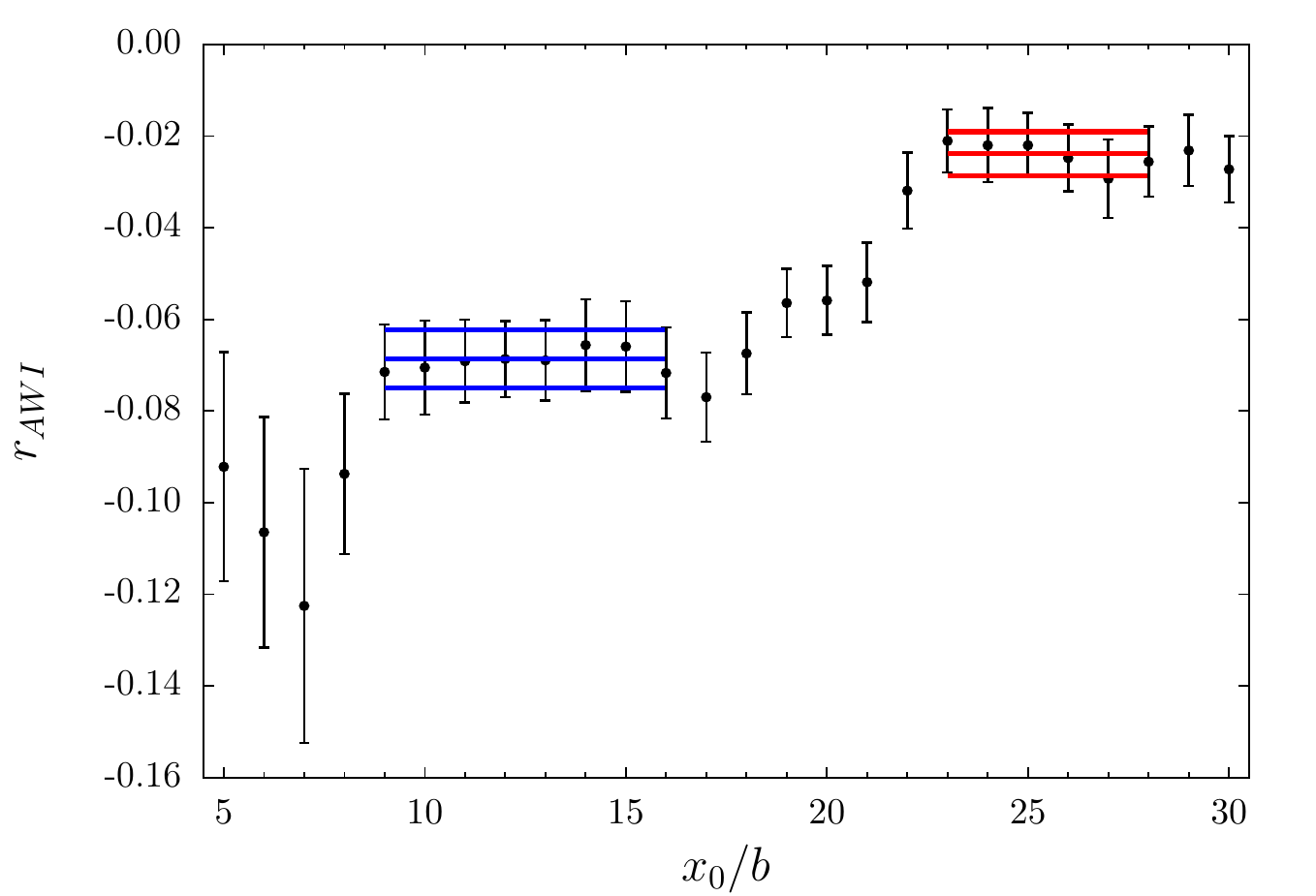}
\end{center}
\vspace{-0.3cm}
\caption{A typical plot of $r_{AWI}$ against $x_0/b$, here for the case of $\eta=-1.1374$, $b\mu=0.0224$ and $\beta=5.85$. The two time regions where $r_{AWI}$ shows a plateau are $A: x_0 \sim[0.9,1.8]$~fm and $B: x_0 \sim[2.7,3.3]$~fm.}
\label{fig:fig_r_AWI_A_B}
\begin{center}
\includegraphics[scale=0.5]{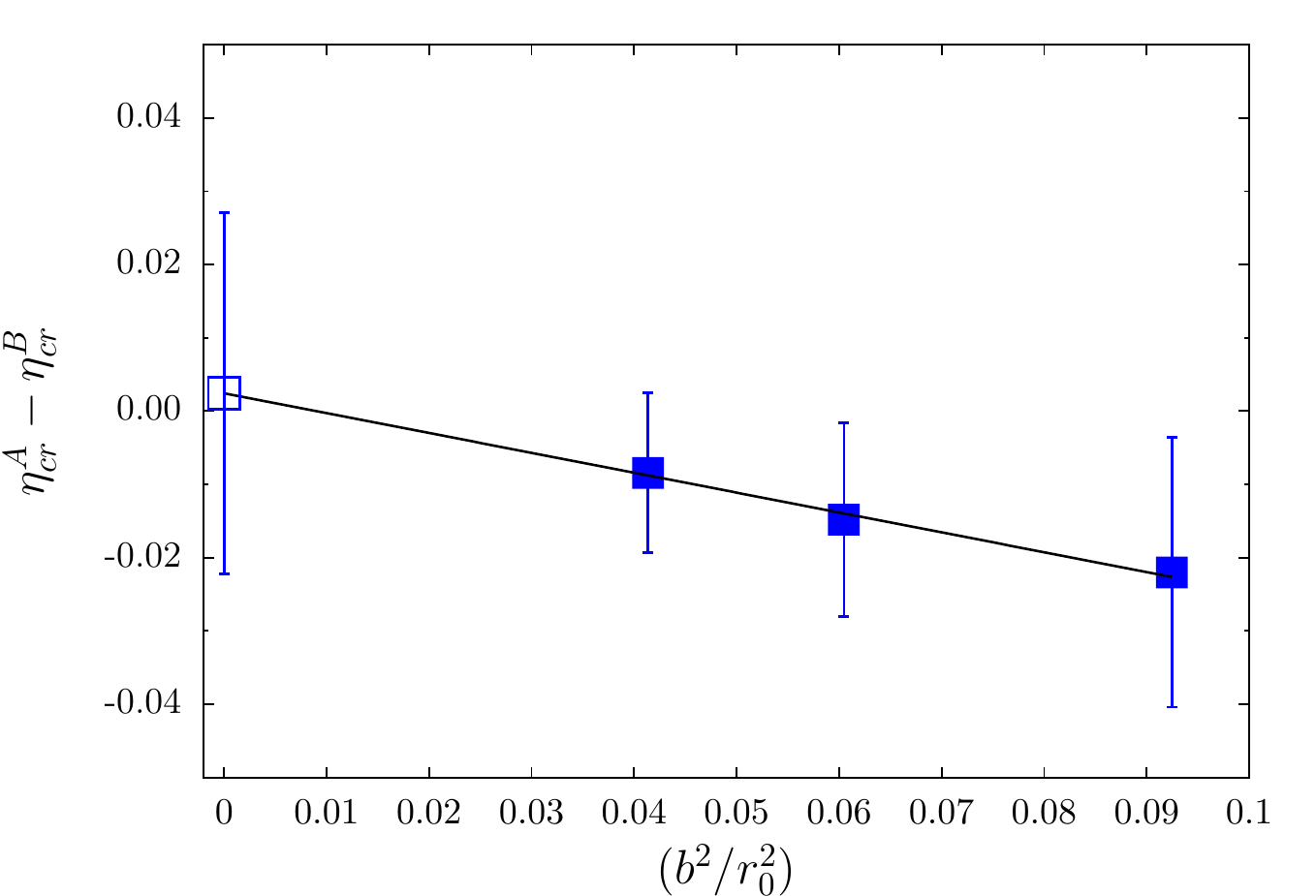}
\end{center}
\vspace{-0.3cm}
\caption{The difference $\eta_{cr}^A - \eta_{cr}^B$ vs.\ $b^2/r_0^2$, where 
$\eta_{cr}^{A(B)}$ is estimated by using the plateau value of $r_{AWI}$ in the 
time region A(B). The value of $\eta_{cr}^A - \eta_{cr}^B$ extrapolated to the 
continuum limit is well compatible with zero.}
\label{fig:eta_cr_AB_vs_b}
\end{figure}

Another technical remark concerns our particular choice of the lattice estimator
for $\eta_{cr}$. By definition this special value of $\eta$ is the one at which
the correlator ratio $r_{AWI}$ (see Eq.~(L15)) vanishes for time separations $x_0$ 
and $y_0-x_0$ both sufficiently away from zero, up to O($b^2$) lattice artifacts. 
Due to the latter, in practice one obtains different lattice estimators of $\eta_{cr}$ 
for different ranges of $x_0$ (while we keep $y_0-x_0 \equiv \tau = 0.6$~fm fixed), because 
different combinations of intermediate states contribute to three-point correlators entering 
$r_{AWI}$. In principle any choice of the $x_0$--range is legitimate, however in order 
to keep lattice artifacts under control we make a choice of $x_0$-range (the region
A discussed in next paragraph) where only intermediate states with low mass 
contribute to the correlators. By considering a second reasonable choice of $x_0$--range 
(the region B below) we also explicitly check that the difference between the two resulting 
lattice estimators of $\eta_{cr}$ vanishes in the continuum limit, as expected. The 
selection of these two $x_0$ ranges is suggested by the fact that the profile of $r_{AWI}$ 
as a function of $x_0$ (Fig.~\ref{fig:fig_r_AWI_A_B}) typically exhibits two 
plateau regions.

The $r_{AWI}$ values shown in Fig.~1 of the main text and the $\eta_{cr}$ results of 
Tab.~\ref{tab:eta_cr} all refer to the time plateau region at $x_0$ smaller than $T/2$ 
that we call region A. In this case 
the $x_0$ range is chosen in such a way that the three-point correlators 
in Eqs.~(L16)-(L17) are dominated by the PS-meson state
at times between 0 and $x_0$, by the one-$\Phi$-particle state at times 
between $x_0$ and $y_0 = x_0+\tau$ and by the vacuum at times larger than $y_0$. 
The time plateau region of $r_{AWI}$ called region B corresponds instead to $x_0$ 
larger than $T/2$. In this situation the same three-point correlators are dominated by 
the vacuum at times between 0 and $x_0$, by a PS-meson plus one-$\Phi$-particle 
state at times between $x_0$ and $y_0$ and by one-PS-meson state at times larger 
than $y_0$.
We check in Fig.~\ref{fig:eta_cr_AB_vs_b} that the difference in the values of $\eta_{cr}$ 
obtained by evaluating $r_{AWI}$ from the two plateau regions A and B vanishes
as an O($b^2$) quantity in the continuum limit. 

In the NG phase we compute $M_{PS}$ and $m_{AWI}$ for the values of $\eta$ 
and $\mu$ listed in Tab.~\ref{tab:sim_NG}. Again strong interactions are 
responsible for the spontaneous breaking of $\tilde\chi$ symmetries (restored 
at $\eta_{cr}$), with the vacuum being polarized by the combined effect of
the twisted mass term ($\propto \mu$) and the NP $\tilde\chi$ breaking 
term(s) (e.g. $\propto c_1\Lambda_S$) in the effective 
Lagrangian (see Eq.~(L13)).
The lattice spatial size $L$, which is kept fixed around $2.4$~fm (if we assume 
$r_0 = 0.5$~fm for the Sommer scale $r_0$) 
at the three lattice spacings, is large enough that the pseudoscalar meson 
mass $M_{PS}$ always satisfies $M_{PS}(\eta,\mu)L \geq 4.7$.

The pseudoscalar (PS) meson mass $M_{PS}$ is straightforwardly evaluated from the
two-point correlator $\sum_{\bf x}\langle P^{1}(x) P^1(0)\rangle $, which at all
our simulation points exhibits a clear exponential dependence on the time separation
$x_0$, with ground PS-state dominance for large $x_0$ (typically $\gtrsim 1$~fm).
In terms of the lattice fermion 
operators~(\ref{latt_PSdens})-(\ref{latt_AXcurr}) $m_{AWI}$ reads (see Eq.~(L11))
\begin{equation} \label{mAWI_latt}
m_{AWI} = \frac{  \partial_{0}^{FW} \!\sum_{\bf x}\langle \tilde A_0^{1,BW}(x) P^1(0)\rangle }{ 2 \sum_{\bf x}\langle P^{1}(x) P^1(0)\rangle }\Big{|}_{\eta_{cr}}  \, .
\end{equation}
We extrapolate these observables to $\eta_{cr}$ and $\mu=0$ with the ansatz 
$M_{PS}^2=Y_0+Y_1\eta+Y_2\mu+Y_3\eta^2+Y_4\mu^2+Y_5\eta\mu\;$ and 
$m_{AWI}=Y_0+Y_1\eta+Y_2\mu+Y_3\mu^2\;$.

In this way our analysis assumes no special value for $c_1\Lambda_S$. In 
particular the chosen fit ansatz for the $\mu$-dependence of our observables 
and the variants described in Sect.~\ref{sec:GLAN} are suitable to 
describe the data even in the case of $c_1\Lambda_S=0$, when 
$m_{AWI}$ should be (owing to lattice artifacts) an analytic function of $\mu$
that at $\eta=\eta_{cr}$ vanishes smoothly as $\mu \to 0$, and in the same
limit the relation $M_{PS}^2 \sim \mu$ is expected to hold -- up to small anomalous
``chiral log'' corrections in the quenched approximation~\cite{Bernard:1992mk,Sharpe:1992ft}.
If instead $c_1\Lambda_S \neq 0$, then the low energy effective theory of the 
critical model must involve both massive (even in the $\mu \to 0$ limit) 
pseudoscalar meson and the Goldstone modes of the scalar field. In this situation 
at $\eta=\eta_{cr}$ our observables will depend on the effective quark mass 
$\propto \sqrt{ (c_1\Lambda_S)^2 + \zeta^2 \mu^2 }$, which in turn admits 
a smooth Taylor expansion around the $\mu=0$ point. Here $\zeta$ denotes a 
relative normalization factor which is immaterial for our discussion. 

The definition of a renormalized counterpart of the matrix element ratio that we 
have suggestively denoted by $m_{AWI}$ (see Eq.~(L11)) requires computing 
both $Z_P^{-1}$, which, 
as discussed in the section {\em ``Lattice study and results''}, is provided by an hadronic 
matrix element evaluated for $(\eta,\mu) \to (\eta_{cr},0)$ in the Wigner phase, and
$Z_{\tilde A} = Z_{\tilde J}|_{\eta_{cr}} = Z_{\tilde V}$. 
Since this last renormalization factor is independent
of the twisted mass parameter $\mu$ appearing in the action term~(\ref{ACT-TM}),
it can be accurately computed by imposing, at $\eta=\eta_{cr}$, the validity of the 
$\tilde\chi$ SDE involving the bare lattice vector current $\tilde V^{2,BW}=\tilde J^{2,BW}_L+\tilde J^{2,BW}_R$, namely 
\begin{eqnarray}
\hspace{-.8cm}&&Z_{\tilde V}\partial_{0}^{FW} \!\sum_{\bf x}\langle \tilde V_0^{2,BW}(x) P^1(0)\rangle\Big{|}_{\eta_{cr}}=\label{ZV}\\
\hspace{-.8cm}&&= 2\mu  \sum_{\bf x} \langle P^1(x)P^1(0)\rangle\Big{|}_{\eta_{cr}} +{\mbox{O}}(b^2)\, .\nonumber
\end{eqnarray}
Indeed from the exact WTI of $\chi_L \times \chi_R$ symmetry that is satisfied by 
the lattice $\chi$-vector current $V_\nu^2$,
 \begin{eqnarray}
\partial_\nu V_\nu^2(x) = 2 \mu  P^1(x)  \, ,\label{CHICUR}
\end{eqnarray}
we learn that the combination $2 \mu P^1$ needs neither to be renormalized nor rescaled by
any factor. In practice the estimates of $Z_{\tilde V}$ that we obtained in the NG phase 
by imposing Eq.~(\ref{ZV}) at the simulation points in Tab.~\ref{tab:sim_NG} exhibit only a 
smooth dependence on $\eta$ and $\mu$, which allowed us to extrapolate them to 
$\eta\to\eta_{cr}$ and $\mu\to 0$ by assuming the polynomial ansatz $a+b\eta+c\mu+d\mu^2$.

\begin{table}[hpt]
\begin{tabular}{|ccc|}
\hline 
 $b$/fm   & $\eta$ & $b\mu$    \\
\hline
0.152   & -1.2714   &0.0050 0.0087, 0.0131, 0.0183, 0.0277   \\
   ($\beta=5.75$)                                            & -1.2656       & 0.0131                 \\
                                                 &  -1.2539          & 0.0183 \\ 
                                                 & -1.2404  &0.0131 \\
                                					  & -1.231   &  0.0087,  0.0183 \\
\hline
0.123
 & -1.2105   & 0.0040, 0.0070, 0.0100, 0.0120   \\
   ($\beta=5.85$)                                                   & -1.2068   & 0.0040, 0.0070, 0.0100              ,0.0224\\
                                                   & -1.2028   & 0.0040, 0.0070, 0.0100, 0.0120, 0.0224      \\ 
                                                   & -1.1949   &  0.0100       \\ 
                                                  & -1.1776   &  0.0070, 0.0100, 0.0140, 0.0224        \\ 
&   -1.1677           &     0.0316    \\                               
\hline
0.102   & -1.1474   & 0.0066, 0.0077, 0.0116, 0.0145, 0.0185   \\
   ($\beta=5.95$)                                                   & -1.1449    & 0.0060, 0.0077, 0.0116, 0.0145                 \\
                                                  & -1.1215 & 0.0077 \\
                                                   & -1.1134     & 0.0077, 0.0108  \\
\hline
\end{tabular}
\caption{Simulation parameters in the NG phase. We used 60 gauge and scalar paired configurations at each simulation point. The spatial extension of the lattice is fixed to $L\sim2.4$~fm, while the time extension is $T=6.0$~fm for the coarsest lattice spacing and $T=4.9$~fm for the two other ones.}\label{tab:sim_NG}
\end{table}

\vspace{-0.3cm}
\subsection{Check of analysis stability and systematics}
\label{sec:GLAN}

\begin{figure}[htbp]
\centerline{\hspace{0.0cm} \includegraphics[scale=0.55,angle=0]{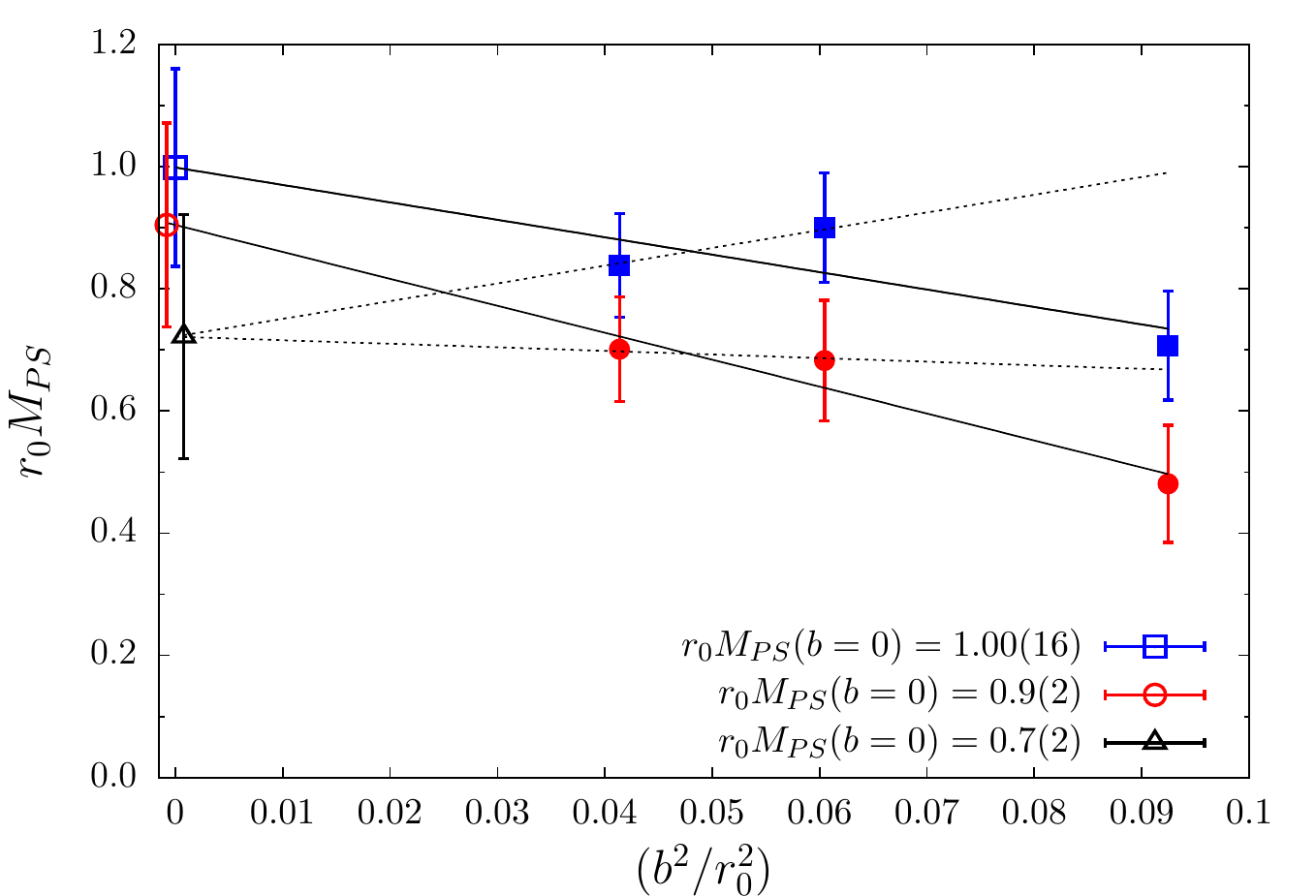} }
\caption{$M_{PS}r_0$ data and continuum limit extrapolations as obtained by working 
at $\eta^A_{cr}$ (blue points) or at $\eta^B_{cr}$ (red points) 
and using data at three lattice spacings as well
as combining data on the two finer lattices at both $\eta^A_{cr}$ and $\eta^B_{cr}$
under the constraint of a common continuum limit. A conservative choice of the time plateau
for $M_{PS}$ and the fit ansatz $M_{PS}^2 =Y_0 + Y_1 \eta + Y_2 \mu +Y_3\eta^2+Y_4\mu^2+Y_5\eta\mu$ 
were adopted.
}
\label{fig:M_PS_3ancomp}
\end{figure}

\begin{figure}[htbp]
\centerline{\hspace{0.0cm} \includegraphics[scale=0.55,angle=0]{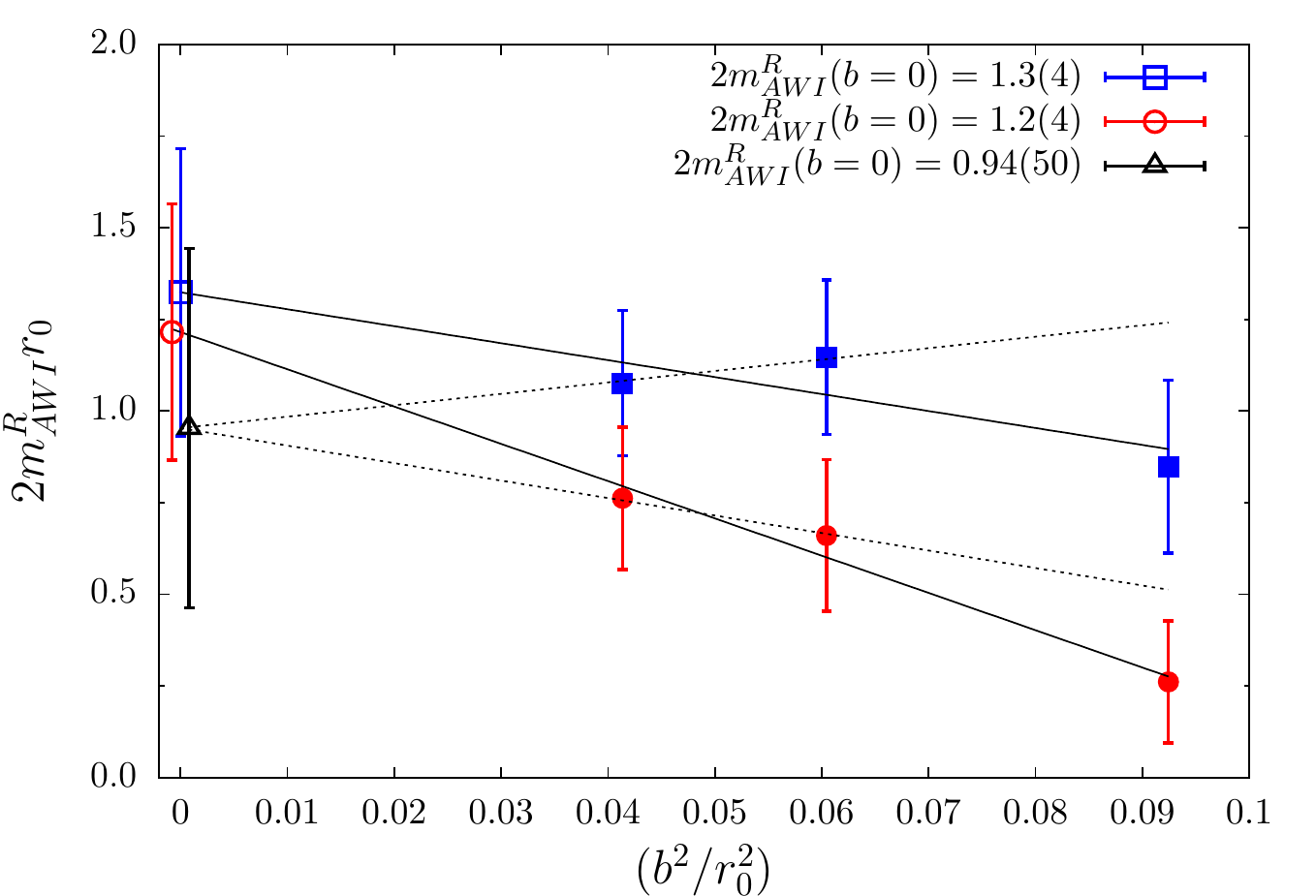} }
\caption{Data and continuum limit extrapolations for $2m_{AWI}^R r_0 \equiv 2r_0 m_{AWI} 
Z_{\tilde V} Z_P^{-1}$ in the same cases as for Fig.~\ref{fig:M_PS_3ancomp}. The
fit ansatz $m_{AWI}=Y_0+Y_1\eta+Y_2\mu+Y_3\mu^2$ was adopted. The large values of
$m_{AWI}^R = Z_P^{-1} m_{AWI}$ as compared to $m_{AWI}$ reflect the large values of 
$Z_P^{-1}$ (see Table~\ref{tab:eta_cr}) in the chosen hadronic renormalization scheme.}
\label{fig:m_AWI_3ancomp}
\end{figure}

In order to check the stability of the results presented in 
the section {\em ``Lattice study and results''} of the main text and the 
underlying analysis of Sect.~\ref{sec:analybasic} we have studied the impact of modifying in turn 
\begin{enumerate}
\item the plateau choice for $r_{AWI}$, which was taken in the time region either A or B of Fig.~\ref{fig:fig_r_AWI_A_B};
\item the fit function for $r_{AWI}$ in the Wigner phase, which was chosen to be either of the form $K_0+K_1\eta+K_2\mu$ or extended by adding the non-linear term $K_4 \mu^2$ (data are very well linear in $\eta$). In addition the mean value over the results from the two fits above was considered;
\item the choice of the time plateau for $M_{PS}$ and $m_{AWI}$, taken to be either $[1.4,2.2]$~fm or $[1.1,2.4]$~fm;
 \item the fit functions for $m_{AWI}$ and $M_{PS}^2$ in the NG phase, which 
 were both either truncated to $Y_0+Y_1\eta+Y_2\mu$ 
 or extended by adding in turn terms non-linear in $\eta$ and/or $\mu$,
 such as $Y_4\mu^2$, $Y_3\eta^2 + Y_4\mu^2$ or $Y_3\eta^2 + Y_4\mu^2 + Y_5\eta\mu$.
\end{enumerate}
Our care in checking the determination of $\eta_{cr}$ at each lattice spacing is due to
the fact that its uncertainty, once propagated onto $m_{AWI}$ and $M_{PS}^2$, turns
out to be the main source of error in the final results~(\ref{MPSM})--(\ref{MAWIM}).
For each combination of the above choices the properly renormalized results of the fits 
to $(\eta,\mu) = (\eta_{cr},0)$, i.e.\ $r_0M_{PS}$ and $2 r_0 m_{AWI} Z_{\tilde V} Z_P^{-1}$,
were extrapolated linearly in $b^2/r_0^2$ to the continuum limit.
To estimate the impact of possible O($b^4$) and higher order lattice artifacts, we have 
also considered a continuum extrapolation excluding the coarsest lattice spacing, but 
imposing that the two choices of taking either $\eta^A_{cr}$ or $\eta^B_{cr}$ yield 
the same continuum value. For a typical choice of the time plateau and the fit function 
in $\eta$ and $\mu$ for $M_{PS}r_0$, in Fig.~\ref{fig:M_PS_3ancomp} we show the 
continuum limit extrapolations obtained by working at $\eta^A_{cr}$ or at $\eta^B_{cr}$
and using data at three lattice spacings, as well as the combined continuum extrapolation
that results from using data on the two finer lattices at both $\eta^A_{cr}$ and $\eta^B_{cr}$.
The analogous comparison for the renormalized ratio $2r_0 m_{AWI} Z_{\tilde V} Z_P^{-1}$ 
is presented in Fig.~\ref{fig:m_AWI_3ancomp} .

The combination of the 
variants discussed above produced 72 different analyses, none of which
gave a result for $m_{AWI}$ or $M_{PS}$ that is compatible with zero 
(i.e.\ with the no-mechanism scenario). The median over all the analyses, 
excluding only  a few ones for $M_{PS}^2$ that yield 
a $\chi^2$ value per degree of freedom larger than 2, gives
\begin{eqnarray}
\hspace{-0.6cm}& r_0M_{PS}  &  = \, 0.93\!\pm \!{0.09}\!\pm\!{0.10} \, ,
\label{MPSM} \\
\hspace{-0.6cm}& 2 r_0 m_{AWI} Z_{\tilde V} Z_P^{-1} &  = \, 1.20\!\pm \!{0.39}\!\pm\!{0.19} \, ,  \label{MAWIM}
\end{eqnarray}
where the first error is statistical and the second systematic. These two error terms 
were estimated by attributing equal weight to all analyses, following the method of 
Ref.~\cite{ETMC_analysis} (see there the text around Eq.~(28)). 

Further strong evidence for a non-zero fermion mass in the NG phase EL comes 
from the following ``{\it reductio ad absurdum}'' argument. 
Let us assume that no fermion mass term is generated.
If this were the case, at $\eta=\eta_{cr}$ 
one should have $M_{PS} = {\mbox{O}}(b^2)$, and hence $M_{PS}^2$
would approach a vanishing continuum limit with only O($b^4$) artifacts.
\begin{figure}[htbp]
{\centerline{\includegraphics[scale=0.50,angle=0]{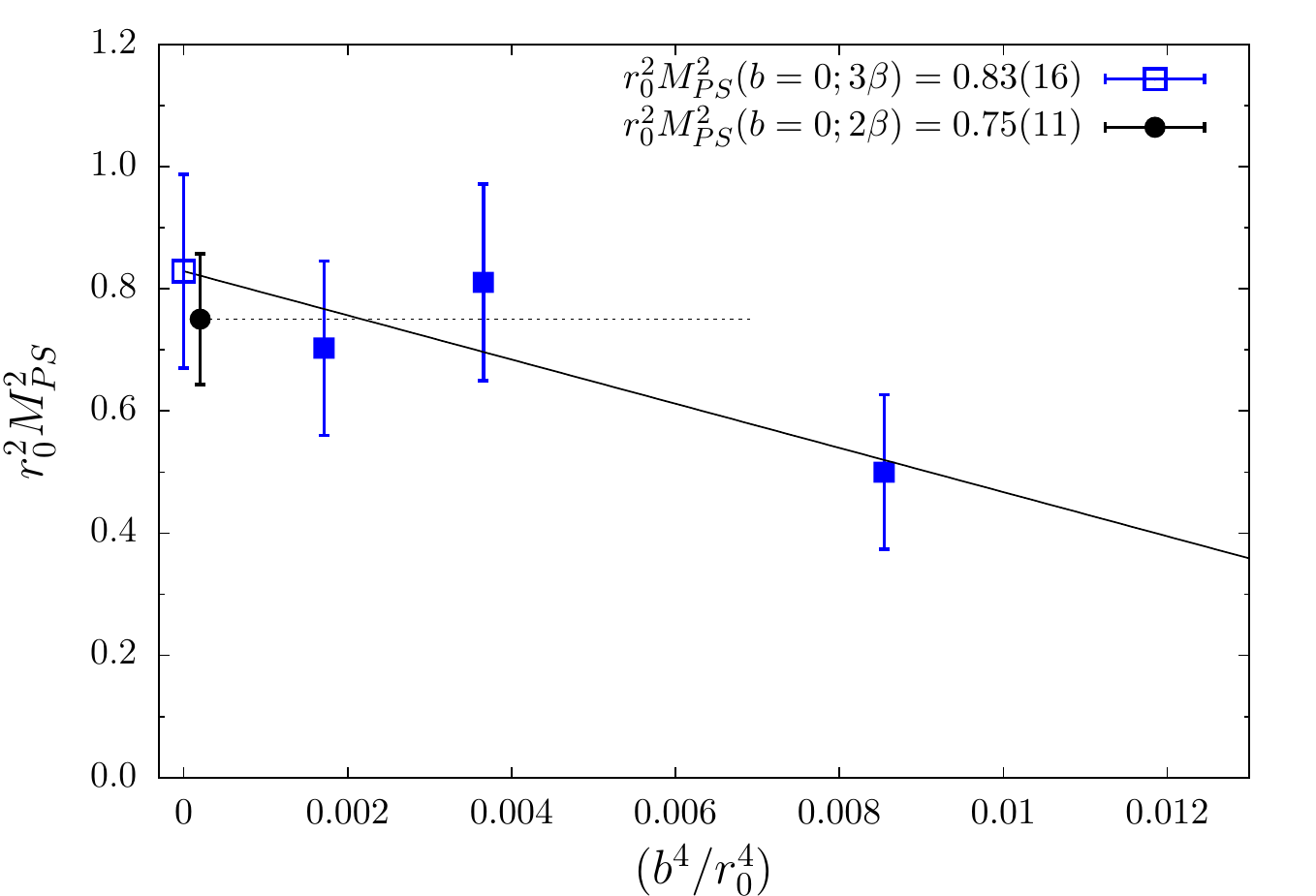}}}
\caption{\small{$r_0^2 M_{PS}^2$ against $(b/r_0)^4$ in the NG phase: 
full (dotted) straight lines show the continuum extrapolation linear in $b^4$ using 
data at all three (only the two finest) lattice spacings.}}
\label{fig:figBQ}
\end{figure}
In Fig.~\ref{fig:figBQ} we plot $r_0^2 M_{PS}^2$ as a function of $(b/r_0)^4$
(the original data are those at $\eta=\eta_{cr}^A$ already shown  
in Fig.~3 of the main text, and a subset of
those in Fig.~\ref{fig:M_PS_3ancomp}).
On a $(b/r_0)^4$ scale the lattice results for $r_0^2 M_{PS}^2$ 
lie very close to the continuum limit, which appears to be non-zero 
by more than five standard deviations, whatever ansatz is taken for
the continuum extrapolation. Thus the hypothesis of ``no mechanism'' 
is not supported by lattice data.

\subsection{Check and irrelevance of finite size effects}
\label{sec:FSE}

Our observables in the NG phase were evaluated on lattices of linear size 
$L = 2.4$~fm. In order to check the impact of finite-size effects (FSE),
for one lattice resolution ($\beta=5.85$, $b \simeq 0.123$~fm) we
carried out simulations in the NG phase at precisely the same bare
parameters as specified in Table~\ref{tab:sim_NG} for two additional 
lattice volumes, namely $L/b=16 \leftrightarrow  L\simeq 2.0$~fm and 
$L/b=28 \leftrightarrow L\simeq 3.4$~fm. The lattice time extension
was kept fixed at $T/b=40$, i.e.\ $T\simeq 4.8$~fm, and we
restricted attention to $\eta=-1.2068$, practically coinciding with 
$\eta_{cr}$, and two twisted mass values, $b\mu=(0.0040,0.0070)$.

\begin{figure}[htbp]
\centerline{\hspace{0.0cm} \includegraphics[scale=0.44,angle=0]{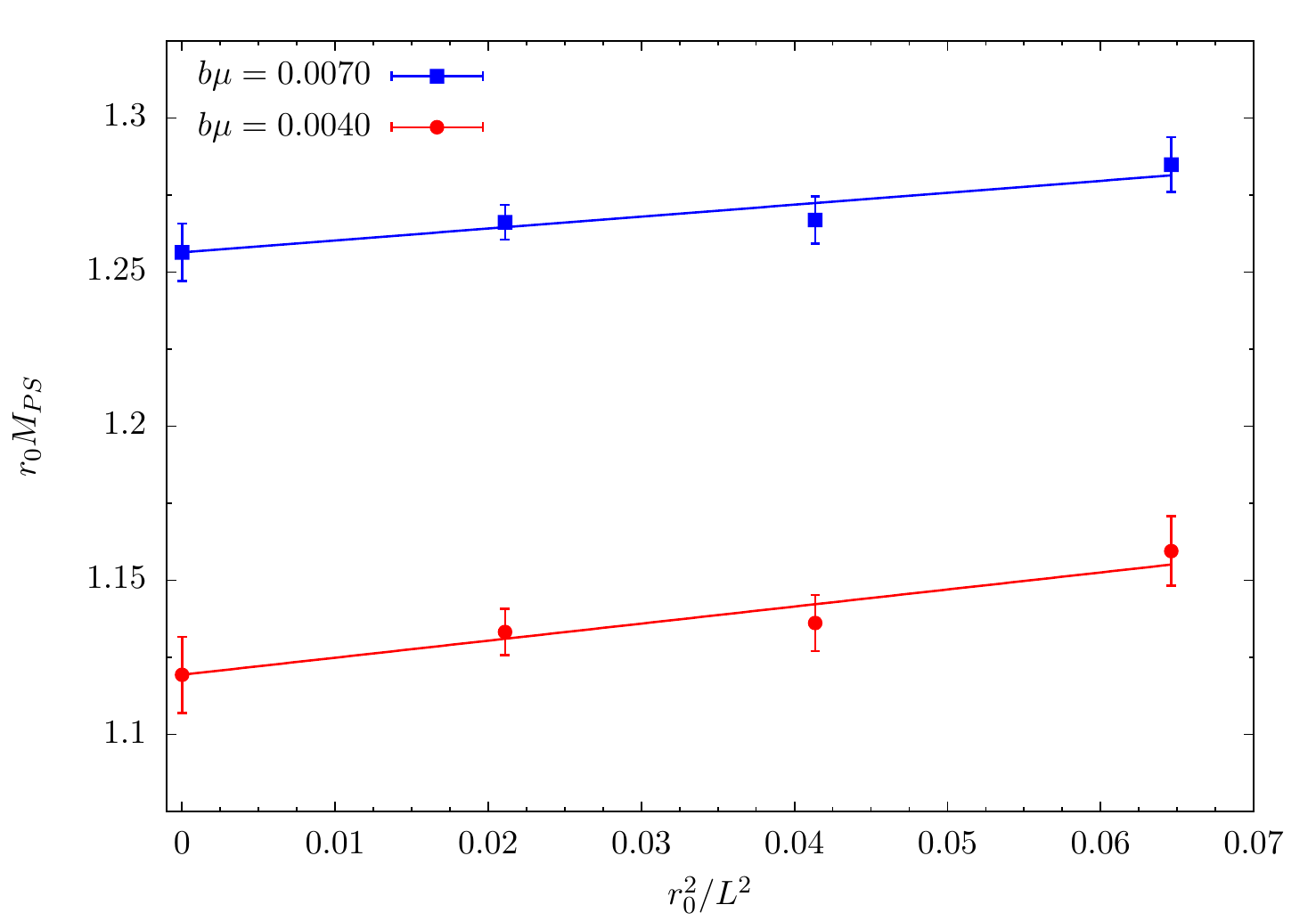} }
\caption{The PS meson mass $r_0M_{PS}$ for different lattice sizes $L$ as a function of 
$(r_0/L)^2$ in the the critical model at $\beta=5.85$ and two values of twisted mass $b\mu$,
together with its extrapolated value at $(r_0/L)^2=0$. The two linear extrapolations have 
very small $\chi^2 < 0.1$.
} 
\label{fig:FSE-MPS}
\end{figure}

\begin{figure}[htbp]
\centerline{\hspace{0.0cm} \includegraphics[scale=0.44,angle=0]{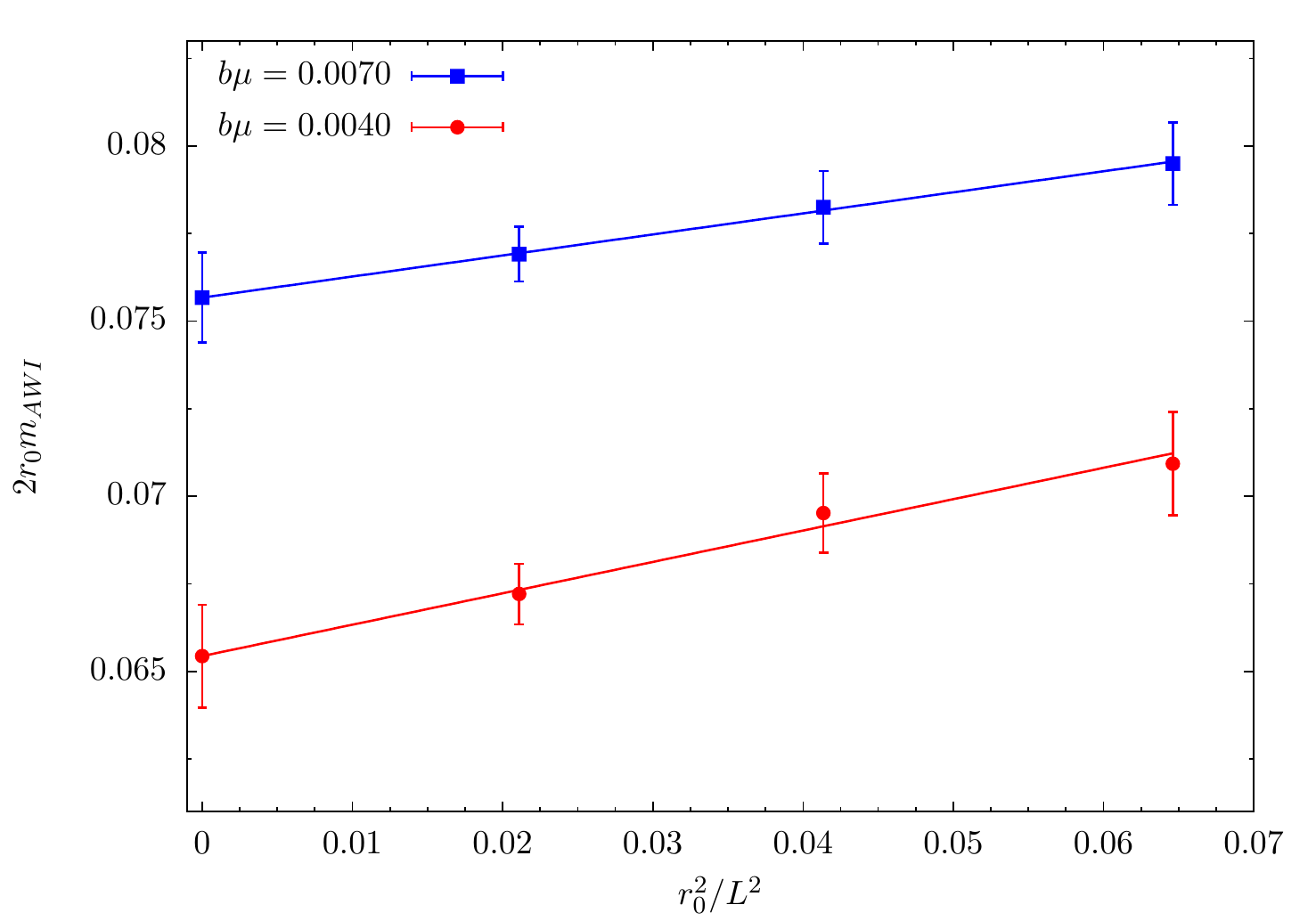} }
\caption{The (unrenormalized) matrix element ratio $2m_{AWI}r_0$ 
for different lattice sizes $L$ as a function of
$(r_0/L)^2$ in the the critical model at $\beta=5.85$ and two values of twisted mass $b\mu$,
together with its extrapolated value at $(r_0/L)^2=0$. The linear extrapolations have 
$\chi^2$ around 0.2 ($b\mu=0.007$) and 0.4 ($b\mu=0.004$). 
}  
\label{fig:FSE-mAWI}
\end{figure}

As the results at $L = 2.4$~fm indicate a clear evidence in favour
of the dynamical generation of a NP terms $\propto c_1\Lambda_S$ in
the effective Lagrangian (L13) of our model, we expect that, owing
to this and analogous NP effective action terms, the Goldstone modes
of the scalar field $\Phi$ are still coupled to the fermion-gauge
sector in spite of working at $\eta_{cr}$. Thus the
low energy effective theory should include both these Goldstone modes
and the massive PS mesons, with interactions giving rise to FSE due
to around-the-world propagation of Goldstone modes and PS mesons.
While the latter will be exponentially suppressed at large $L$ (at 
$L= 2.4$~fm for $b\mu=0.0040$ we have $M_{PS}L \simeq 5.8$), the 
former kind of FSE are expected to decrease only as $1/L^2$ and
will hence be dominating. For this reason in Figs.~\ref{fig:FSE-MPS}
and~\ref{fig:FSE-mAWI} we plot as a function of $(b/L)^2$ our results 
for $bM_{PS}$ and $bm_{AWI}$, together with their linear extrapolation 
to infinite $L$. The use of the simple extrapolation ansatz $K_0 +
K_1 L^{-2}$ is justified a posteriori by the relative smallness
of the observed FSE, well below 3\% for $M_{PS}$ and 10\% for $m_{AWI}$
in the whole $L^{-2}$--range. The error bars on the simulated data
points are merely statistical, plus those entailed by the extrapolation 
in $L^{-2}$ for the infinite volume estimates. 

Although given these errors the statistical significance for FSE is
not impressive (especially in the case of $M_{PS}$), still we do
observe a uniform trend for both $M_{PS}$ and $m_{AWI}$, which is
consistent with a slight linear decrease as a function of $L^{-2}$.
As anticipated in Sect.~\ref{Sec:Gauge-Scal}, the deviation of $M_{PS}$ 
and $m_{AWI}$ evaluated at $L\simeq 2.4$~fm from their infinite volume 
counterparts  can be conservatively estimated to be around 1-2\% and 5-7\%, 
respectively. 

FSE corrections of this size are (in particular for the matrix element ratio 
named $m_{AWI}$, see Eq. (L17)) a bit larger than those one should observe 
(see e.g.\ Ref.~\cite{Colangelo:2005gd}) in a QCD-like theory with scalars 
completely decoupled, but seem to be of a reasonable magnitude for a model with 
Goldstone modes non-perturbatively coupled to the fermions owing to a non-zero
$c_1$. The latter can be roughly estimated to be of order 0.1. In fact on the basis of 
the $M_{PS}$ results quoted in Eq.~(\ref{MPSM}) $c_1$ should describe a NP fermion
mass close to one third of the strange quark mass (if we calibrate our model
with $r_0$ from QCD). 

Anyway the observed finite-$L$ corrections on the results for $M_{PS}$ and 
$m_{AWI}$ at $L\simeq 2.4$~fm are well below the uncertainties shown in Figs.~2 
and~3 of the main text, or those quoted in Eqs.~(\ref{MPSM})--(\ref{MAWIM}). 

A more detailed understanding of the FSE on specific hadron masses and operator matrix
elements is beyond the scope of the present work. It is however important to remark that 
the evidence for a NP terms $\propto c_1\Lambda_S$ in the effective Lagrangian of
our critical model can be perfectly well established in finite volume, provided
the system volume ($L^3 T$) is large enough to allow for vacuum orientation under the
phenomenon of (dynamical) spontaneous breaking of the $\tilde\chi$ flavour symmetries.

In our case we can check the large volume condition by looking at the value
of $M_{PS}L$ for $L=2.4$~fm, or equivalently $L/r_0 = 4.8$, and $M_{PS} \sim 
0.93/r_0$, as obtained in the limits $\mu \to 0$ and $b^2\to 0$, see Eq.~(\ref{MPSM}).
In this way we find $M_{PS}L \sim 4.5$, a confortably large value which guarantees
that the vacuum of the system under the (strong interaction driven) spontaneous breaking 
of the $\tilde\chi$ flavour symmetries gets actually polarized by the dominating
effective action terms breaking explicitly these symmetries. For this reason, in the
results of Eqs.~(\ref{MPSM})--(\ref{MAWIM}), which establish the mass generation mechanism at 
$L=2.4$~fm, we can safely ignore the finite size corrections, which also turn out to be  
much smaller than other errors.

\subsection{Studies at larger values of $\rho$ and $v$}
\label{sec:LRHO}

Once a fermion mass term breaking $\tilde\chi_L \times \tilde\chi_R$ symmetry is found 
in the continuum limit of the critical theory in its NG phase, as 
conjectured in Ref.~\cite{Frezzotti:2014wja}, it is 
important to check the behaviour of this mass as a function of $\rho$ and the scalar field
expectation value $v$.

Since as $\rho \to 0$ also $\eta_{cr}$ vanishes~\cite{Frezzotti:2014wja}, at $\rho=0$ 
the $\tilde\chi_L \times \tilde\chi_R$ transformations become exact symmetries of the 
lattice model~(\ref{ACTLAT}) and $m_{AWI}$ vanishes, as well as the infinite volume 
value of $M_{PS}$. One thus expects
that $m_{AWI}$ and $M_{PS}^2$ are increasing functions of $\rho$, and arguments were 
given in Ref.~\cite{Frezzotti:2014wja} according to which for small $\rho$ they should increase 
at least as O($\rho^2$). The main results of this work were produced at $\rho = 1.96$
and $r_0 v_R = 1.207(3)$, for lattice spacings where the combination $\rho b v$, which roughly
speaking plays here a role analogous to the one of the $r$-parameter for Wilson fermions as far as the magnitude of the 
$\tilde\chi_L \times \tilde\chi_R$ breaking is concerned, ranges from 0.5 to 0.7. 

In order to check the $\rho$ dependence of $m_{AWI}$ and $M_{PS}^2$ we have thus  studied
the case of a 1.5 times larger $\rho$ value, namely $\rho = 2.94$, 
for the intermediate lattice spacing ($\beta=5.85$, $a \sim 0.123$~fm) and
for the same scalar sector parameters as specified in Sect.~\ref{Sec:Gauge-Scal}.

The list of simulations we performed at $\rho=2.94$ is given in Tab.~\ref{tab:big_rho}.
We limited ourselves to a few simulations since due to the larger $\rho$ value the CPUtime
effort for inverting the Dirac matrix for similar volumes and twisted mass values increased
by a factor 5 to 10 with respect to the case of $\rho=1.96$.  

\begin{table}[!h]
\begin{center}
\begin{tabular}{|cccc|}
\hline
phase  &  $(L^3 \times T)/b^4$ & $\eta$      &  $b\mu$    \\
\hline
Wigner       &  $16^3 \times 32$ &  -1.765          & 0.0140, 0.0210, 0.0280    \\
&     &    -1.7160        &      0.0210            \\
                          &     &    -1.6670       &      0.0210\\
\hline
NG      &  $20^3 \times 40$ &       -1.8376 & 0.0100\\
    & &  -1.8197    &   0.0050   0.0100    0.0150     \\
 &  &  -1.7808    & 0.0100  \\
 \hline
\end{tabular}
\end{center}
\caption{Simulation parameters at $\rho=2.94$, $\beta=5.85$. We use 480 gauge and scalar paired configurations in the Wigner phase and 60 in the NG phase at each simulation point.}
\label{tab:big_rho}
\end{table}

In the Wigner phase, using the data analysis procedure of Sect.~\ref{sec:analybasic},
we found that changing $\rho$ from $1.96$ to $2.94$ leads to a substantial change
of $\eta_{cr}$, from $-1.207(8)$ to $-1.838(13)$. Notice that the increase of $\eta_{cr}$
is almost perfectly linear in $\rho$, in line with the theoretical remark~\cite{Frezzotti:2014wja} 
that $\eta_{cr}$ is an O($\rho$) quantity, and an odd function of $\rho$.

\begin{figure}[hpt]
\begin{center}
\includegraphics[scale=0.5]{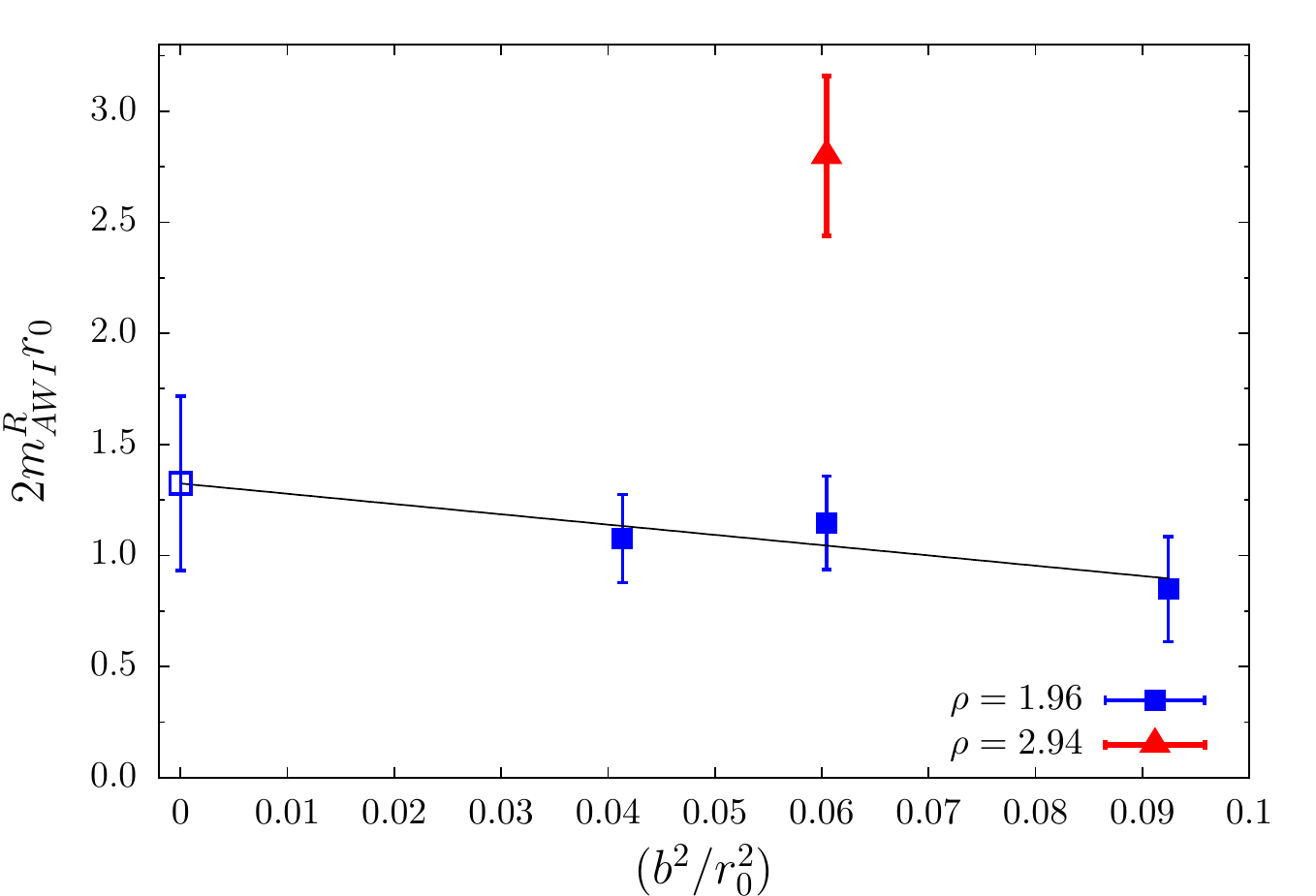}
\end{center}
\vspace{-0.3cm}
\caption{For $\eta=\eta_{cr}(\rho)$ the renormalized ratio $2m_{AWI}^R r_0 \equiv 2r_0 m_{AWI} Z_{\tilde V} Z_P^{-1}$ is plotted versus $(b/r_0)^2$. Open blue squares refer to $\rho=1.96$ and the red triangle to $\rho=2.94$.}
\label{fig:m_AWI_big_rho}
\end{figure}

In the NG phase we observed a significant increase of 
$2r_0m_{AWI} Z_{\tilde A} Z_P^{-1}$ from 
$1.15(0.21)$ to $2.80(0.36)$
(see Fig.~\ref{fig:m_AWI_big_rho}) as well as a similar increase of $M_{PS}^2 r_0^2$. These findings are in nice
agreement with theoretical expectations from Ref.~\cite{Frezzotti:2014wja}. For their interpretation in the
context of a revised concept of universality in more realistic models 
based on the NP mass mechanism under
discussion, see the remarks at the end of 
the section {\em ``Lattice study and results''} and 
in Sect.~\ref{sec:TOPHEN}. 

Due to CPU time limitations we leave for a future study the investigation of the behaviour of $m_{AWI}$ and $M_{PS}^2$ in the critical theory as a function of the scalar field expectation value $v$, which is motivated by the remarks in the main text (section {\em ``Nambu--Goldstone phase and NP anomaly''}) on the dependence of $c_1$ on $\hat\mu_\phi^2$.

\subsection{Summary of numerical results}
\label{sec:SNR}

For completeness we summarize in the Tables~\ref{W_values} (for the Wigner phase) 
and~\ref{NG_values} (for the NG phase) all the simulation data we have obtained, 
which were used to produce the results and the figures presented in the paper. In
the Table~\ref{tab:QCD_values} we also provide the data from the additional simulations
at $\eta=\rho=0$ that were carried out in view of the consistency checks discussed in
the section~\ref{sec:eta0_rho0}. 

\section{$m_{\phi}^2$ independence of $\eta_{cr}$}
\label{INDEP-ETACR} 

The critical value of the bare Yukawa coupling, $\eta_{cr}$, is defined as the value of $\eta$
that solves the equation $\eta -\bar\eta(\eta; g_0^2, \rho, \lambda_0)=0$. The dimensionless
coefficient function $\bar\eta(\eta; g_0^2, \rho, \lambda_0)$ characterizes the UV 
power divergent mixing of the $d=6$ operator ${O}_6^{Li}$ with the $d=4$
density $ \tilde{D}_L^i$. Here $\tilde{D}_L^i$ and ${O}_6^{Li}$ stand for the 
operators that are obtained from the variation of  the ${\cal L}_Y$ and ${\cal L}_W$ 
terms in Eq.~(\ref{ACTLAT}), respectively, under an isotriplet (index $i$) 
$\tilde\chi_L$ transformation. The expression of $\tilde{D}_L^i$ is
given in Eq.~(L6), while the detailed form of ${O}_6^{Li}$ can be found 
in Sect.~III~D of Ref.~\cite{Frezzotti:2014wja}.

In what follows we explain why the dimensionless mixing coefficient $\bar\eta$, and consequently also $\eta_{cr}$, is independent of the scalar squared mass parameter $m_{\phi}^2$, or 
more precisely of its subtracted counterpart $\mu_\phi^2 \equiv m_{\phi}^2-m_{cr}^2$ (we 
neglect here a dimensionless renormalization factor, as it is irrelevant for our arguments). This 
property is a consequence of renormalizability, locality and straightforward dimensional 
considerations. 
We start by observing that in a correlator of renormalized fields with one insertion of 
${O}_6^{Li}$ the {\em leading} UV divergency is given by quadratically divergent Feynman 
diagrams and is insensitive to the shift $\mu_{\phi}^2 \to \mu_{\phi}^2+\delta m^2$. Indeed, 
if we imagine expanding in $\delta m^2$ these diagrams, we see that for dimensional reasons 
the terms in the expansion beyond the first one will be 
necessarily suppressed by at least one relative power of $b^2\delta m^2$. 
Hence they do not contribute to $\bar\eta$.

At the NP level one might worry about the possibility that $\mu_\phi^2$ 
enters $\bar \eta$ scaled by $\Lambda_S^2$. We now prove that the presence of either positive 
or negative powers of the ratio $\mu_{\phi}^2/\Lambda_s^2$ in $\bar \eta$ would violate the 
general principles of renormalizability in Quantum Field Theories, which are assumed 
to hold in our ${\cal L}_{toy}$ model.

1) {\bf Positive mass powers} - 
Terms of the kind $\mu_\phi^2/\Lambda_s^2$, i.e.\ terms displaying inverse powers 
of the RGI scale of the theory, cannot appear in a renormalizable field theory.  
Apart from the fact that this kind of dependence would completely ruin the Feynman diagrams 
power counting, from the argument given above and renormalizability to the NP level in $g_0^2$ 
it follows that corrections $\sim \mu_\phi^2/\Lambda_s^2$ to $\bar\eta$ can never occur. 

In fact, one explicitly checks that any correlator of renormalized fields 
with one insertion of ${O}_6^{Li}$ can be expanded in powers of $\eta$, $\rho$ and
$\lambda_0$ yielding a series where each term corresponds to a diagram consisting  
only of renormalized effective vertices with external legs and/or scalar lines 
sewn with a number of scalar field propagators. 
In all terms of this series the $\mu_\phi^2$ dependence comes only from the scalar
propagators, since the renormalized effective vertices, which by construction here
resum diagrams with gluon and fermion (but no scalar) internal lines, are independent 
of $\mu_\phi^2$, up to immaterial O($b^2$) artifacts. 
Hence, under the shift $\mu_{\phi}^2 \to \mu_{\phi}^2+\delta m^2$, 
a correction to the correlator of the kind $\delta m^2/\Lambda_s^2$ 
does not occur. Indeed, were such a correction present, it could only
arise as the product of a relative O($b^2 \delta m^2$) correction 
from the scalar propagators times an O($b^{-2}\Lambda_S^{-2}$) 
contribution from the $\mu_\phi^2$-independent renormalized 
effective vertices. But, as the renormalized effective vertices are by definition 
UV finite, this is manifestly impossible, thereby ruling out terms of 
the kind $\mu_\phi^2/\Lambda_s^2$ in the correlator.

2) {\bf Negative mass powers} - Terms of the kind $\Lambda_s^2/\mu_\phi^2$, i.e.\ terms displaying inverse powers of mass parameters cannot show up in mixing coefficients. In fact, IR divergencies, revealed in the limit of vanishingly small mass, can only appear in matrix elements and not in the expression of the renormalized operators in terms of the bare ones. This is the consequence of the fact that one can determine the mixing coefficients by only using correlators at non-exceptional momenta that do not develop IR divergencies in the massless limit. 

\begin{table}[hpt]
\begin{center}
\begin{tabular}{|cccc|}
\hline
 $\eta_{cr}^A$ & $\eta_{cr}^B$ & $b^2\mu_\phi^2 $ &  $r_0^2\mu_\phi^2$ \\
\hline
 ~~ -1.145(6) ~~ & ~~ -1.136(6) ~~ & ~~ 0.0504(10) ~~ & ~~ 1.22(3) ~~    \\
\hline
 ~~ -1.138(2) ~~ & ~~ -1.132(2) ~~ & ~~ 0.0171(10) ~~ & ~~ 0.41(3) ~~   \\
\hline
\end{tabular}
\caption{The values of $\eta_{cr}^A$ and $\eta_{cr}^B$ at fixed bare couplings
$g_0^2 = 6/5.95$, $\rho=1.96$, $\lambda_0=0.6022$ for the values of
$\mu_\phi^2 \equiv (m_\phi^2 - m_{cr}^2)$ given in the 3$^{rd}$ (4$^{th}$) column 
in lattice ($r_0$) units.}
\label{tab:eta_cr_mphismall}
\vspace{-0.5cm}
\end{center}
\end{table}

Finally, as a check of consistency, we repeated the numerical computation of 
$\eta_{cr}$ at $\beta=5.95$ reducing the subtracted squared scalar mass, 
$r_0^2 \mu_\phi^2 \equiv r_0^2 (m_\phi^2 - m_{cr}^2)$, from $1.22$ to $0.41$, see 
Table~\ref{tab:eta_cr_mphismall}.
For this smaller value of the subtracted scalar mass we worked at the same 
bare parameters and lattices as employed for the larger one,
see the block of Tab.~\ref{tab:sim_Wigner} corresponding to $\beta=5.95$.
Adopting the same analysis procedure as discussed in Sect.~\ref{sec:analybasic},
we find that the change in the estimated value of $\eta_{cr}$ observed upon
reducing $\mu_\phi^2$ by a factor of three is not significant within 
our statistical errors. Moreover the change in the central $\eta_{cr}$ value
obtained with the two different values of $\mu_\phi^2$ in 
Table~\ref{tab:eta_cr_mphismall} but the same choice of time plateau for $r_{AWI}$ 
is at a given  $\mu_\phi^2$ as small as the central value of $\eta_{cr}^A - \eta_{cr}^B$,
which is certainly an O($b^2$) lattice artifact in 
the computation of $\eta_{cr}$. 
These findings altogether are nicely consistent with 
$\eta_{cr}=\eta_{cr}(g_0^2,\rho,\lambda_0)|_{g_0^2=6/5.95,\rho=1.96,\lambda_0=0.6022}$
being independent of the value of $\mu_\phi^2$ up to
O($b^2$) lattice artifacts.

\vspace*{-0.4cm}
\section{Lattice artifacts ${\rm O}(b^{2n})$}
\label{sec:RADA} 

We show in this section that in the lattice model~(\ref{ACTLAT}) only discretization
errors of order ${\rm O}(b^{2n})$ ($n$ positive integer) can show up in the continuum
limit of renormalized quantities. This property is a consequence of the
exact invariance of the lattice action under the global $\chi_L \times \chi_R$
transformations, the form of which is given by Eqs.~(L3)-(L4),
with $\Omega_L$ and $\Omega_R$ independent SU(2) matrices. 

The proof is based on the observation that for symmetry reasons the Symanzik local 
effective Lagrangian~\cite{Symanzik:1983dc,Symanzik:1983gh,Luscher:1996sc} 
describing the lattice artifacts of the model~(\ref{ACTLAT}), 
with the subtracted scalar squared mass $\mu_\phi^2 \equiv m_\phi^2 - m_{cr}^2$ 
set to an O($b^0$) value, cannot involve operator terms with odd engineering mass 
dimension, thereby taking the form
\begin{equation} 
 L_{\;\rm toy}^{Sym} = L_4 + \sum_{n=1}^{\infty} b^{2n} L_{4+2n} \; .
\label{SYMLAG} \end{equation}
In fact, by properly combining parity ($P$), time reversal ($T$), a finite $\chi_L 
\times \chi_R$ transformation and a finite vector U(1) transformation, which all 
correspond to exact lattice symmetries, one can define a lattice transformation 
which, in analogy with the Wilson Lattice QCD case discussed in 
Ref.~\cite{Frezzotti:2003ni}, we 
call ${\cal D}_d$. The action of ${\cal D}_d$ on the lattice fields reads 
\begin{eqnarray}
& Q(x) \to i Q(x^{PT}) \; , \quad & \bar Q(x) \to i \bar Q(x^{PT}) \; , \nonumber \\
& \Phi(x) \to -\Phi(x^{PT}) \; , \quad & U_\mu(x) \to  U_\mu^\dagger(x^{PT}) \, , 
\label{DdDEF} \end{eqnarray}
where the SU(3) gauge link transformation implies for the gauge 
potentials $A_\mu(x) \to - A_\mu(x^{PT})$, $\mu=0,1,2,3$. 
One can also check that the gauge
covariant derivatives transform so that e.g.\ $\widetilde\nabla_\mu Q(x) 
\to  - i \widetilde\nabla_\mu Q(x^{PT})$, while for the plaquette $U_{\mu\neq\nu}(x)$
we have $U_{\mu\neq\nu}(x) \to U_{\mu\neq\nu}^\dagger(x^{PT})$.

We thus see that all the fields entering the basic lattice action~(\ref{ACTLAT})
under the operation ${\cal D}_d$ get transformed in their counterparts localized 
at the $PT$-reflected coordinate $x^{PT} = -x$ times a phase $(-1)^{d_F}$, 
where $d_F$ is the engineering mass dimension 
of the field $F$, namely $d_A = d_{\Phi} =-1$ for gluons and scalars, $d_Q = d_{\bar Q}
= (-1)^{3/2} = i$ for fermion fields. By the arguments of Ref.~\cite{Luscher:1996sc}, 
the invariance of the lattice action~(\ref{ACTLAT}) under ${\cal D}_d$ implies an 
analogous invariance of the formal Symanzik effective Lagrangian 
$L_{\;\rm toy}^{Sym}$ under the continuum counterpart of ${\cal D}_d$,  
which in turn rules out from it all the local terms with odd engineering mass 
dimension, leading to Eq.~(\ref{SYMLAG}).

By using again the exact ${\cal D}_d$ invariance and the standard arguments of 
Ref.~\cite{Luscher:1996sc} about the lattice artifact corrections to local operators,
it is also clear that the discretization errors on the matrix elements of a local
operator $O(x)$ taken between states at momenta parametrically much smaller than $b^{-1}$
can be fully described in terms of the Symanzik effective Lagrangian $L_{\;\rm toy}^{Sym}$
and the effective local operator $O_0(x) + \sum_{n=1}^{\infty} b^{2n} O_{2n}$. 

The same kind of arguments can be straightforwardly extended to prove that
the power divergent mixings of a local operator can only involve even powers 
of $b^{-1}$.

\vspace*{-0.4cm}
\section{Further consistency checks}
\label{sec:COCHE}

After discussing in Sects.~\ref{sec:data_analysis} to~\ref{sec:RADA}
several technical aspects concerning the numerical 
evidence for the NP mass generation mechanism under investigation, here we present
evidence that no such a phenomenon occurs if the fermionic chiral transformations
$\tilde\chi_L \times \tilde\chi_R$ (see Eqs. (L3)--(L4)) are exact in the UV regulated
lattice model, i.e.\ for the case of $\rho=0$ and $\eta=\eta_{cr}(g_0^2,0,\lambda_0)=0$. 
Since the scalar $\Phi$ field is then completely decoupled, this case actually
corresponds to quenched lattice QCD with na\"ive fermions. 

We also show that in the Wigner phase of the critical model corresponding to
$\rho=1.96$ the pseudoscalar (PS) meson squared mass in the combined chiral 
and continuum limits vanishes within errors, as expected from the arguments 
given in the main text (Sect.~2) and in Ref.~\cite{Frezzotti:2014wja}.

\subsection{The critical model at $\rho=\eta=0$}
\label{sec:eta0_rho0}

In this section we present a study of the model obtained by taking $\eta=0$, $\rho=0$ with the purpose of showing that in the absence of chirally breaking terms the dynamical mass generation phenomenon identified in this work does not take place. Indeed, in this situation we get quenched lattice QCD with the scalars completely decoupled from (na\"ive) fermions and gluons. We thus expect both $M_{PS}$ and $m_{AWI}$ to vanish in the infinite volume limit. 
As in the case of the toy-model~(\ref{ACTLAT}) discussed before, reliable simulations at $\eta=0$, $\rho=0$ require introducing a twisted mass term, like Eq.~(\ref{ACT-TM}), that can be safely removed by taking the limit $\mu\to 0$ at fixed lattice spacing. 


For the sake of this check we will limit our investigation to the finest lattice 
resolution at our disposal, namely $\beta=5.95$, corresponding to $b \simeq 
0.102(6)$~fm, and consider seven (small) values of the twisted quark mass $\mu$. 
Details about simulation parameters (chosen as in the study
of the model with non-zero $\tilde\chi$ symmetry breaking) and
numerical results are given in Table~\ref{tab:QCD_values}.

\begin{figure}[hpt]
\begin{center}
\includegraphics[scale=0.5]{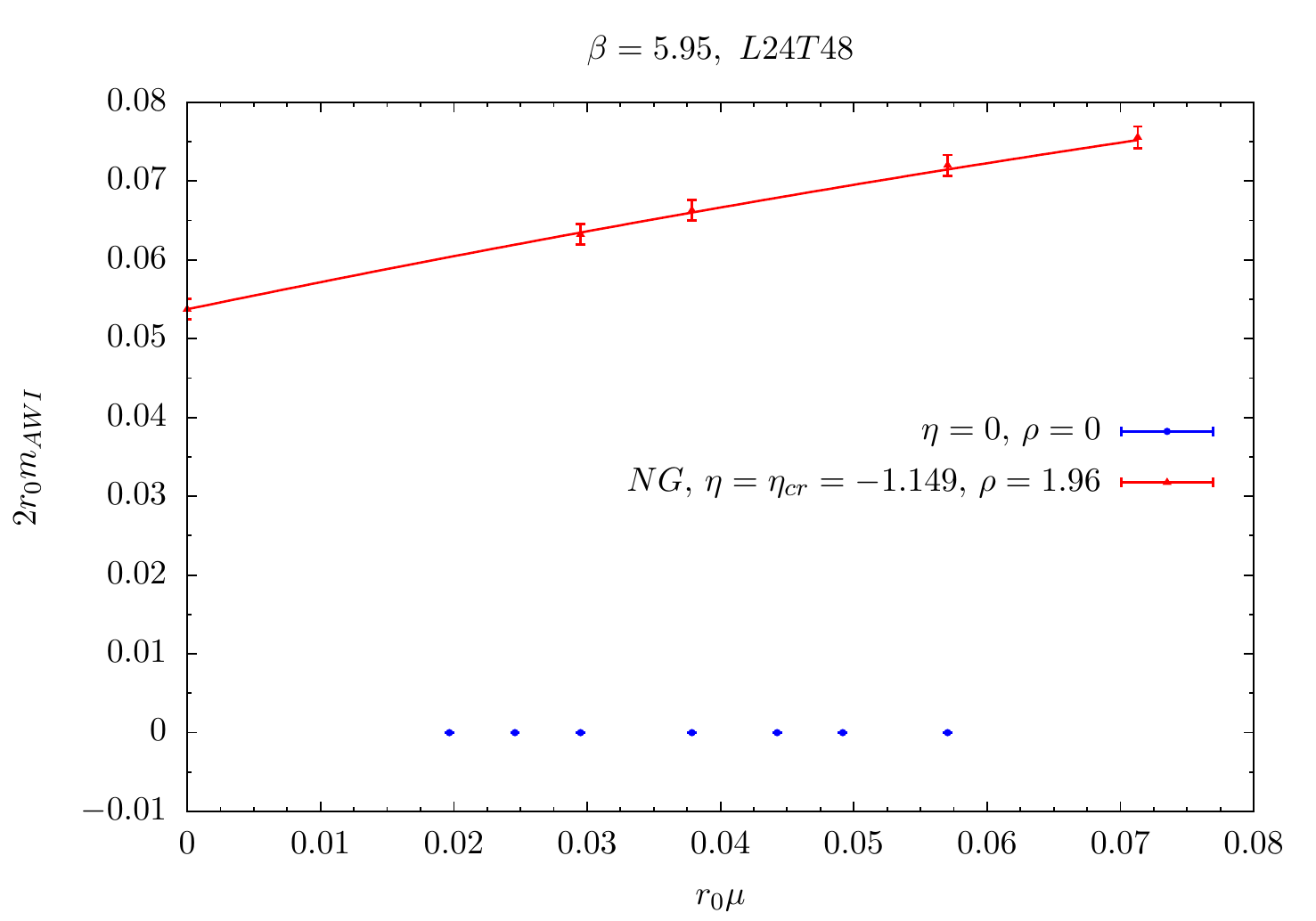}
\end{center}
\vspace{-0.3cm}
\caption{The values of $2 r_0 m_{AWI}$  in the NG phase at $\beta=5.95$ at $\eta=\eta_{cr}$, $\rho=1.96$ (red curve) and their counterparts at $\eta=0$, $\rho=0$ (blue curve). 
In this comparison at fixed lattice spacing we use the unrenormalized quantity $m_{AWI}$.}
\label{fig:m_AWI_eta0_rho0}
\end{figure}

\begin{figure}[hpt]
\begin{center}
\includegraphics[scale=0.5]{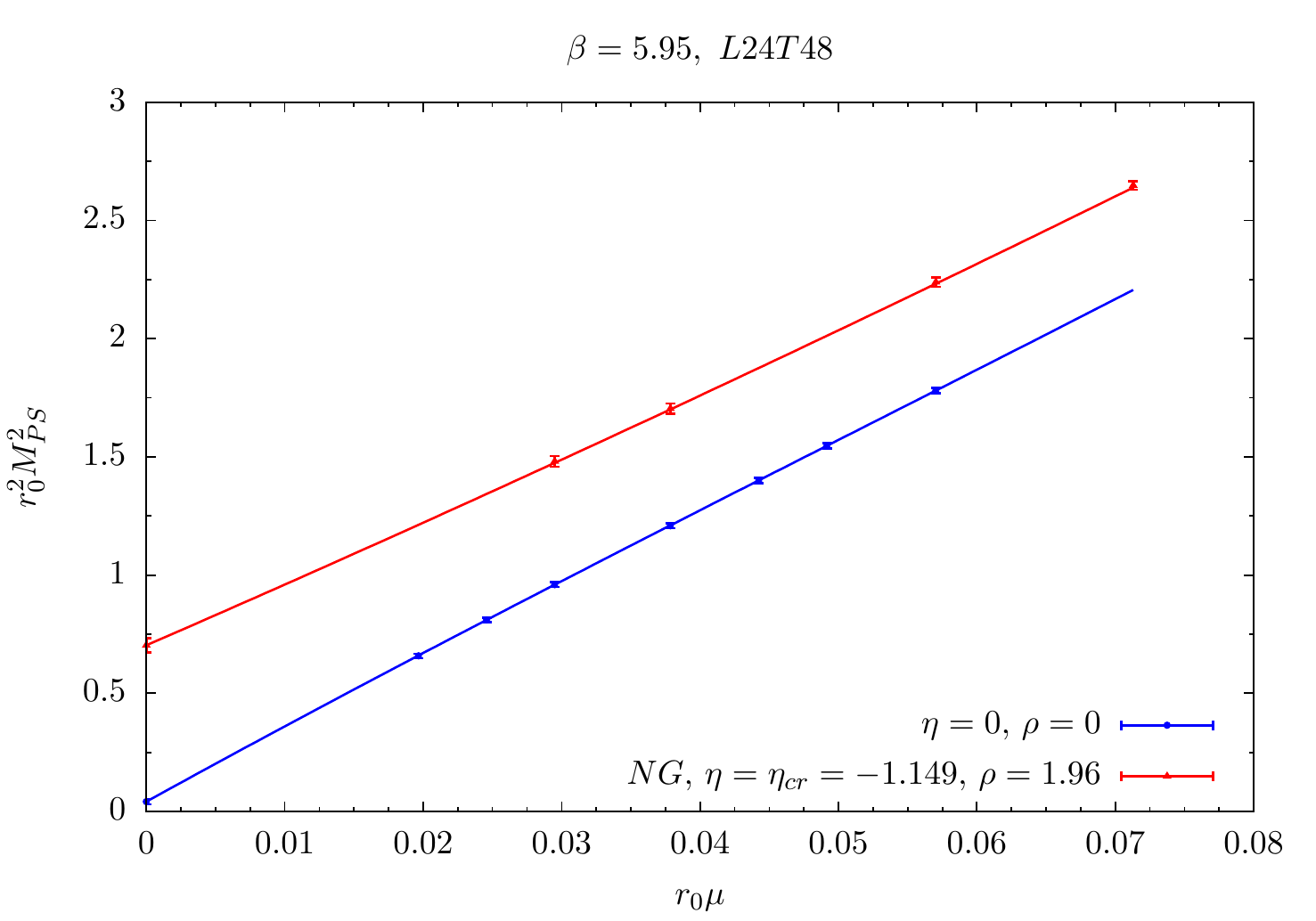}
\end{center}
\vspace{-0.3cm}
\caption{Comparison of the values of $r_0^2M_{PS}^2$ at $\beta=5.95$ in the NG phase at $\eta=\eta_{cr}$ and $\rho=1.96$ (red curve) with what one finds at $\eta=0$, $\rho=0$ (blue curve).}
\label{fig:M_PS_eta0_rho0}
\end{figure}

In Figs.~\ref{fig:m_AWI_eta0_rho0} and~\ref{fig:M_PS_eta0_rho0} we compare the values of $2r_0m_{AWI}$ and $r_0^2M_{PS}^2$, respectively, at $\rho=\eta=0$ with the values obtained for the same lattice resolution ($\beta=5.95$) at $\rho=1.96$ and $\eta=\eta_{cr}$ in the NG phase (Tables~\ref{tab:sim_NG} and~\ref{NG_values}). 

First of all we see that in the case $\eta=0$, $\rho=0$ $m_{AWI}$ vanishes for all values of $b\mu$. This is a somewhat trivial result, however, because actually the twisted mass term~(\ref{ACT-TM}) does not contribute to the SDE associated with the $\tilde \chi_L\times \tilde \chi_R$ transformations~(L11) for $i=1$, from which $m_{AWI}$ is extracted at non-zero $\mu$--values (see Eq.~(\ref{mAWI_latt})). 
More interesting is the case of $M_{PS}^2$. The value of $r_0^2 M_{PS}^2$ at $\rho=\eta=0$ extrapolated to $\mu=0$ is $\sim 20$ times smaller than the value we found at $\eta=\eta_{cr}$, $\rho=1.96$. 

While in the latter case the data from simulations at several $\eta$- and $\mu$-values
at $\beta=5.95$ (Table~\ref{tab:sim_NG}) were exploited and fit according to the 
ansatz $M_{PS}^2=Y_0+Y_1\eta+Y_2\mu+Y_3\eta^2+Y_4\mu^2+Y_5\eta\mu\,$,
for the case of $\eta=\rho=0$ the extrapolation to $\mu=0$ yields
$r_0^2 M_{PS}^2|_{\mu=0;\beta=5.95} = 0.041(11)$ and is performed by fitting our data 
(with $M_{PS}L \geq 4$) to a third order polynomial in the twisted mass parameter $\mu$.
Following several authors~\cite{Butler:1994em,Bernard:1998db,Blum:2000kn} who studied the 
meson spectrum in the quenched approximation with similar lattice sizes and statistics,
we choose for $M_{PS}^2$ as a function of the quark mass $\mu$ a polynomial fit ansatz
that reproduces the LO behaviour expected in Chiral Perturbation Theory and effectively
describes higher order chiral terms and lattice artifacts proportional to $b^2\mu^2\Lambda_S$
or $b^2\mu^3$. We do not try to disentangle quenched and/or standard ``chiral log'' effects,
a task that is beyond the scope of our work and typically requires higher statistics than
we have. For instance, using an effective statistics about 30 times larger (due to more
gauge configurations and bigger lattices) than here the authors of Ref.~\cite{Aoki:1999yr} 
could estimate only with a 50--60\% relative uncertainty the parameter $\delta \sim 0.1$ 
of the anomalous quenched ``chiral log''~\cite{Bernard:1992mk,Sharpe:1992ft} 
behaviour  $M_{PS}^2 \sim (2B\mu)^{1/(1+\delta)}$.

Although lattice symmetries here imply that in infinite volume one should get 
$M_{PS}^2 \to 0$ as $\mu\to 0$, the marginally significant deviation from 
zero that we observe in $r_0^2 M_{PS}^2|_{\mu=0;\beta=5.95}= 0.041(11)$  is likely 
to arise from a combination of statistical uncertainties and small residual 
finite size and/or quenched ``chiral log'' effects. Our result appears anyway
to be well in line with what was obtained (with larger errors on a bit smaller
lattices) in previous studies of quenched twisted mass lattice QCD,
see e.g.\ Refs.~\cite{Jansen:2005gf,Jansen:2005kk}, which for $L=1.6$~fm at fixed (very similar) lattice
spacing find $r_0^2 M_{PS}^2|_{\mu=0;\beta=6} = 0.014(58)$.

Most importantly for the goals of the present work, if one assumes a systematic 
absolute uncertainty of $0.04$ in the extrapolation of $r_0^2 M_{PS}^2$ to $\mu=0$
the resulting error on  $r_0 M_{PS}$ amounts to only $\sim 2.5\%$ of our final result
for $r_0 M_{PS}$ in the NG phase, see Eq.~(\ref{MPSM}),  
and is thus four times below the relative systematic error quoted there. 
The present study of $m_{AWI}$ and $M_{PS}^2$ in the model at $\rho=\eta=0$
provides hence a nice consistency check of our results in the main text and 
Sect.~\ref{sec:data_analysis}.

The findings above are also in agreement with the numerical evidence 
presented in Sect.~\ref{sec:LRHO} that, when the non-zero parameter 
$\rho$ that controls the explicit breaking of the $\tilde\chi$ symmetries is 
increased by a factor 1.5, the NP fermion mass that is dynamically generated
in the NG-phase of the critical model increases by a factor $\sim 2.4(3)$,
i.e.\ nearly like $\rho^2$, as expected from arguments in \cite{Frezzotti:2014wja}.

\begin{figure}[hpt]
\begin{center}
\includegraphics[scale=0.5]{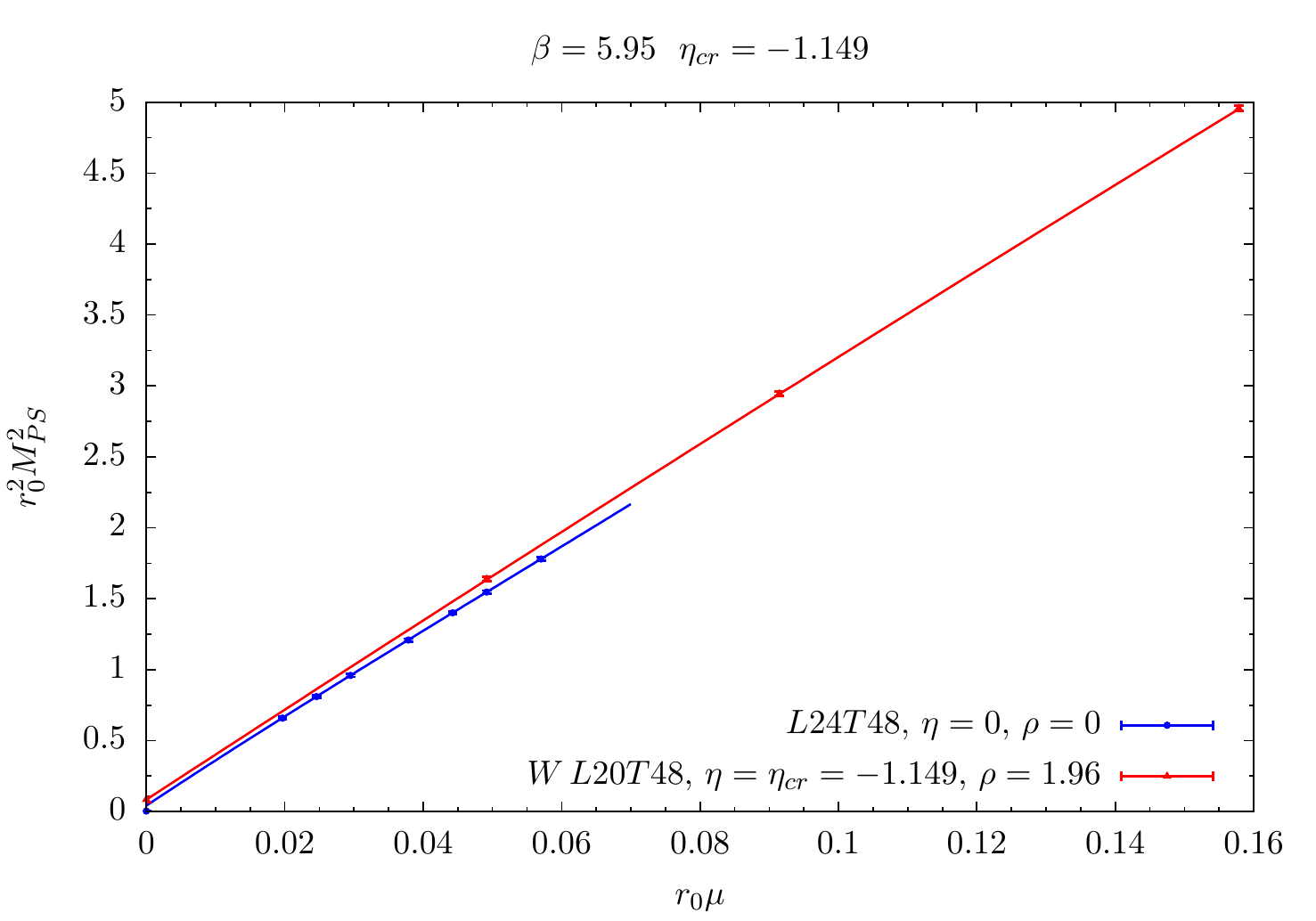}
\end{center}
\vspace{-0.3cm}
\caption{Comparison of the values of $r_0^2M_{PS}^2$ at $\beta=5.95$ in the Wigner phase at $\eta=\eta_{cr}$, $\rho=1.96$ (red curve) with their counterparts at $\eta=0$, $\rho=0$ (blue curve).}
\label{fig:M_PS_eta0_rho0_W}
\end{figure}

It is finally worth comparing the model at $\eta=\rho=0$ with the
model at $\rho=1.96$ and $\eta=\eta_{cr}$ in the Wigner phase. 
These two choices of parameters are expected to describe the same model in the continuum limit, but at fixed lattice spacing they will differ by lattice artifacts. Indeed Fig.~\ref{fig:M_PS_eta0_rho0_W} shows that at $\beta=5.95$ the extrapolated values of $r_0^2M_{PS}^2$ at $\mu=0$ do not significantly differ, while the small misalignment of the blue ($\eta=0$, $\rho=0$) and red curves ($\eta=\eta_{cr}$, $\rho=1.96$) should be ascribed to lattice artifacts and/or small (differences in) finite
 size effects.

\subsection{$M_{PS}$ in the Wigner phase of the critical model}
\label{sec:MPS_Wig}

Here we show evidence that in the Wigner phase of the quenched critical 
model at $\rho=1.96$ the quantity $r_0^2M_{PS}^2$ approaches zero within
errors in the combined continuum and $\mu \to 0$ limits. Our final error
on the extrapolated value of $r_0^2M_{PS}^2$ is not much smaller than it was in 
the quenched twisted mass lattice QCD study of Refs.~\cite{Jansen:2005gf,Jansen:2005kk},
because in the present work the simulations in the Wigner phase are rather expensive 
due to $\Phi$-field induced fluctuations of the fermion propagators and have been 
thus limited to only three or four values of $\mu$ (times three values of $\eta$) 
per each lattice spacing. This was enough for our main goal of
determining $\eta_{cr}$. Although $r_0^2M_{PS}^2$ should vanish as $\mu \to 0$
even at finite lattice spacing, combining data at three different lattice resolutions
effectively increases the statistical information and allows for a better control of 
the $\mu$-dependence in the available PS-meson mass range. 

In this consistency check we employ PS-meson mass data coming from our 
Wigner phase simulations (satisfying $M_{PS}L \geq 5.3$), which for each 
lattice spacing are extrapolated to 
$\eta=\eta_{cr}$ linearly in $(\eta-\eta_{cr})^2$. Even if performing an 
extrapolation linear in $(\eta-\eta_{cr})$ would yield equivalent numerical
results within our present errors, from a theoretical viewpoint in the phase 
with zero vacuum expectation value of $\Phi$ the two--points PS--meson correlator 
and all derived quantities are expected to admit a Taylor expansion in powers of 
$(\eta-\eta_{cr})$ with only terms of even integer degree. 

\begin{figure}[hpt]
\begin{center}
\includegraphics[scale=0.5]{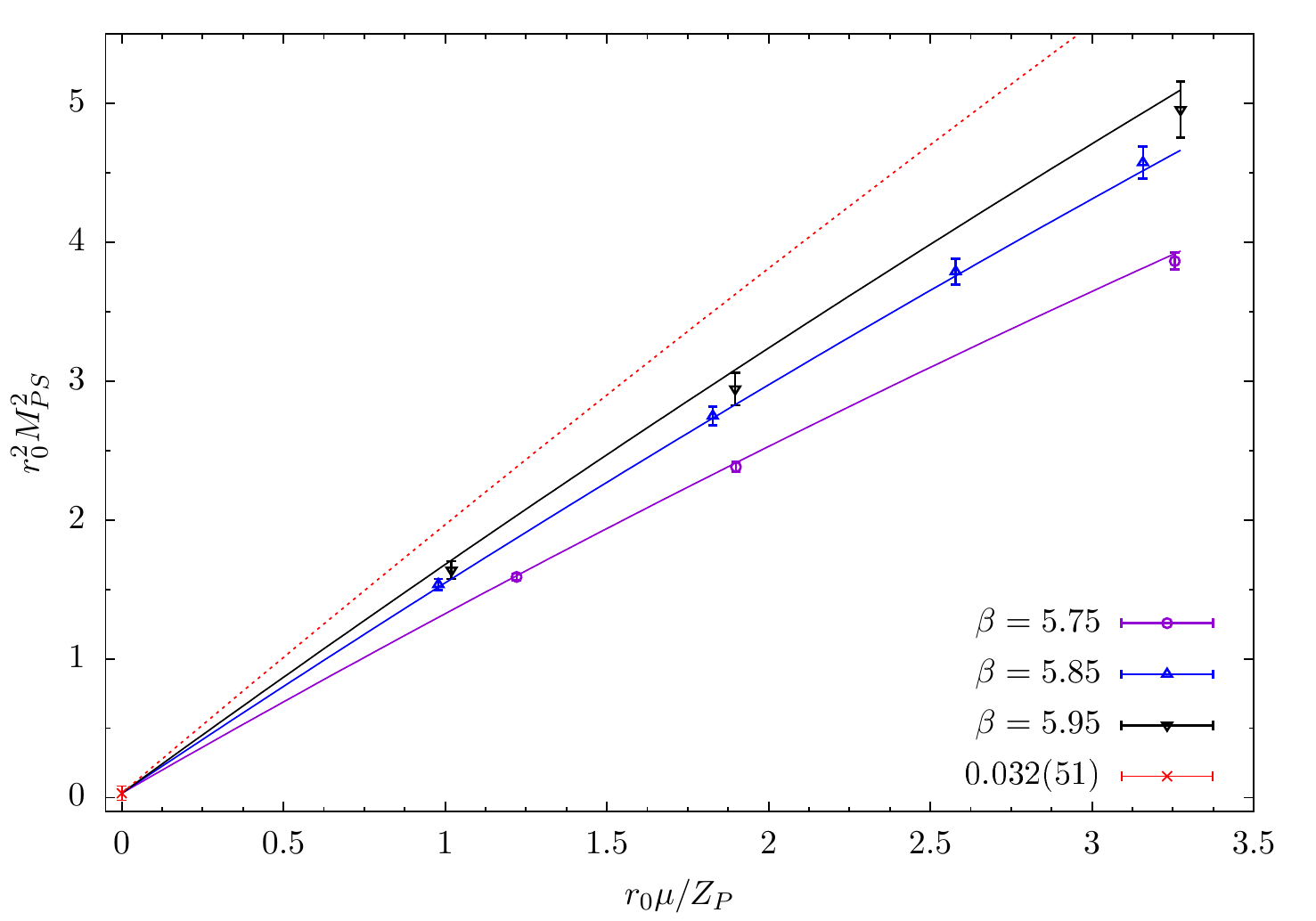}
\end{center}
\vspace{-0.3cm}
\caption{The values of $r_0^2M_{PS}^2$ in the Wigner phase at $\eta=\eta_{cr}$, $\rho=1.96$,
are shown as functions of $\mu_R r_0$ together with the curves at fixed $\beta$ obtained 
from the combined fit (see text) in $b^2$ and $\mu$. The result of the continuum and 
$\mu \to 0$ extrapolation is denoted by a red cross symbol.} 
\label{fig:MPS2_W_allbeta}
\end{figure}

In Fig.~\ref{fig:MPS2_W_allbeta} we plot
the critical model lattice data for $r_0^2 M_{PS}^2$ at the three lattice spacings 
and several values of the twisted mass as a function
of $r_0 \mu_R = r_0 \mu Z_P^{-1}$, where $\mu_R$ is the renormalized twisted mass
and $Z_P^{-1} = G_{PS}^W r_0^2 = \langle 0 | P^1 | {\rm PS~meson} \rangle^W$ (see main
text, Sect.~{\em ``Lattice study and results''}).
The $r_0^2 M_{PS}^2$ data are then fit to the ansatz $K_0 + K_1 \mu_R r_0 + 
\tilde{K}_1 \mu_R b^2 r_0^{-1} + K_2 \mu_R^2 r_0^2$, while the errors
on $bM_{PS}$, $r_0/b$ and $Z_P^{-1}$ at each $\beta$ are propagated taking
correlations into account through a jackknife analysis. 

The resulting best fit describes well the data and is shown for each lattice resolution
(continuous curves) and in the continuum limit (dotted curve), where at $\mu_R =0$
we obtain $r_0^2 M_{PS}^2 = 0.032(51)$. The deviation from zero is statistically
not significant, but does not exclude small finite size effects and/or quenched 
``chiral log'' corrections (see also the discussion in Sect.~\ref{sec:eta0_rho0}). 



\section{Towards a realistic model}
\label{sec:TOPHEN}

We extend here the discussion started in the 
section {\em ``Conclusions and outlook''} of the main text 
about how one can possibly build particle physics models based on the NP mass 
generation mechanism discussed and identified 
in this paper. In order to get a phenomenologically sensible theory one has to extend
the model~(L1) by introducing, besides EW interactions,
also a ``Tera-strong'' force and a set of ``Tera-fermions'' (we take this 
name from Ref.~\cite{Glashow:2005jy}) that
communicate with ordinary matter (quarks or leptons) via strong and/or
EW interactions~\cite{FR18}. The ``Tera-strong'' force should be given
by a new non-Abelian gauge interaction that becomes strong at the scale
$\Lambda_T\gg\Lambda_{QCD}$, where $\Lambda_T$ can be roughly estimated
to be in the few TeV range~\cite{Frezzotti:2014wja,FR18}. 

Models of this type without tree-level flavour changing neutral currents can be
easily formulated, if SU(2)$_L \times$U(1)$_Y$ invariant EW interaction terms are 
introduced in the same way as in the SM, e.g.\ by coupling SU(2)$_L$ gauge bosons
only to left-handed Tera-fermions arranged in doublets of definite hypercharge.
By then taking all right-handed Tera-fermions to be 
weak isospin singlets with given hypercharge
the usual relation $Q_{em} = T_3 + Y/2$
between electric charge $Q_{em}$, third weak isospin 
component $T_3$ and hypercharge $Y$ is always fulfilled. Owing to the sizeable NP 
mass of the ``Tera-fermions'', the lightest Tera-mesons are expected to be rather heavy, 
i.e.\ with a mass around 2-3 times $\Lambda_T$, implying that these extended models 
have a good chance of passing the EW precision tests, in particular those concerning 
the $S$-parameter~\cite{Peskin:1990zt,Peskin:1991sw}.
By treating the chiral SU(2)$_L \times$U(1)$_Y$ interactions in perturbation theory and 
carrying out unquenched lattice simulation of models including QCD and ``Tera-
strong'' gauge forces, it is in principle possible to relate with controllable errors 
the NP mass of the top quark to the mass of the lightest ``Tera-mesons'', which should be 
seen as resonances in accelerator experiments at sufficiently high energy.
 
Lattice methods can also be useful to check and possibly establish
the existence in the $WW+ZZ$ channel of one composite Higgs state with mass 
$M_h \sim 125$~GeV, which is expected to form owing to the interactions 
of weak gauge bosons with ``Tera-fermions'' that
are subjected to EW and ``Tera-strong'' interactions.
For a first step in this direction see Ref.~\cite{Romero-Lopez:2018zyy}. 
If the existence of such a composite Higgs state is confirmed, then 
perturbative unitarity of the (longitudinal) $WW$ scattering amplitude at 
momentum scales well below $\Lambda_T$, for which the Higgs state plays
a crucial role~\cite{Lee:1977eg}, would make very plausible that the 
Higgs couplings are close to the present experimental findings and deviate
from their SM value only by small O($p^2/\Lambda_T^2$) effects, with $p^2$ 
standing for $M_h^2$ or the typical momentum scale of the process where the 
Higgs couplings are measured.

A striking and interesting feature of our mass generation mechanism is the observed
dependence of low energy physical quantities on the parameters of the $d \geq 6$ 
$\tilde\chi$ breaking Lagrangian terms. To all orders in the perturbative expansion
in the gauge coupling these
terms would represent irrelevant details of the UV completion of the critical model.
But at the NP level, in the NG phase of the critical theory, the $d \geq 6$
$\tilde\chi$ breaking operators appearing in the UV regulated classical Lagrangian
produce relevant O($\Lambda_{UV}^0$) effects, because the symmetries of 
our model allow for the occurrence in the EL of RGI finite NP terms, such
as those with coefficients $c_1$ and $c_2$ in Eq.~(L13). As first
pointed out in Ref.~\cite{Frezzotti:2014wja} and discussed in the section 
{\em ``Nambu--Goldstone phase and NP anomaly''}, 
this dynamical phenomenon is due to a subtle interplay 
between the residual O($b^2$) terms violating $\tilde\chi_L \times \tilde\chi_R$ and 
the strong interactions, which are responsible for spontaneous breaking of the 
approximate $\tilde\chi_L \times \tilde\chi_R$ symmetries. Of course no such a
phenomenon would be observed if the critical model were defined in a UV 
regularization preserving exactly $\tilde\chi_L \times \tilde\chi_R$ invariance,
e.g. at $\rho = 0$ and $\eta=\eta_{cr}|_{\rho = 0}=0$. This remark, which is 
indeed consistent with our numerical finding that $c_1$ increases upon increasing 
$\rho$, see Sect.~\ref{sec:data_analysis}~C, also implies that non-Abelian gauge 
models including fermions and scalars in general fall into different universality 
classes depending on whether the UV regularization breaks or possibly preserves
the purely fermionic chiral symmetries. 

In other words, the ``NP anomaly'' to the recovery of broken fermionic chiral
symmetries that we demonstrated in our work appears to require a non-trivial 
revision of the  concept of universality, as compared to its standard formulation 
which can be proven to all orders of perturbation theory (only). This important
point certainly deserves a more thorough investigation.  
At the moment it seems to us conceivable that 
the kind of universality that is realized in the context of maximal 
$\tilde\chi$ restoration carries over also to the NP mass terms in the EL~\cite{FR18}.
In this new NP framework universality should thus mean that a set of UV complete models 
endowed with different $\tilde\chi$ breaking Lagrangian terms lead to the same 
low energy observables, once the conditions of maximal $\tilde\chi$ restoration 
are fulfilled at correspondingly 
different critical values of the relevant bare parameters. The low energy physics
should thus be controlled only by the gauge couplings, the number of
fermions and their transformation properties under the gauge groups,
plus possibly a finite number of dimensionless coefficients parametrizing
the relative strength of the $\tilde\chi$ breaking terms for the different fermion 
generations and having a magnitude limited by the conditions of maximal 
$\tilde\chi_L$ restoration.

We finally note that a beyond the SM model supporting the 
mass generation mechanism we have discussed and endowed with a reasonable 
content of (standard and Tera) matter fields, if a suitable, non standard 
choice of hypercharges for Tera-fermions is made,
can yield unification of EW and strong
(and perhaps even Tera-strong) gauge couplings~\cite{Frezzotti:2016bes} 
at an energy scale around $10^{18}$~GeV and to a precision level that is
as good as for the unification achieved in the Minimal SUSY Model.


\begin{table*}[]\centering\small
\setlength\extrarowheight{1.2pt}
\setlength{\tabcolsep}{10pt}
\begin{tabular}{cccccc}
\hline
$b\mu$ & $\eta$ &  $r_{AWI}^A$    &  $r_{AWI}^B$ &   $b^2G_{PS}^W$ & $b^2M_{PS}^2$ \\ 
\hline
\multicolumn{6}{c}{$\beta=5.75$,\,
$\lambda_0=0.5807$,\, $b^2m_{\phi}^2=-0.4150$,\, 
 $\rho=1.96$,\, $L/b = 16$, $T/b=32$}\\
\hline
0.018  & -1.1505   &  -0.1108(60)   & -0.0748(44)    & 2.085(58) & 0.14672(86)\\    
0.028  & -1.1505   &  -0.1063(64) &   -0.0631(45)    & 2.188(56) & 0.21952(87)\\        
0.048  & -1.1505  &  -0.0998(79)  &   -0.0410(60)    & 2.376(53) & 0.35619(90) \\       
0.018  & -1.1898  &  -0.0718(59)  &   -0.0355(44)    & 2.095(59) & 0.14710(81)\\        
0.028  & -1.1898  &  -0.0675(64)  &   -0.0233(45)    & 2.196(56) & 0.21999(82)\\        
0.048  & -1.1898  &  -0.0609(79)  &   -0.0015(61)    & 2.382(53) & 0.35671(87)\\           
0.018  & -1.3668  &   0.1017(61)  &    0.1389(44)    & 2.065(58) & 0.14808(85)\\         
0.028  & -1.3668  &   0.1063(66)  &    0.1537(49)    & 2.176(56) & 0.22072(87)\\ 
0.048  & -1.3668  &   0.1137(81)  &    0.1757(70)    & 2.373(53) & 0.35730(96)\\ 
\hline
\multicolumn{6}{c}{$\beta=5.85$, 
$\lambda_0=0.5917$, $b^2m_{\phi}^2=-0.4615$  
 $\rho=1.96$,\, $L/b = 16$, $T/b=40$}\\
\hline
 0.0224 & -1.0983  & -0.1090(65) & -0.0613(48)   & 1.482(34)& 0.1652(11) \\  
0.0316  & -1.0983  & -0.1078(75) & -0.0509(57)   & 1.597(32)& 0.2280(11) \\  
0.0387  & -1.0983  & -0.1063(82) & -0.0432(63)   & 1.684(31)& 0.2755(11) \\  
0.012   & -1.1375  & -0.0693(56) & -0.0372(40)   & 1.364(38)& 0.0924(09) \\
0.0224 & -1.1375   & -0.0703(66) & -0.0238(48)   & 1.491(34)& 0.1658(11) \\  
0.0316 & -1.1375   & -0.0690(75) & -0.0132(57)   & 1.605(32)& 0.2287(11) \\  
0.0387 & -1.1375   & -0.0674(82) & -0.0055(63)   & 1.692(32)& 0.2762(11) \\  
0.0224 & -1.2944   &  0.0849(74) &  0.1266(52)   & 1.483(35)& 0.1664(12) \\  
0.0387 & -1.2944   &  0.0879(86) &  0.1456(65)   & 1.690(34)& 0.2766(13) \\   
\hline
\multicolumn{6}{c}{$\beta=5.95$,\, 
$\lambda_0=0.6022$,\, $b^2m_{\phi}^2=-0.4956$,\, 
 $\rho=1.96$,\, $L/b = 20$, $T/b=48$}\\
\hline
0.0186 &  -0.9761  &  -0.1681(41) & -0.1377(37)& 1.027(37)  & 0.11941(74) \\
0.0321 &  -0.9761  &  -0.1671(45) & -0.1255(35)& 1.181(36)  & 0.20216(82) \\\
0.0186 &  -1.0354  &  -0.1094(37) & -0.0798(36)& 1.044(36)  & 0.12033(68) \\\
0.0321 &  -1.0354  &  -0.1087(42) & -0.0676(34)& 1.194(35)  & 0.20340(77) \\
0.01   &  -1.0771  &  -0.0675(35) & -0.0478(36)& 0.955(37)  & 0.06673(65) \\
0.0186 &  -1.0771  &  -0.0680(36) & -0.0390(36)& 1.051(36)  & 0.12072(66) \\
0.0321 &  -1.0771  &  -0.0677(41) & -0.0268(35)& 1.200(35)  & 0.20396(75) \\
\hline
\multicolumn{6}{c}{$\beta=5.95$,\, 
$\lambda_0=0.6022$,\, $b^2m_{\phi}^2=-0.5289$,\, 
 $\rho=1.96$,\, $L/b = 20$, $T/b=48$}\\
 \hline
0.0186 & -0.9761  &-0.1572(18) & -0.1392(16)    & 1.005(033)     & 0.11813(91) \\    
0.0321 & -0.9761  &-0.1561(24) & -0.1302(17)    & 1.170(033)     & 0.20087(99) \\  
0.0186 & -1.0354  &-0.0996(18) & -0.0811(15)    & 1.034(034)     & 0.11957(76) \\  
0.0321 & -1.0354  &-0.0983(24) & -0.0722(17)    & 1.191(034)     & 0.20276(85) \\
  0.01 & -1.0771  &-0.0592(14) & -0.0465(16)    & 0.946(035)     & 0.06619(65)   \\   
0.0186 & -1.0771  &-0.0596(19) & -0.0401(15)    & 1.046(035)     & 0.12019(71) \\
0.0321 & -1.0771  &-0.0581(24) & -0.0316(19)    & 1.199(034)     & 0.20360(82) \\
\hline\hline
\multicolumn{6}{c}{$\beta=5.85$, 
$\lambda_0=0.5917$, $b^2m_{\phi}^2=-0.4615$,  
 $\rho=2.94$,\, $L/b = 16$, $T/b= 40$}
 \\\hline
0.021 & -1.667     & -0.1527(100)& -0.0861(73)& 1.587(36)  & 0.16168(100) \\   
0.021 & -1.716     & -0.1073(100)& -0.0417(74)& 1.604(37)  & 0.16264(97) \\ 
0.014 & -1.765      & -0.0634(100)& -0.0160(70)& 1.518(40)  & 0.11239(88) \\  
0.021 & -1.765     & -0.0622(110) &  0.0026(77)& 1.615(37)  & 0.16601(80)\\ 
0.028 & -1.765     & -0.0598(100) &  0.0168(89)& 1.709(36)  & 0.21274(100) \\
\hline
\end{tabular}
\caption{The values of $r_{AWI}^{A(B)}$, $b^2G_{PS}$ and $b^2M_{PS}^2$ measured 
in the Wigner phase at $\rho=1.96$ and $\rho=2.94$.  
The values of $r_{AWI}$ in the time plateau region $A(B)$ are denoted by $r_{AWI}^{A(B)}$
(see Fig.~\ref{fig:fig_r_AWI_A_B} and Sect.~\ref{sec:analybasic}).
We recall $Z_P^{-1} = [G_{PS}^W r_0^2]_{\mu\to 0}^{\eta \to \eta_{cr}}$.}
\label{W_values}
\end{table*}


\begin{table*}[]\centering\small
\setlength\extrarowheight{1.2pt}
\setlength{\tabcolsep}{10pt}
\begin{tabular}{ccccc}
\hline
$b\mu$ & $\eta$ &  $b^2M_{PS}^2$  & $2bm_{AWI}$ & $Z_V$  \\
\hline 
\multicolumn{5}{c}{$\beta=5.75$,\,
$\lambda_0=0.5807$,\, $b^2m_{\phi}^2=-0.5941$,\, 
  $\rho=1.96$, \, $L/b = 16$, $T/b=40$}\\
\hline
0.005  & -1.2715   & 0.0882(22)& 0.01738(41)&  0.9904(240)  \\
0.0087 & -1.2715   & 0.1214(15)& 0.02206(34)&  0.9752(93)  \\
0.0131 & -1.2715   & 0.1561(13)& 0.02611(32)&  0.9764(56)  \\
0.0183 & -1.2715   & 0.1945(12)& 0.02989(32)&  0.9791(43)  \\
0.0277 & -1.2715   & 0.2600(12)& 0.03541(33)&  0.9834(33)  \\
0.0131 & -1.2657   & 0.1489(12)& 0.02359(31)&  0.9735(52)  \\
0.0183 & -1.2540   & 0.1757(10)& 0.02212(29)&  0.9730(35)  \\
0.0131 & -1.2404   & 0.1244(09)& 0.01357(26)&  0.9630(38) \\
0.0131 & -1.2277   & 0.1161(08)& 0.00894(24)&  0.9592(33)  \\
0.0183 & -1.2277   & 0.1557(08)& 0.00894(24)&  0.9658(26)  \\
\hline
\multicolumn{5}{c}{$\beta=5.85$,\,
$\lambda_0=0.5917$,\, $b^2m_{\phi}^2=-0.5805$\, 
  $\rho=1.96$, \,  $L/b = 20$, $T/b=40$}\\
  \hline
0.004  & -1.2106   & 0.0851(20)& 0.01894(44) &  0.9420(190)  \\
0.007  & -1.2106   & 0.1032(17)& 0.02099(39) &  0.9518(100) \ \\
0.01   & -1.2106   & 0.1220(16)& 0.02271(39) &  0.9577(70)  \\
0.012  & -1.2106   & 0.1348(15)& 0.02371(39) &  0.9609(57)  \\
0.004  & -1.2069   & 0.0785(19)& 0.01717(40) &  0.9422(170)  \\
0.007  & -1.2069   & 0.0974(17)& 0.01930(37) &  0.9517(95) \\
0.01   & -1.2069   & 0.1169(15)& 0.02104(37) &  0.9577(65)  \\
0.0224 & -1.2067   & 0.1981(13)& 0.02609(39) &  0.9701(30)  \\
0.004  & -1.2028   & 0.0717(18)& 0.01529(36) &  0.9420(150)  \\
0.007  & -1.2028   & 0.0914(16)& 0.01749(35) &  0.9517(87) \\
0.01   & -1.2028   & 0.1115(14)& 0.01924(36) &  0.9577(61) \\
0.012  & -1.2028   & 0.1250(14)& 0.02024(36) &  0.9607(50)  \\
0.0224 & -1.2028   & 0.1943(13)& 0.02432(38) &  0.9697(29) \\
0.01   & -1.1950   & 0.1022(13)& 0.01591(33) &  0.9578(52)  \\
0.007  & -1.1777   & 0.0636(10)& 0.00756(25) &  0.9538(56)  \\
0.01   & -1.1777   & 0.0861(10)& 0.00896(28) &  0.9580(39)\\
0.014  & -1.1777   & 0.1153(10)& 0.01048(30) &  0.9620(28)  \\
0.0224 & -1.1777   & 0.1746(10)& 0.01309(32) &  0.9674(20)  \\
0.0316 & -1.1677   & 0.2325(10)& 0.01103(33) &  0.9706(15) \\
\hline
\multicolumn{5}{c}{$\beta=5.95$,\,
$\lambda_0=0.6022$,\, $b^2m_{\phi}^2=-0.5756$,\,
  $\rho=1.96$,\, $L/b = 24$, $T/b=48$}\\
  \hline
0.006  & -1.1475   & 0.0615(11) & 0.01380(33)&  0.9725(110)   \\
0.0077 & -1.1475   & 0.0705(11) & 0.01444(32)&  0.9729(84) \\
0.0116 & -1.1475   & 0.0922(11) & 0.01565(31)&  0.9741(54) \\
0.0145 & -1.1475   & 0.1089(11) & 0.01640(30)&  0.9749(43)  \\
0.0185 & -1.1475   & 0.1325(11) & 0.01732(30)&  0.9757(34)  \\
0.006  & -1.1450   & 0.0589(11) & 0.01279(32)&  0.9713(110)   \\
0.0077 & -1.1450   & 0.0681(11) & 0.01344(31)&  0.9720(80)  \\
0.0116 & -1.1450   & 0.0902(11) & 0.01463(30)&  0.9735(51)  \\
0.0145 & -1.1450   & 0.1071(11) & 0.01538(30)&  0.9744(41)  \\
0.0077 & -1.1216   & 0.0522(09) & 0.00462(23)&  0.9671(47)\\
0.0108 & -1.1134   & 0.0691(09) & 0.00224(21)&  0.9669(34)\\
0.0077 & -1.1134   & 0.0496(08) & 0.00166(21)&  0.9642(44) \\
\hline\hline
\multicolumn{5}{c}{$\beta=5.85$,\
$\lambda_0=0.5917$,\ $b^2m_{\phi}^2=-0.5805$,\ 
  $\rho=2.94$, \, $L/b = 20$, $T/b=40$}\\
  \hline
0.01 & -1.8376	   & 0.2044(30)& 0.04219(65)&  0.922(17) \\ 
0.005& -1.8197	   & 0.1512(33)& 0.03254(69)&  0.847(35) \\ 
0.01 & -1.8197	   & 0.1785(28)& 0.03560(59)&  0.916(16) \\ 
0.015& -1.8197	   & 0.2085(25)& 0.03869(55)&  0.942(10) \\ 
0.01 & -1.7808	   & 0.1292(22)& 0.02232(48)&  0.915(12) \\ 
\hline
\end{tabular}
\caption{The values of $b^2M_{PS}^2$, $2 b m_{AWI}$ and $Z_V$ measured in the NG phase at $\rho=1.96$ and $\rho=2.94$.}
\label{NG_values}
\end{table*}


\begin{table*}[]\centering\small
\setlength\extrarowheight{1.2pt}
\setlength{\tabcolsep}{10pt}
\begin{tabular}{ccccc}
\hline
$b\mu$ & $\eta$ &  $b^2M_{PS}^2$  & $2bm_{AWI}$ & $1-Z_{\tilde{V}}$  \\
\hline
\multicolumn{5}{c}{$\beta=5.95$,\,
$\lambda_0=0.6022$,\, $b^2m_{\phi}^2=-0.5756$,\,
  $\rho=0$,\, $L/b=24$, $T/b=48$}\\
\hline
0.0040  &  0   & 0.02729(39) & -18(6)$\times10^{-12}$& 1(13)$\times10^{-12}$\\
0.0050  & 0    & 0.03352(41) & -2(8)$\times10^{-12}$ & 15(14)$\times10^{-12}$\\
0.0060  &  0   & 0.03967(42) & -3(6)$\times10^{-12}$ & 10(12)$\times10^{-12}$\\
0.0077  &  0   & 0.05005(45) & 2(9)$\times10^{-12}$ &  -1(11)$\times10^{-12}$\\
0.0090  &  0   & 0.05794(47) & -6(7)$\times10^{-10}$&  -7(7)$ \times10^{-9}$\\
0.0100  &  0   & 0.06399(48) & -4(7)$\times10^{-10}$&  18(13)$\times10^{-12}$\\
0.0116  &  0   & 0.07363(49) & -6(8)$\times10^{-12}$& 14(10)$\times10^{-12}$\\
\hline
\end{tabular}
\caption{The values of $b^2M_{PS}^2$, $2 b m_{AWI}$ and $Z_{\tilde{V}}$ measured at 
$\rho=0$ and $\eta=0$. Boundary conditions were the same as in the study 
of the model with non-zero $\rho$ and $\eta$.}
\label{tab:QCD_values}
\end{table*}


\bibliography{bibliography}